\documentclass[letterpaper,english,preprint,aps,nofootinbib]{revtex4-1}
\usepackage[T1]{fontenc}
\usepackage[utf8]{inputenc}
\usepackage{amsmath}
\usepackage{amssymb}
\usepackage{xcolor}
\usepackage{graphicx,subfig}

\makeatletter

\pdfpageheight\paperheight
\pdfpagewidth\paperwidth

\providecommand{\LyX}{L\kern-.1667em\lower.25em\hbox{Y}\kern-.125emX\@}

\makeatother

\usepackage{babel}
\begin{document}

\title{Holographic boiling and generalized thermodynamic description beyond
local equilibrium}

\author{Xin Li}
\affiliation{School of Physical Sciences, University of Chinese Academy of Sciences,
Beijing 100049, China}
\affiliation{School of Science, Kunming University of Science and Technology, Kunming 650500, China}

\author{Zhang-Yu Nie}
\email{Corresponding Author (E-mail address: niezy@kmust.edu.cn)}
\affiliation{School of Science, Kunming University of Science and Technology, Kunming 650500, China}

\author{Yu Tian}
\email{Corresponding Author (E-mail address: ytian@ucas.ac.cn)}
\affiliation{School of Physical Sciences, University of Chinese Academy of Sciences,
Beijing 100049, China}
\affiliation{Center for Theoretical Physics, Massachusetts Institute of Technology, Cambridge, MA 02139, USA}
\affiliation{Institute of Theoretical Physics, Chinese Academy of Sciences, Beijing
100190, China}

\date{\today}

\begin{abstract}
Tuning a very simple two-component holographic superfluid model, we
can have a first order phase transition between two superfluid phases in the probe limit.
Inspired by the potential landscape discussion, an intuitive physical picture for systems with first order phase transitions is provided. We stress that holography perfectly offers a generalized thermodynamic
description of certain strongly coupled systems even out of local equilibrium, which enables us to carefully study domain wall structures of the
system under first order phase transitions, either static or in real time dynamics.
We numerically construct the 1D domain wall configuration and compute the surface tension of the domain wall from its generalized grand potential. {We also numerically simulate the real time dynamics of a 2D bubble nucleation process (holographic boiling).} The surface tension of the 1D domain wall nicely matches the final state of the 2D bubble nucleation process when the bubble radius is large enough.
\end{abstract}



\maketitle

\section{\label{sec:level1}Introduction}

The gauge/gravity duality\cite{Maldacena:1997re,Gubser:1998bc,Witten:1998qj} provides a useful tool to investigate the strongly coupled field theory. Since the success of holographic modeling of superconductor phase transition\cite{Gubser:2008px,Hartnoll:2008vx}, this duality has been applied to various condensed matter systems\cite{Zaanen:2015oix}. Besides realization of static solutions, it is proved to be even more powerful in simulating non-equilibrium processes (see, e.g. \cite{ACL,CGL,DNTZ,LTZ,Liu,GKLTZ,rotate,XZZNTL,YXZZ} as an incomplete list and \cite{Sonner} for a review).

First order phase transitions are not only quite common in our daily life, but are also important phenomena in superfluids such as Helium-3\cite{Volhardt-Wolfle-1990} and condensed matter systems with multi-condensates\cite{Kuklov:2004prl,Liu:2017prl,Zhang:2020jpa}. First order phase transitions from normal phase to superconductor phase have also been realized holographically in Refs.~\cite{Franco:2009yz,Ammon:2009xh}. An important feature accompanied by first order phase transitions is phase separation which usually implies the creation and evolution of bubbles and domain walls. {Such structures also play an important role in the evolution of the early universe and in gravitational wave physics\cite{Hawking:1982ga,Kosowsky:1992vn,Cai:2017tmh}.}

However, it is quite difficult to accurately describe first order phase transitions, because dynamical processes of first order phase transitions are usually far from (local) equilibrium. Such a dynamical process in quantum many body systems is extremely complicated in general, so certain theoretical description of a general dynamical process would be very important and helpful. Such kind of efforts for holographic systems are made recently in \cite{TWZ}, where some generalized thermodynamic (hydrodynamic) description beyond local equilibrium turns out to be available and a generalized grand potential (free energy) can be well defined accordingly. We then would like to see some concrete application of this description in, e.g. superfluid systems with first order phase transitions mentioned above, where it can be verified and hopefully even developed further.

In a recent paper \cite{Janik:2017ykj}, the authors realized a dynamic process of domain wall formation holographically. There are also some other progresses in this topic\cite{Attems:2019yqn,Bellantuono:2019wbn}. However, in these studies the holographic systems are effectively 1D, so more realistic structures like bubbles (at least 2D) cannot be considered. In order to study the more realistic and more complicated structures in first order phase transition, it is wise to consider a holographic model with first order phase transition in the probe limit, where back-reaction of the matter field to the bulk space-time is neglected and complex numerical relativity problems can be avoided. Such a system with first order phase transition in the probe limit can be easily built from a holographic model with two competing orders, as in Ref.~\cite{Nishida:2014lta}.

In this paper, we study the nontrivial configurations and dynamical processes in the first order phase transition between superfluid phases holographically. In fact, our model here is {probably} the simplest for holographic study of first order phase transitions, and it turns out that essential physics can already be observed quite well in our system. We first realize the first order phase transition between two s-wave superfluid phases holographically, and give a concrete picture to explain how the first order phase transition occurs from a ``grand potential landscape'' point of view.

This grand potential landscape point of view is supposed to be true not only in the holographic context, but also in general thermodynamics, offering an intuitive picture of the stability of phases and the phase structures. Actually, in thermodynamic systems with first order phase transitions (like the van der Waals gas-liquid system), as well as in many holographic models including ours here, there are typically three equilibrium states involved, the stable one, the meta-stable one and the unstable one. It can be argued that the unstable state is a saddle point in the landscape and acts as the minimal potential barrier between the stable and meta-stable states, which will also be verified in our discussions of inhomogeneous configurations (mixture states of different phases).

For the inhomogeneous configurations with local structures like domain wall, which go beyond local equilibrium, we can use the grand potential of the unstable state to estimate the maximum of the grand potential density of the domain wall (bubble) structures, from the grand potential landscape point of view. A more quantitative discussion of such configurations out of local equilibrium needs to be based on the generalized thermodynamic description introduced in \cite{TWZ}, where it turns out that the potential (including the generalized free energy and grand potential) landscape point of view is also crucial. To show the practicability of this point of view and the generalized thermodynamic description more concretely, we numerically build some typical inhomogeneous configurations.
In the numerical work, we first construct the domain wall between the two superfluid phases in a simple 1D setup of our model. We can calculate the surface tension of the domain wall from its generalized grand potential. We then study the dynamical process of bubble nucleation after a local quench to a homogeneous meta-stable state in 2D and find that the surface tension with ``generalized balance conditions'' for the bubble configuration matches well with the above calculation for the 1D domain wall configuration.

This paper is organized as follows. We firstly realize a first order phase transition between two s-wave superfluid phases holographically, and give the concrete picture of first order phase transitions from a ``grand potential landscape'' point of view in Sec.~\ref{sec:model}. We then give a domain wall configuration in Sec.~\ref{sec:DomainWall}, and study the dynamical process of bubble nucleation in Sec.~\ref{sec:Bubble}. We finally take some conclusions and discussions in Sec.~\ref{sec:C&D}. Moreover, we include four appendices:  the quasi-normal modes are calculated in App.~\ref{sec:QNM}, equations of state are shown numerically in App.~\ref{sec:EOS}, the scheme for time evolutions as well as related numerical details are discussed in App.~\ref{sec:time_evolution}, and the concrete form of grand potential is presented in App.~\ref{sec:GrandP}.

\section{Holographic superfluid model with a first order phase transition}
\label{sec:model}

We use the holographic s+s model to realize a first order phase transition in the probe limit, under which the gravitational background is fixed, and the numerical work can be greatly simplified. The gravitational background can be taken as the 4 dimensional asymptotic AdS black brane
\begin{equation}\label{AdS}
\text{d}s^{2}=\frac{L^2}{z^2}(-f(z)\text{d}t^{2}+\frac{\text{d}z^{2}}{f(z)}+\text{d}x^{2}+\text{d}y^{2}),
\end{equation}
with the AdS radius $L=1$ for convenience and
\begin{equation}
f(z)=1-z^3.
\end{equation}
The boundary superfluid system is in a thermal bath with the same temperature as the Hawking temperature
\begin{equation}
	T=\frac{3}{4\pi}
\end{equation}
of the bulk black brane.

The Lagrangian for the matter fields is
\begin{eqnarray}\label{Lmatter}
L_{m}=\frac{1}{e_2^2}\Big(-\frac{1}{4}F^{ab}F_{ab}-|D_1 \Psi_1|^2-|D_2 \Psi_2|^2 \\ \nonumber
-m_1^2|\Psi_1|^2-m_2^2|\Psi_2|^2{-\lambda_{12}|\Psi_1|^2|\Psi_2|^2}\Big)~,\label{eq:Lagrangian}
\end{eqnarray}
where $D_1=\partial - i (e_1/e_2)A$~ and $D_2=\partial - i A$~.

The probe limit can be attained consistently by taking the large charge limit, and a scaling symmetry implies that only the ratio $e_1/e_2$ is important. Therefore we set $e_2=1$ in the rest of this paper.

With the standard procedure, we can get the solutions dual to superfluids on the boundary field theory. Because we have two orders dual to $\Psi_1$ and $\Psi_2$ respectively, we can get three different kinds of solutions: Solution S1, Solution S2 and Solution S1+S2. In order to realize the first order phase transition, we introduce the interacting term with coefficient $\lambda_{12}$ in \eqref{Lmatter}. We will introduce more details and show the phase diagram in the following subsection.

\subsection{Engineering a first order phase transition}
\label{model}
We take the following ansatz for the scalar and gauge fields in order to obtain the static solutions:
\begin{equation}\label{ansatz}
  \Psi_1=\Psi_{1}(z),~ \Psi_2=\Psi_{2}(z),~ A_\mu dx^\mu=A_t(z) dt.
\end{equation}
The equations of motion are then simplified to be
\begin{eqnarray}
\Psi_1''+\frac{f'}{f} \Psi_1'+\left(\frac{e_1^2 A_t^2}{f^2}-\frac{m_1^2}{z^2 f}-\frac{\lambda_{12} \Psi_2^2}{z^2 f} \right) \Psi_1 =0,\label{eq:homogeneous-psi1}\\
\Psi_2''+\frac{f'}{f} \Psi_2'+\left( \frac{ A_t^2}{f^2}-\frac{m_2^2}{z^2 f}-\frac{\lambda_{12} \Psi_1^2}{z^2 f} \right) \Psi_2 =0,\\
A_t''-2\frac{e_1^2 \Psi_1^2 + \Psi_2^2}{z^2 f} A_t=0.\label{eq:homogeneous-At}
\end{eqnarray}

We can take the standard procedure to solve these equations numerically to get the solutions dual to superfluid phases on the boundary theory. There are three such solutions:
We call the one with $\Psi_2=0$ and $\Psi_1\neq 0$ Solution S1, the one with $\Psi_1\neq 0$ and $\Psi_2=0$ Solution S2, and the one with both the two scalar fields nonzero as Solution S1+S2.

To Solve these solutions, the three fields can be expanded near the black hole horizon as
\begin{equation}\label{horizon}
\Psi_1=\Psi_{10}+\Psi_{11} (1-z)+\cdots,~~\Psi_2=\Psi_{20}+\Psi_{21} (1-z)+\cdots,~~ A_t=\phi_1 (1-z)+\cdots,
\end{equation}
where only $\Psi_{10},~\Psi_{20},~\phi_1$ are independent parameters.

Near the boundary, the three fields have the following asymptotic behavior
\begin{equation}
\Psi_1=\Psi_{1-} z^{\Delta_{1-}}+\Psi_{1+} z^{\Delta_{1+}}+\cdots,~~\Psi_2=\Psi_{2-} z^{\Delta_{2-}}+\Psi_{2+} z^{\Delta_{2+}}+\cdots,~~ A_t=\mu - \rho z+\cdots.
\end{equation}
where $\Delta_{i\pm}=(3\pm\sqrt{9+4 m_i^2 L^2})/2$ ($i=1,2$), $\mu$ and $\rho$ are the chemical potential and charge density of the system respectively. 
For each scalar field, either $\Psi_+$ or $\Psi_-$ can be chosen as the expectation value of scalar operator in the boundary field theory. We set source free condition for both the two scalars to get superfluid phases that spontaneously break the U(1) gauge symmetry. In order to further simplify the numerical work, we set
\begin{equation}
m_1^2=m_2^2=-2,
\end{equation}
and set $\Psi_{1-}=0$ for $\Psi_1$ and $\Psi_{2+}=0$ for $\Psi_2$. In this way, we have chosen different quantizations for the two scalar fields, the standard quantization for $\Psi_1$ and the alternative quantization for $\Psi_2$. We define the boundary operator dual to $\Psi_1$ as $O_1$ and the boundary operator dual to $\Psi_2$ as $O_2$. In this way, we get a dimension $2$ operator $O_1$ and a dimension $1$ operator $O_2$.

For the two scalar operators with different conformal dimension, the condensed solution will get ``parallel'' thermodynamic potential curves if the two orders also have the same charge. We can therefore tune the ratio $e_1/e_2$ to make the two thermodynamic potential curves have an intersection point. Near this intersection point, an s+s coexistent solution appears. For convenience, we can choose this ratio to be $e_1/e_2=e=4.5$.

\begin{figure}
	\centering
	\includegraphics[width=0.4\textwidth] {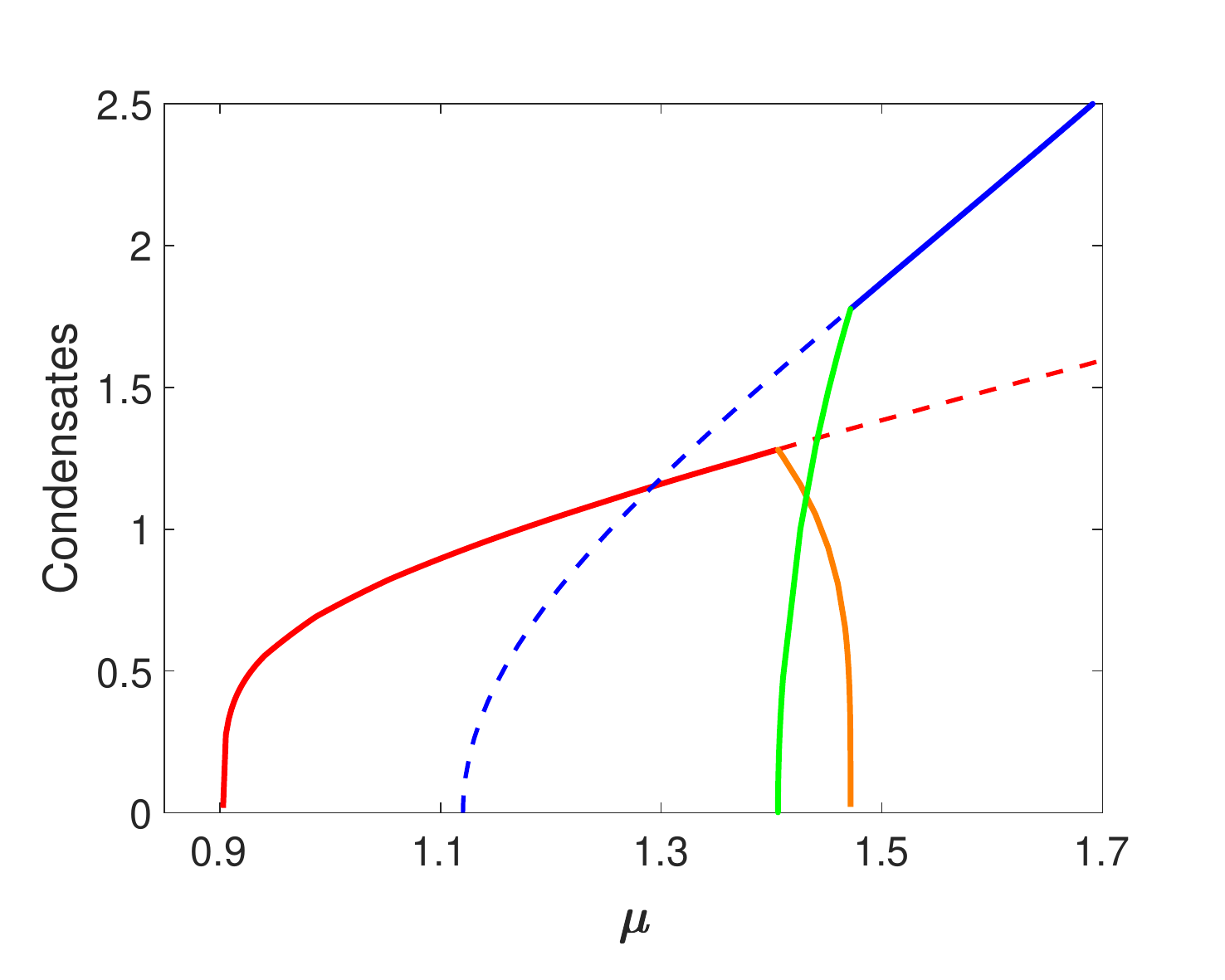}
	\qquad
	\includegraphics[width=0.4\textwidth] {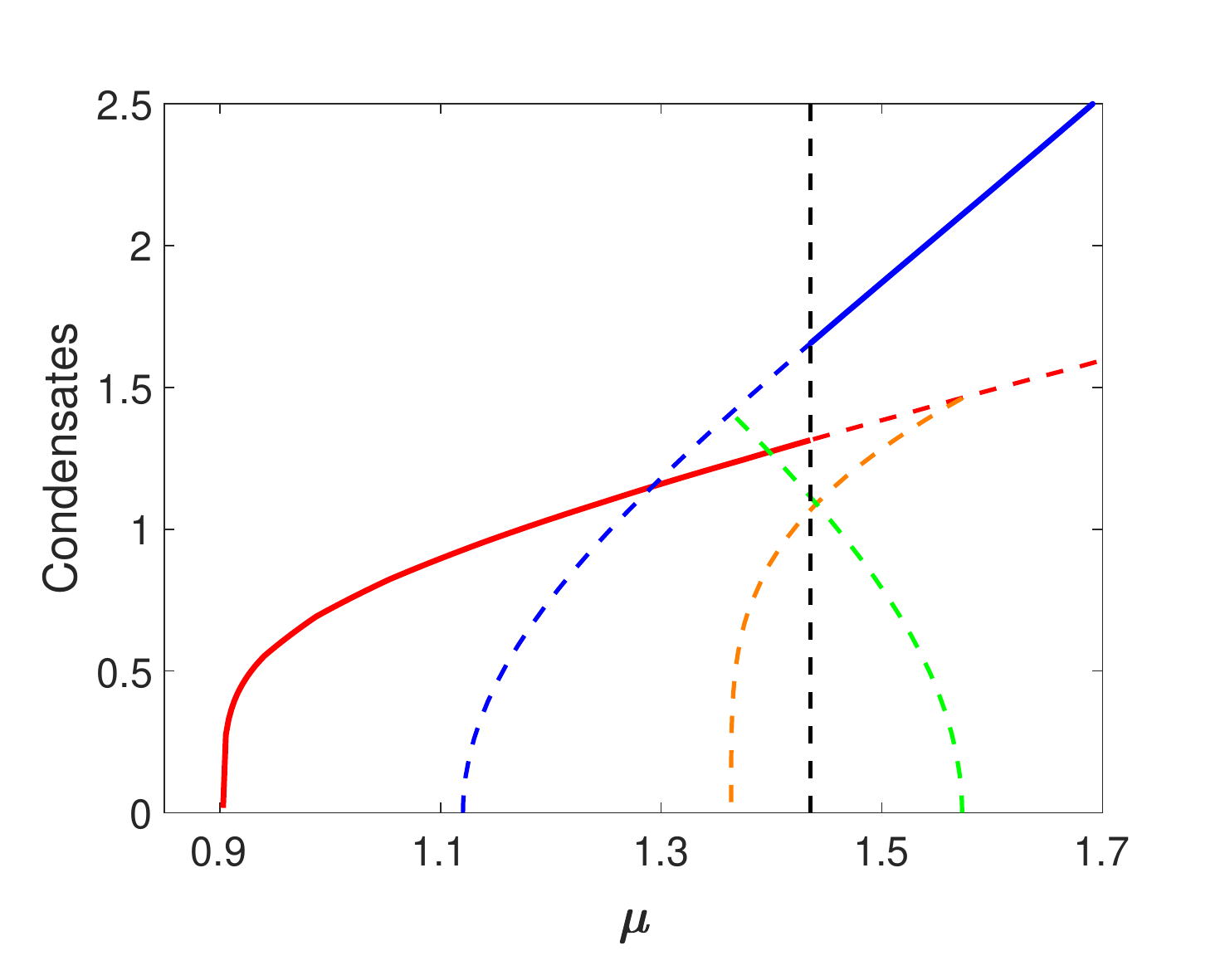}
	\caption{\label{Condensates}Condensates of two orders in the case with $\lambda_{12}=0$ (left) and $\lambda_{12}=0.4$ (right). Red line denotes the condensate of $O_1$ in Solution S1, blue line denotes the condensate of $O_2$ in Solution S2, while orange and green lines denote the condensate of $O_1$ and $O_2$ in Solution S1+S2 respectively. The dashed lines denote the condensate value in unstable or meta-stable sections.}
\end{figure}

Without considering the interaction term (or set $\lambda_{12}=0$ equivalently), the s+s solution will emerge near the intersection point of the two thermodynamic potential curves, and connect the two solutions with single condensate. It is the one with lowest thermodynamic potential  in the region where it exists. In this case, the system undergoes two second order phase transitions from the stable region of Solution S1 to the stable region of Solution S2, showing a typical ``x-type'' figure for condensates as in the left panel of Fig.~\ref{Condensates}.

In order to get a first order phase transition, we can tune the value of $\lambda_{12}$~. In Ref.~\cite{Nishida:2014lta}, the influence of a such kind of interacting parameter in a holographic system with multiple orders is already shown in the s+d system. Here in the holographic s+s model, the qualitative influence of $\lambda_{12}$ on the phase diagram is the same. It is easy to see that this term will not change the condensate as well as the thermodynamic potential of the superfluid solutions with single condensate. It will only change Solution S1+S2 and has a quite obvious influence.

Generally, if we decrease the value of $\lambda_{12}$,  Solution S1+S2 will exist in a larger region, and become stabler, while if we increase the value of $\lambda_{12}$,  Solution S1+S2 will firstly shrink to be in a smaller region, and become less stable. When we keep increasing $\lambda_{12}$, at a certain point, Solution S1+S2 will become totally unstable. After that, the region of  Solution S1+S2 will then increase, but its thermodynamic potential is still increasing, making this coexisting solution more and more unstable.
With some numerical work, we have confirmed that when $\lambda_{12}>\lambda_c\approx 0.0962$~, the Solution S1+S2 becomes totally unstable, and there is a first order phase transition between {S1} and {S2}.

\begin{figure}
	\centering			
	\includegraphics[width=0.4\textwidth]{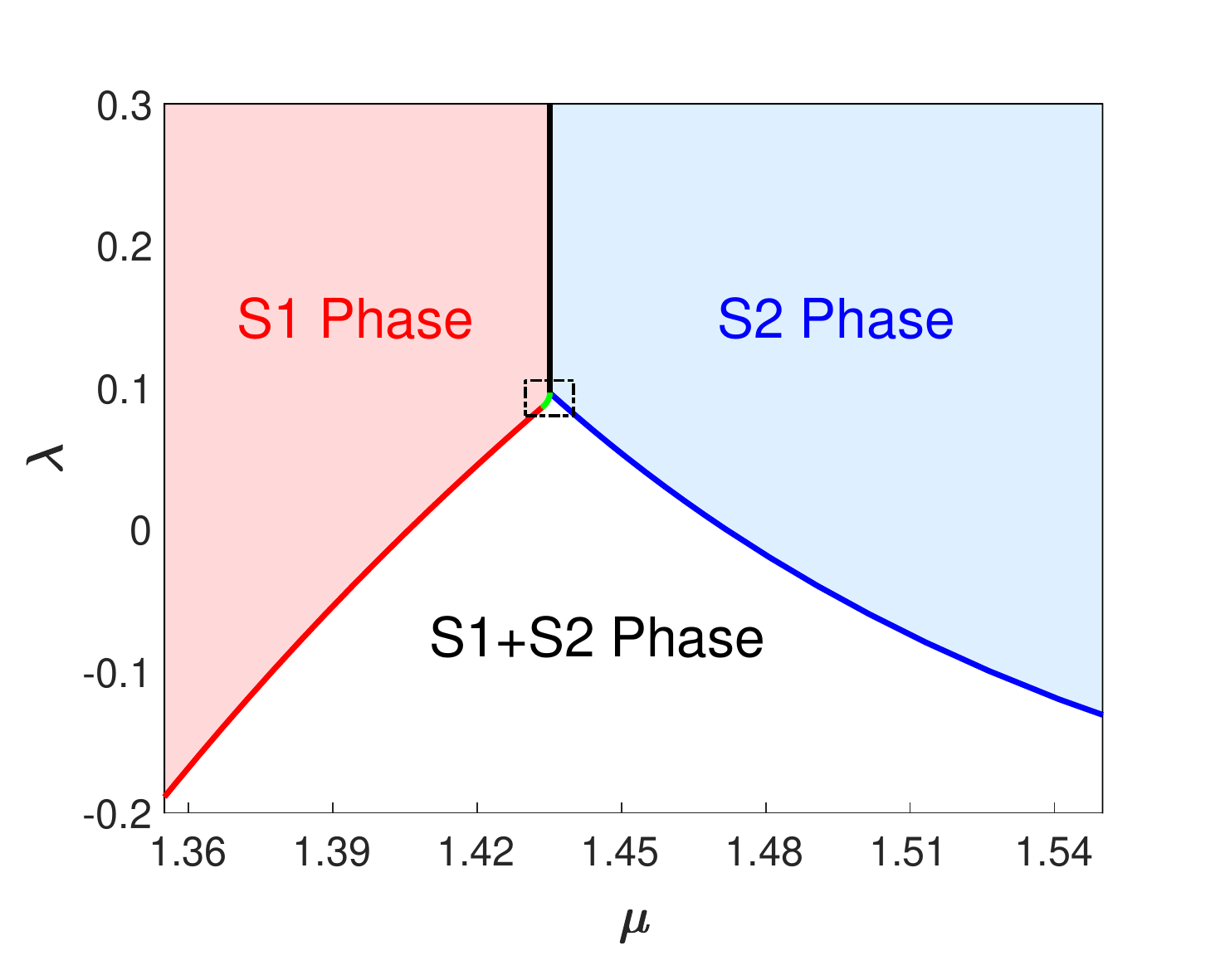}
	\qquad
	\includegraphics[width=0.4\textwidth]{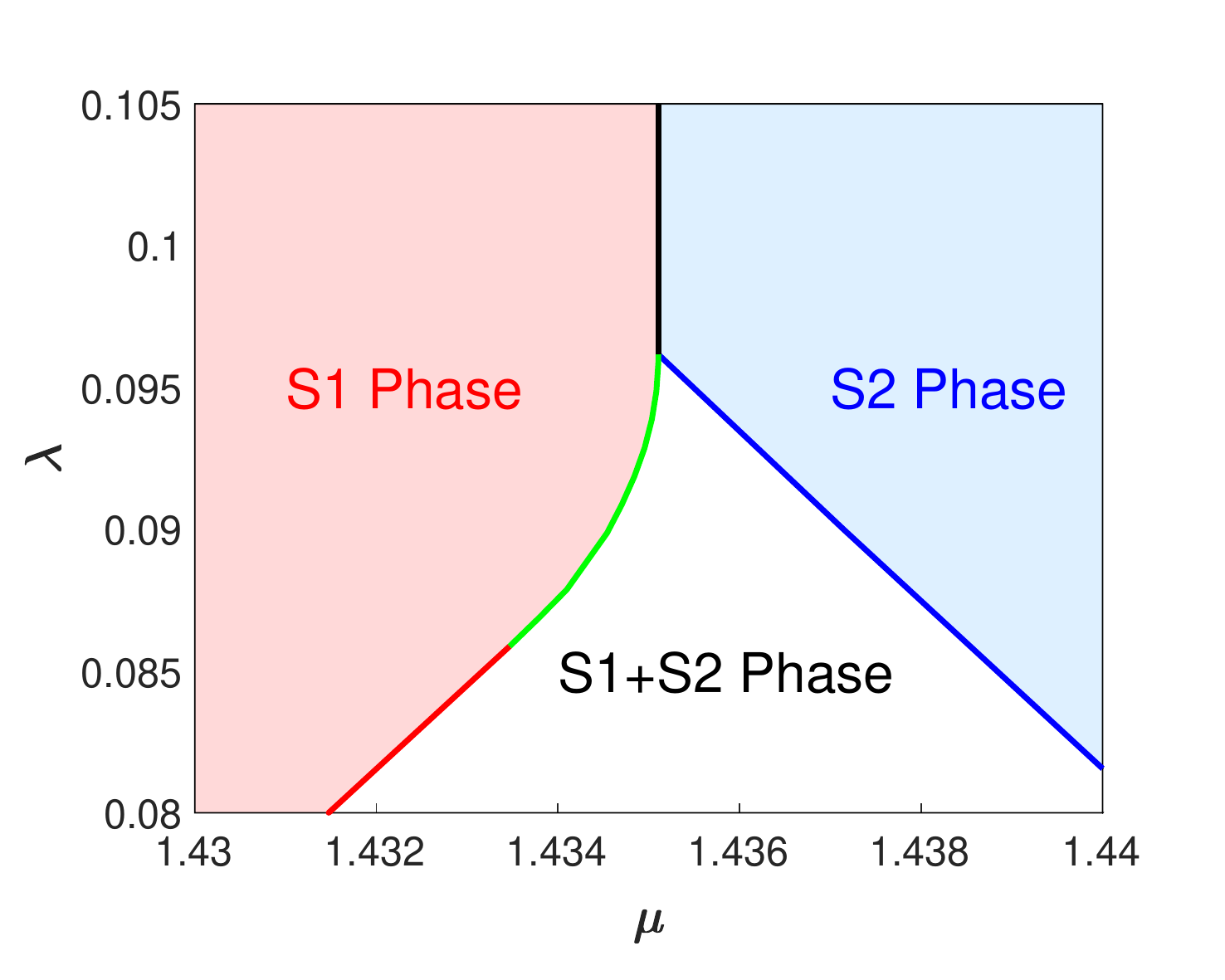}
	\caption{\label{PhaseDiagram}$\lambda_{12}-\mu$ phase diagram. The right plot is the enlarged version of the dashed rectangular region in the left one.}
\end{figure}

To make the influence of $\lambda_{12}$ more concrete, we draw a $\lambda_{12}-\mu$ phase diagram in Figure~\ref{PhaseDiagram}. We choose $\mu$ rather than $\rho$ as the horizontal axis because it is more convenient to work in the grand canonical ensemble. In this phase diagram, the red and blue lines for second order phase transitions are made up of critical points of second order phase transitions. It is a bit more complex to get the green and black lines for first order phase transitions, because we need to compare the thermodynamic potential of two solutions to get the first order phase transition point\cite{Nie:2014qma}.

In the phase diagram, we can see that there is a triple critical point among the S1 phase, S2 phase and the S1+S2 phase. Above the triple critical point, there is a first order phase transition between the S1 phase and S2 phase. Below the triple point, an S1+S2 phase exist between S1 phase and S2 phase. Further numerical calculation tells that in this system, the phase transition between S1+S2 phase and S2 phase is always second order, while the phase transition between S1+S2 phase and S1 phase is second order at a lower value of $\lambda_{12}$ (related to the solid red line) and is first order at a higher value of $\lambda_{12}$ (related to the solid green line).

To study the detail of a first order phase transition, we can also choose $\lambda_{12}$ to be in the region related to the green line. However, we choose $\lambda_{12}>\lambda_c$ in order to make the transition clearer. The stable phase on one side is Solution S1 with only order $O_1$ nonzero, and the stable phase on the other side is Solution S2 with only order $O_2$ nonzero. In the next subsection, we give the condensate behavior and grand potential of the three homogeneous superfluid solutions, which is helpful to understand the picture of first order phase transition.

\subsection{Homogeneous solutions} \label{sec:Homogeneous-Solution}
When $\lambda_{12}>\lambda_c$, the s+s solution becomes totally unstable, and we can get first order phase transition between S1 phase and S2 phase. In order to get the numerical solution with a thin bubble wall, we wish to get a larger value of potential barrier, this is related to a larger value of $\lambda_{12}$. Thus in the next section, we set
\begin{equation}
e_1=4.5,\qquad e_2=1,\qquad\lambda_{12}=0.4,
\end{equation}
to get the inhomogeneous numerical solution dual to a mixed state on the boundary field theory.

\begin{figure}
	\centering
	\includegraphics[width=11cm] {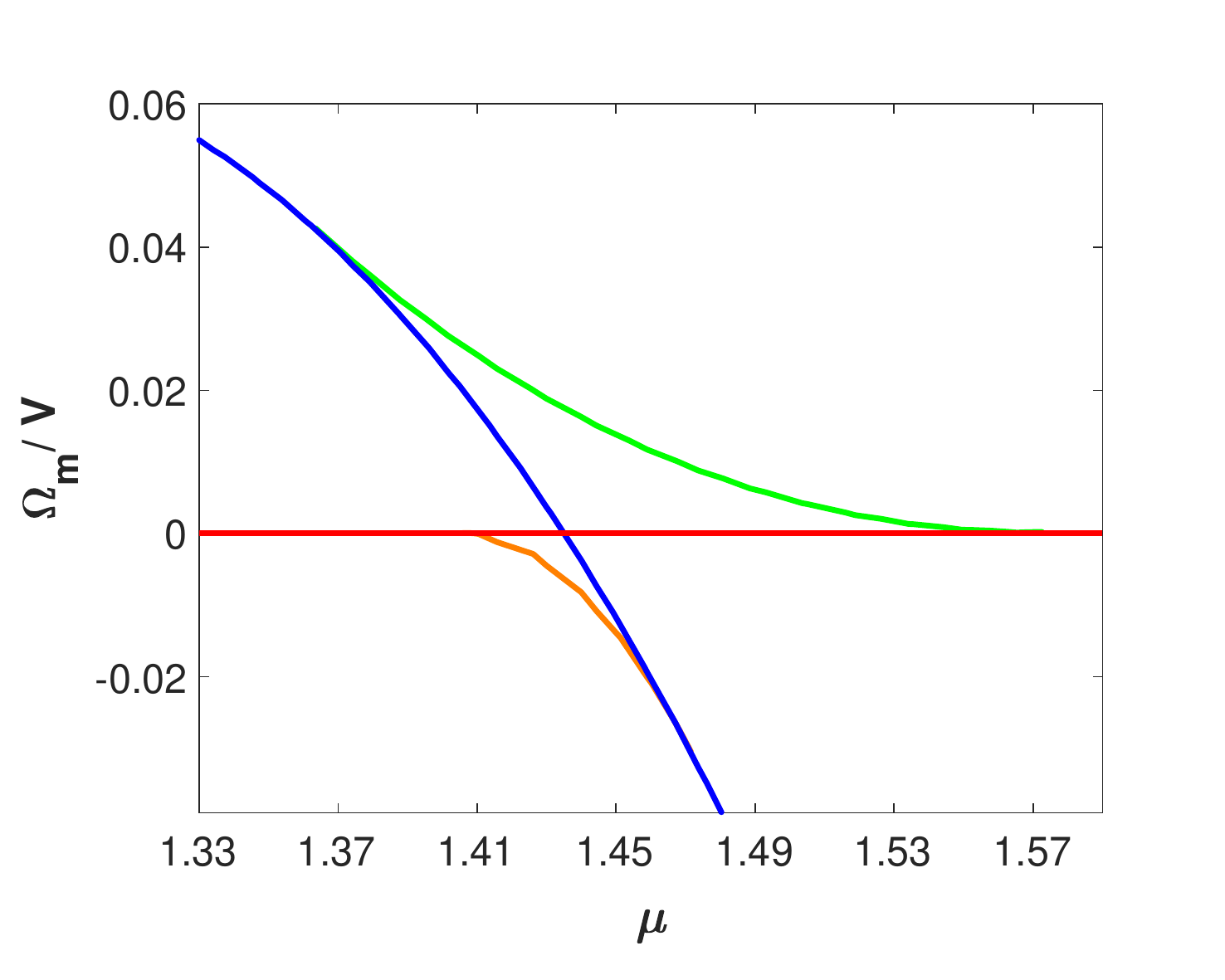}
	\caption{\label{FreeE}Grand potential of Solution S2 (blue) and Solution S1+S2 (green) with respect to that of Solution S1 (red). The orange line is for comparison, which denotes the grand potential of Solution S1+S2 with $\lambda_{12}=0$.}
\end{figure}

In the rest of this section, we show the condensate behavior and grand potential of the static homogeneous solutions with the above choice of $\lambda_{12}$. The condensate behavior is plotted in the right panel of Figure~\ref{Condensates}, and the grand potential curves are plotted in Figure~\ref{FreeE}.

{In the AdS/CFT dictionary, the grand potential of the three solutions can be calculated from the Euclidean on-shell action on the bulk side. Because we work in the probe limit, the contribution from the gravitational part is the same for different solutions, and we only need to calculate the contribution from the matter part (\ref{Lmatter}) to the grand potential\cite{Nie:2013sda}:}
\begin{equation}
\Omega_m=T S_{\text{ME}},
\end{equation}
where $S_{\text{ME}}$ denotes the Euclidean on-shell action of matter fields in the black brane background. With the ansatz and equations of motion, we can finally get
\begin{equation}\label{freeEs1s2}
\Omega_m={V}\Big[-\frac{1}{2}\mu\rho+\int_1^0 \Big(-\frac{A_t^2}{z^2 f}(e^2\Psi_1^2+\Psi_2^2)+\lambda_{12}\frac{\Psi_1^2\Psi_2^2}{z^4}\Big)dz \Big]
\end{equation}
with $V$ the total volume of the space. Since we only consider the probe limit in this paper, we will suppress the subscript $m$ of the grand potential (and other thermodynamic quantities) in the rest of this paper.

\subsection{A picture for first order phase transitions}
In this section, we describe a concrete picture for first order phase transitions, which is consistent with the swallow tail shape of thermodynamic potential curves as well as our numerical results for the inhomogeneous domain wall and bubble configurations in the following sections.

The general picture of first order phase transitions tell us that when a first order phase transition occurs, the system changes from a meta-stable state (which is a local minimum in the configuration space) to a stable state (which is a global minimum in the configuration space). We must stress that in order to better understand the first order phase transition, we need also involve the third state, which is an unstable state and is a saddle point in the configuration space. 

Another feature of the first order phase transition is that the thermodynamic potential usually form a ``swallow'' tail shape as in Figure~\ref{FreeE}, where we can see that the ``swallow tail'' is formed by three part: the red curve, the blue curve and the green curve. The red and blue curves are related to the stable and meta-stable states, with the lower one more stable, while the green curve is related to the unstable state.

In a homogeneous solution of a thermodynamic system, we have the (static) equation of state, the solutions of which make the thermodynamic potential functional get extremal values. In another word, the equation of state can be obtained from the variation of the thermodynamic potential functional. When we apply the variational principle to the thermodynamic potential functional, not only the stable and meta-stable states, but also the unstable state satisfies the extremal condition. Therefore we can get the complete swallow tail shape with solutions of the equation of state. In holography, the extremal condition of thermodynamic potential functional is dual to the (static) equations of motion in the bulk (see the discussion in Sec.~\ref{landscape}).

In order to clarify the physical picture, we can consider the full configuration space, which includes not only the states that satisfies the equation of state (equations of motion in the bulk), but also the non-equilibrium or more general states that do not satisfy the equation of state. In this configuration space the stable solution, the meta-stable solution and the unstable solution can all be connected continuously through some general states. If we draw the thermodynamic potential in this space, we can get a landscape of the potential (see Sec.~\ref{landscape} for details), where the stable and meta-stable solutions are the local minimum points, and the unstable solution is a saddle point.

\begin{figure}
	\centering
	\includegraphics[width=15cm] {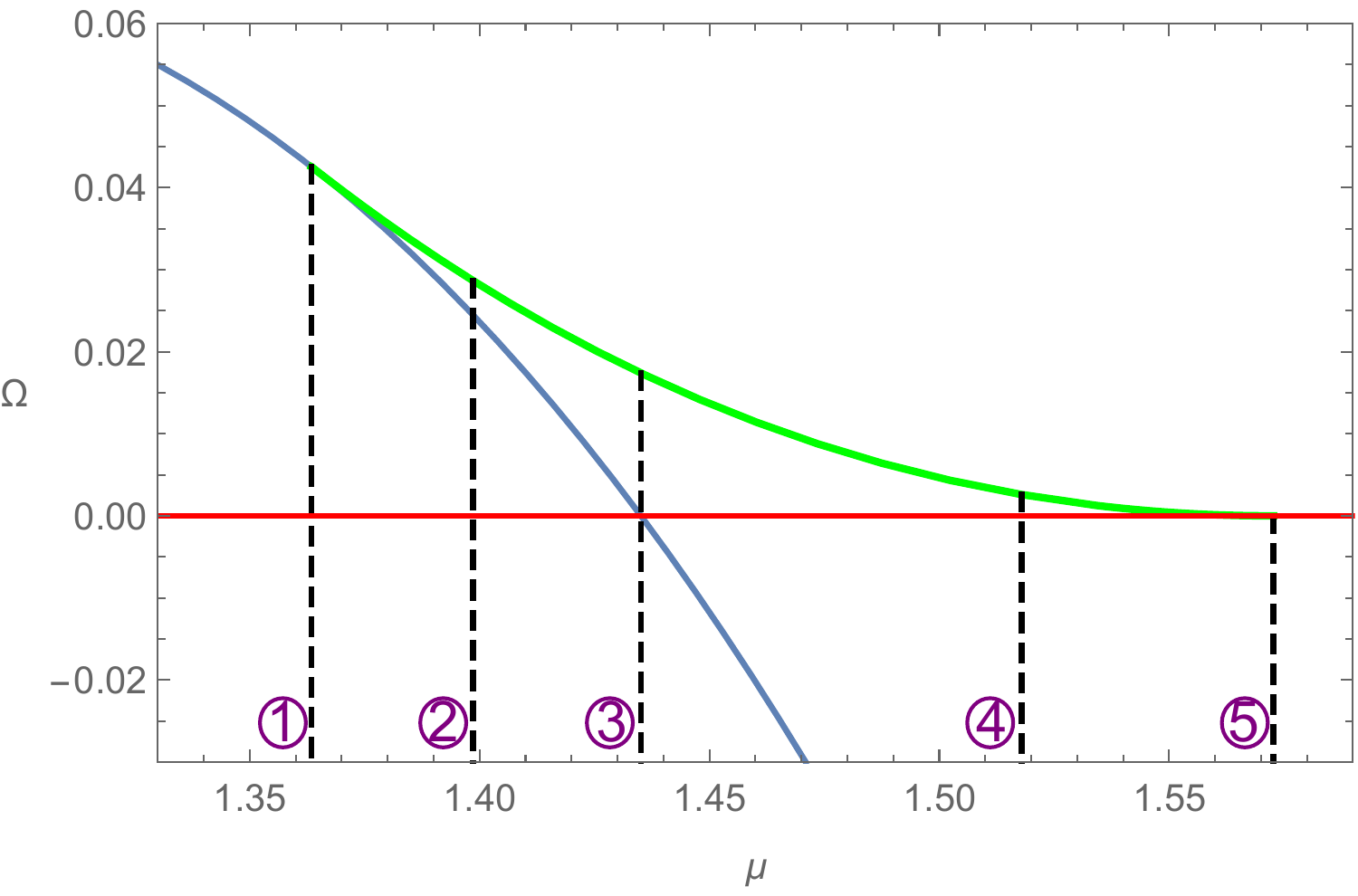}
	\caption{\label{SwallowTail0}A schematic picture showing the grand potential near the first order phase transition, where we can see a clear swallow tail shape.}
\end{figure}

\begin{figure}
	\centering
	\includegraphics[width=0.19\textwidth] {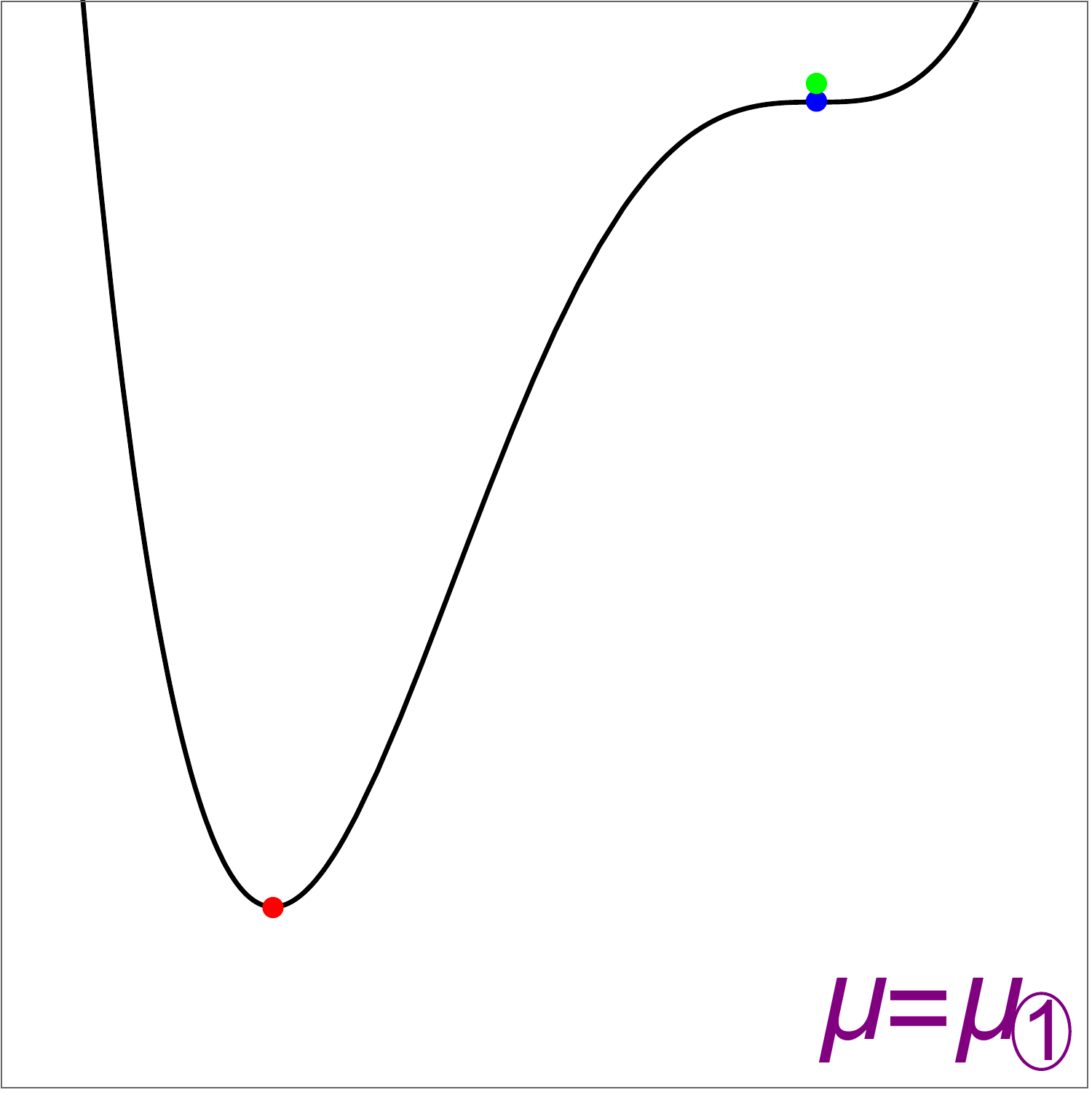}
	\includegraphics[width=0.19\textwidth] {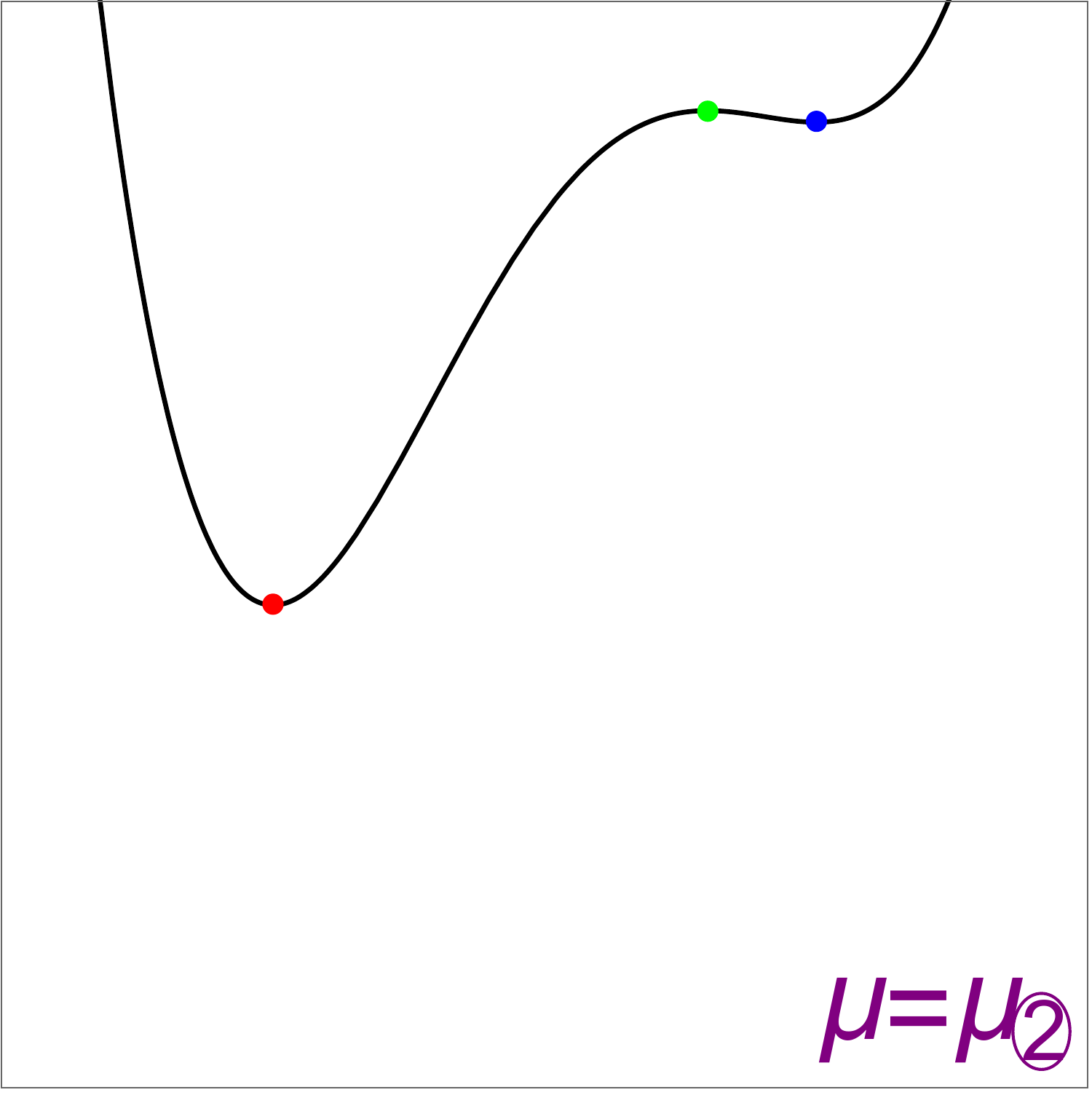}
	\includegraphics[width=0.19\textwidth] {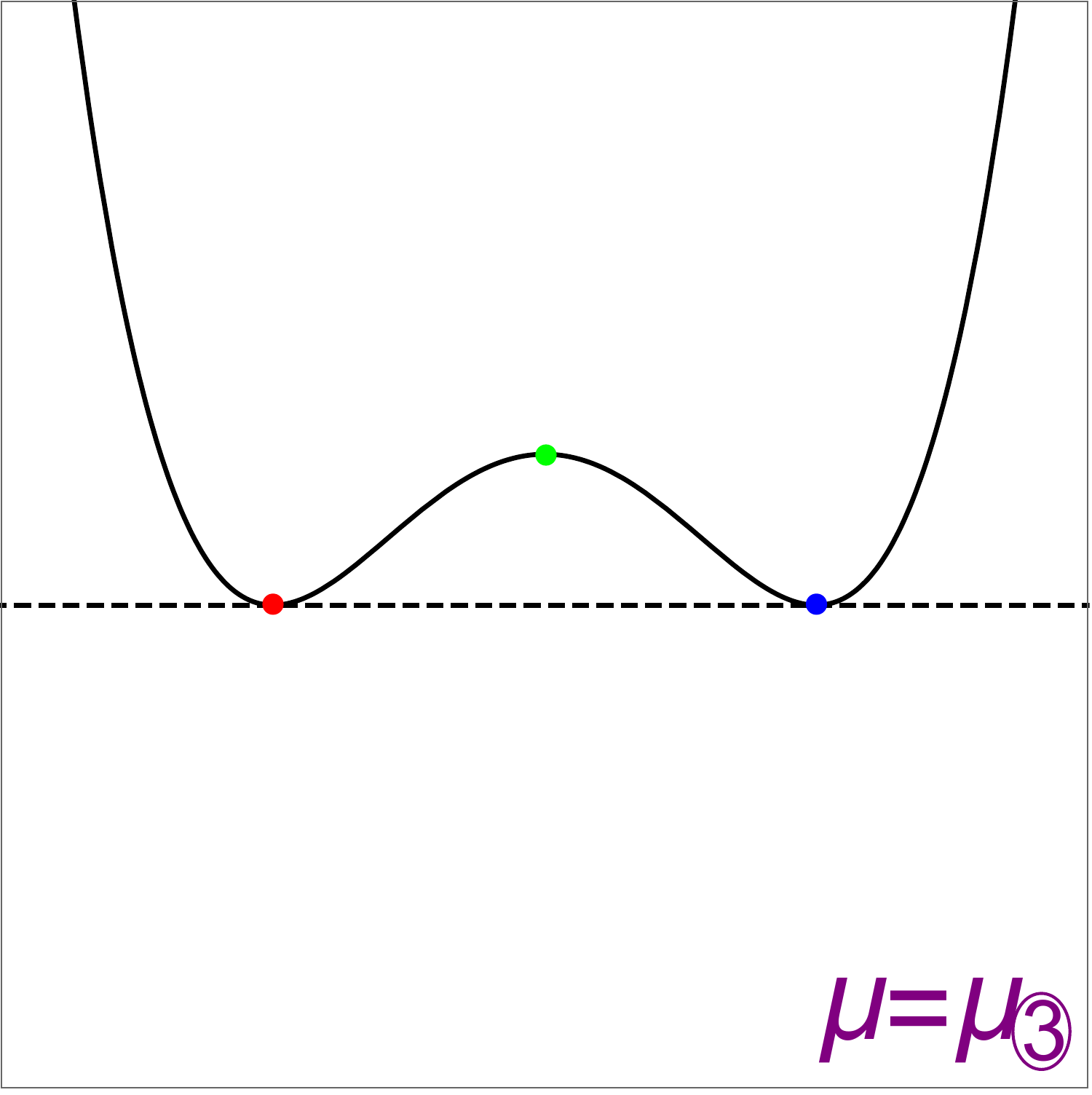}
	\includegraphics[width=0.19\textwidth] {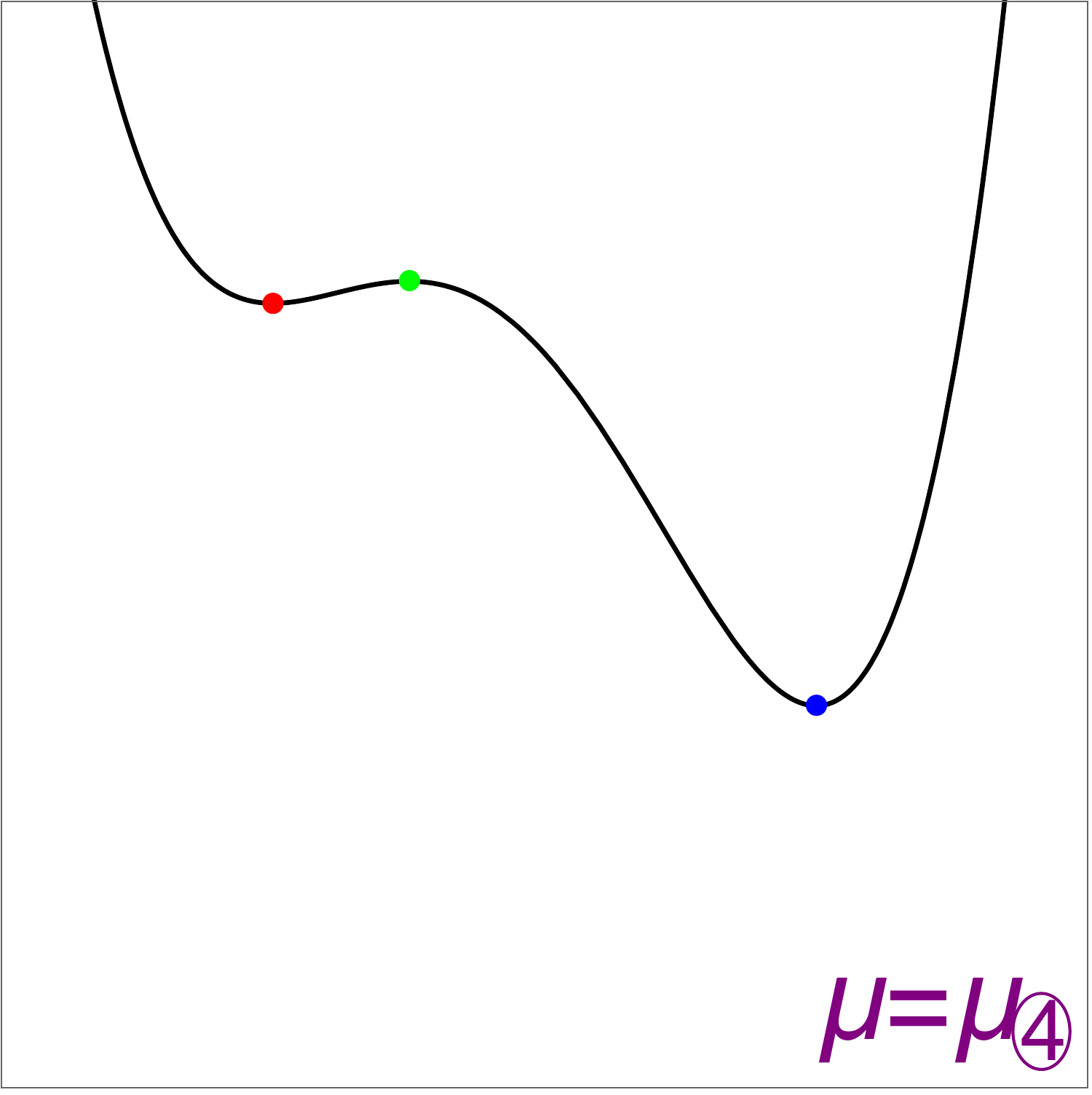}
	\includegraphics[width=0.19\textwidth] {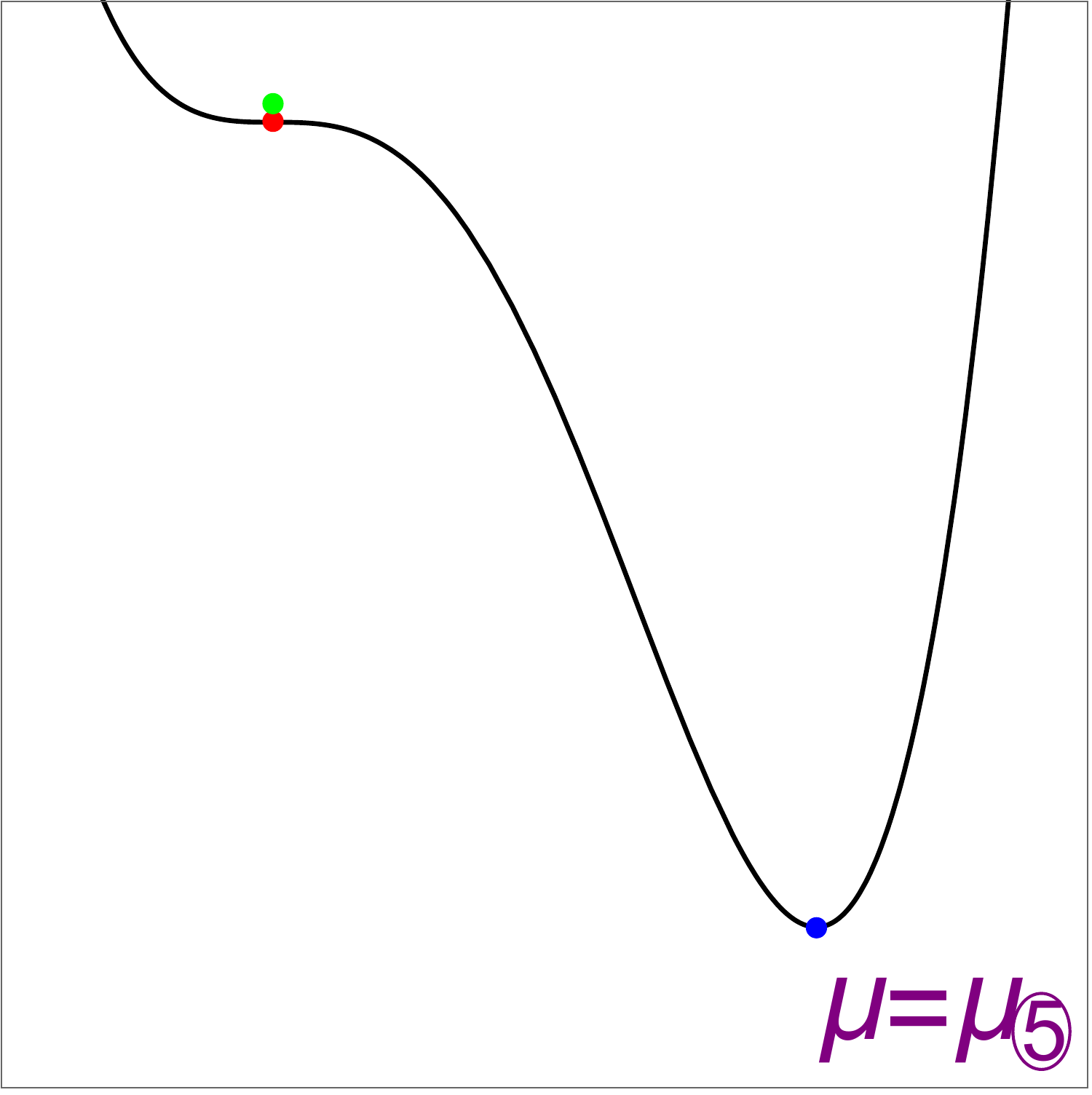}
\caption{\label{5Points}A schematic picture showing the grand potential near the first order phase transition, where the green points are the unstable state.}
\end{figure}

We draw the thermodynamic potential curve with swallow tail shape in Figure~\ref{SwallowTail0}, where we mark five typical points and for each point we show a special curve in the grand potential landscape as in Figure~\ref{5Points}. For Point \textcircled{1}, the S1+S2 solution just emerge from the S2 solution. We can see that the blue and green points coincide at a stationary point. At Point \textcircled{2}, the S1+S2 solution, which is a saddle point, is separated away from the S2 solution, which is a local minimum point. Point \textcircled{3} denote the transition point for the first order phase transition, where the thermodynamic potential for the two minimum are equal. Point \textcircled{4} is similar to Point \textcircled{2}, where the S1+S2 solution is also a saddle point. The difference is that in this situation, the S2 solution has a lower value of grand potential. Point \textcircled{5} is similar to Point \textcircled{1}, which are both at the tip of the swallow tail, where the S1+S2 merge with the S1 solution.

With the above picture, we can see that to make a first order phase transition occur, the meta-stable state needs to skip over a barrier and go to the stable state. The unstable state is right at the point through which the thermodynamic potential barrier is lowest. The reason is the following. Imagine that we draw all lines connecting the stable state and the meta-stable state in the landscape. Then we look at the highest point of each line. The lowest one of all such highest points should be a saddle point, because it is the highest point in the line direction (as shown in Fig.~\ref{5Points}) and at the same time the lowest point in the complementary directions. Assuming that there is only one saddle point in the landscape,\footnote{There can be more than one saddle point, which means that we can have more than one unstable static solution. Then the construction here automatically picks out the saddle point with the lowest potential, which corresponds to the most relevant unstable equilibrium state, so our argument here still works for this most relevant unstable state.} we see that this saddle point is just the unstable state, which is by design the lowest barrier between the stable and unstable states. Therefore the unstable state is very useful to estimate the (nonlinear) stability of meta-stable state qualitatively.

\subsection{Domain walls and bubbles}
\label{wall}
In first order phase transitions, a characteristic phenomenon is phase separation, where there is an interface between different phases, the domain wall. In higher (larger than one) dimensions, the domain wall can have various topologies and shapes, typically a bubble. The two homogeneous regions separated by the wall are in different stable (or meta-stable) states, and the wall region is where the system change from one state to another. Therefore, the wall region is supposed to have higher thermodynamic potential density than the homogeneous region, and one should take into account this excess amount of thermodynamic potential, which is well known as the surface tension and will be further discussed in the following sections, to understand the mechanics of bubbles.

With the above picture of first order phase transition, we can see that through the wall region, the local state must go up the thermodynamic potential barrier. Because the lowest barrier is right on the unstable state, the point with maximum thermodynamic potential in the wall would just be very similar to the unstable state on the swallow tail, which means that they are very close in the configuration space mentioned in the last subsection. The route in configuration space related from the points on the bubble wall may not pass the saddle point because of the complex shape of the landscape. Another difference is that the points on the wall involve additional spacial gradient.

We shall obtain configurations with bubble wall in two different ways. First we build a domain wall solution with only one inhomogeneous spatial dimension. And later, we obtain two dimensional bubble configurations using a quench method.

\subsection{The landscape of free energy surface}
\label{landscape}
To make our previous arguments for the first order phase transition clearer and quantitative, we shall discuss here the landscape picture of the free energy (grand potential) of holographic systems, taking the simplest holographic superfluid model as an example.

Although numerical evolution of holographic dynamics is usually (and most simply) done under the Eddington-Finkelstein coordinates (\ref{Eddington}), in order to make our discussion as general as possible, we will consider a class of coordinates (including the Eddington-Finkelstein one as a special case), under which the time evolution of holographic systems can be done in principle.\footnote{An example of such kind of coordinates other than the Eddington-Finkelstein one (\ref{Eddington}) is the well-known Eddington coordinates, which for the Schwarzschild-AdS black brane (\ref{AdS}) or (\ref{Eddington}) reads
\begin{equation}\label{non-Finkelstein}
ds^{2}=-\frac{1}{z^{2}}\left(-dt^{2}+dz^2+d\vec{x}^{2}+z^3(dt+dz)^2\right).
\end{equation}} In this class of coordinates, $\xi=\partial_t$ is a time-like Killing vector field. The future-directed orthonormal co-vector to a constant $t$ surface is
\begin{equation}
m_{\nu}=-(-g^{tt})^{-1/2}(1,0,\cdots,0),
\end{equation}
and so the total flux of the energy current across a constant $t$ surface $\Sigma$ (null for the Eddington-Finkelstein case or space-like for more general cases like (\ref{non-Finkelstein})) is
\begin{equation}\label{flux}
\int T_{\mu}^{\nu}\xi^{\mu}m_{\nu}\sqrt{\det(g_{ij})}d^{d}x=-\int T_{t}^{t}\sqrt{-g}d^{d}x.
\end{equation}
Note that in holographic systems at finite temperature, the surface $\Sigma$ is bounded by the horizon and the asymptotic AdS conformal boundary.

Let us first illustrate the picture by a scalar field, the simplest case. The Lagrangian density is
\begin{equation}
L=-\frac{1}{2}g^{\mu\nu}\nabla_{\mu}\phi\nabla_{\nu}\phi-\frac{1}{2}m^{2}\phi^{2},
\end{equation}
so we have
\begin{eqnarray}
-T_{t}^{t}&=&-g^{t\nu}\partial_{\nu}\phi\partial_{t}\phi+\frac{1}{2}(g^{\rho\nu}\partial_{\rho}\phi\partial_{\nu}\phi+m^{2}\phi^{2})\nonumber\\
&=&-g^{tt}(\partial_{t}\phi)^{2}-g^{ti}\partial_{i}\phi\partial_{t}\phi+\frac{1}{2}[g^{tt}(\partial_{t}\phi)^{2}+2g^{ti}\partial_{t}\phi\partial_{i}\phi+g^{ij}\partial_{i}\phi\partial_{j}\phi+m^{2}\phi^{2}]\nonumber\\
&=&-\frac{1}{2}g^{tt}(\partial_{t}\phi)^{2}+\frac{1}{2}(g^{ij}\partial_{i}\phi\partial_{j}\phi+m^{2}\phi^{2}),
\end{eqnarray}
which is obviously a sum of the kinetic energy density and the potential density. Note that here we achieve this nice expression under a coordinate system that is generally not time orthogonal. Actually, the potential density
\begin{equation}
V=-L(\partial_{t}\to0)=\frac{1}{2}g^{ij}\partial_{i}\phi\partial_{j}\phi+\frac{1}{2}m^{2}\phi^{2},
\end{equation}
and we can define the static free energy (grand potential) as
\begin{equation}\label{Omega_s}
\Omega_s=\int V\sqrt{-g}d^d x.
\end{equation}
On the other hand, the generalized free energy (grand potential) is just the total flux (\ref{flux}):
\begin{equation}
\Omega=-\int T_{t}^{t}\sqrt{-g}d^{d}x=-\int\frac{1}{2}g^{tt}(\partial_{t}\phi)^{2}\sqrt{-g}d^{d}x+\Omega_{s}.
\end{equation}
We see that
\begin{equation}\label{inequality}
\Omega\ge\Omega_s
\end{equation}
due to the positivity of the kinetic term, where the equality holds when the configuration is static or $g^{tt}=0$ (i.e. the Eddington-Finkelstein case). Locally, in holographic systems we can define
\begin{equation}
\omega_s=\int V\sqrt{-g}dz,\qquad\omega=-\int T_{t}^{t}\sqrt{-g}dz\ge\omega_{s}
\end{equation}
as the static and generalized free energy (grand potential) densities, respectively.

The full equation of motion of the scalar field can be written as
\begin{equation}\label{full}
\partial_{t}(\sqrt{-g}g^{t\mu}\partial_{\mu}\phi)+\partial_{i}(\sqrt{-g}g^{it}\partial_{t}\phi)=\frac{\delta\Omega_{s}}{\delta\phi},
\end{equation}
while the static equation of motion, i.e. the equation of motion for static configurations,
\begin{equation}
\partial_{i}(\sqrt{-g}g^{ij}\partial_{j}\phi)=\sqrt{-g}m^{2}\phi
\end{equation}
is just the extreme of the variation
\begin{equation}\label{static}
\delta\Omega_{s}=\delta\int(\frac{1}{2}g^{ij}\partial_{i}\phi\partial_{j}\phi+\frac{1}{2}m^{2}\phi^{2})\sqrt{-g}dzd^{d-1}\vec{x}=0.
\end{equation}
The time evolution of the generalized free energy (grand potential) can be computed as
\begin{eqnarray}
\frac{d\Omega}{dt}&=&\int[\sqrt{-g}g^{ti}\partial_{t}\partial_{i}\phi+\partial_{i}(\sqrt{-g}g^{it}\partial_{t}\phi)]\partial_{t}\phi d^{d}x+\int\partial_{i}(\sqrt{-g}g^{ij}\partial_{j}\phi\partial_{t}\phi)d^{d}x\nonumber\\
&=&\int\partial_{i}(\sqrt{-g}g^{it}\partial_{t}\phi\partial_{t}\phi)d^{d}x+\int\partial_{i}(\sqrt{-g}g^{ij}\partial_{j}\phi\partial_{t}\phi)d^{d}x\nonumber\\
&=&\int\sqrt{-g}g^{z\mu}\partial_{\mu}\phi\partial_{t}\phi d^{d-1}\vec{x}|_{z_{h}}^{0}=\int T_{t}^{z}\sqrt{-g}d^{d-1}\vec{x}|_{z_{h}}^{0}.
\end{eqnarray}
Under the ingoing Eddington-Finkelstein coordinates or more general ingoing coordinates the only non-vanishing $g^{z\mu}|_{z_h}$ is $g^{zt}|_{z_h}<0$, so the generalized free energy (grand potential) decreases monotonically if there is no source on the AdS conformal boundary, i.e. $T_{t}^{z}|_{z=0}=0$, which can be proved in more general holographic systems with the null energy condition\citep{TWZ}. Physically, this decrease (the energy flux across the horizon) is interpreted as the energy dissipation in this dynamical process\cite{Liu,TWZ1,TWZ2}.

For a Maxwell field, the situation is subtler. The Lagrangian density is
\begin{equation}
L=-\frac{1}{4}F^{\mu\nu}F_{\mu\nu},
\end{equation}
so we have
\begin{eqnarray}
-T_{t}^{t}&=&-(\frac{1}{2}g^{t\mu}g^{i\nu}F_{\mu\nu}F_{ti}-\frac{1}{4}g^{i\mu}g^{j\nu}F_{\mu\nu}F_{ij})\nonumber\\
&=&-\frac{1}{2}(g^{tt}g^{ik}-g^{tk}g^{it})F_{tk}F_{ti}+\frac{1}{4}g^{ik}g^{jl}F_{kl}F_{ij}.
\end{eqnarray}
On the other hand, the flux of the Noether current\cite{TWZ} (or called the canonical current) is given by the Hamiltonian density. The canonical momentum density is
\begin{equation}
p^{i}=\frac{\partial L}{\partial\partial_{t}A_{i}}=\partial^{i}A^{t}-\partial^{t}A^{i},
\end{equation}
so the Hamiltonian density is
\begin{eqnarray}
H&=&p^{i}\partial_{t}A_{i}-L=-(g^{tt}g^{ik}-g^{tk}g^{it})F_{tk}\partial_{t}A_{i}-g^{kt}g^{ji}F_{kj}\partial_{t}A_{i}\nonumber\\
&&+\frac{1}{2}g^{tt}g^{ik}F_{tk}F_{ti}+\frac{1}{2}g^{tk}g^{it}F_{kt}F_{ti}+g^{kt}g^{ji}F_{ti}F_{kj}+\frac{1}{4}g^{ik}g^{jl}F_{kl}F_{ij}\nonumber\\
&=&-\frac{1}{2}(g^{tt}g^{ik}-g^{tk}g^{it})F_{tk}(\partial_{t}A_{i}+\partial_{i}A_{t})-g^{kt}g^{ji}F_{kj}\partial_{i}A_{t}+\frac{1}{4}g^{ik}g^{jl}F_{kl}F_{ij}\nonumber\\
&=&-\frac{1}{2}(g^{tt}g^{ik}-g^{tk}g^{it})(\partial_{t}A_{i}\partial_{t}A_{k}-\partial_{i}A_{t}\partial_{k}A_{t})-g^{kt}g^{ji}F_{kj}\partial_{i}A_{t}+\frac{1}{4}g^{ik}g^{jl}F_{kl}F_{ij}.
\end{eqnarray}
Then it is easy to check that
\begin{eqnarray}
T_{t}^{t}+H&=&-(g^{tt}g^{ik}-g^{tk}g^{it})(\partial_{t}A_{k}\partial_{i}A_{t}-\partial_{i}A_{t}\partial_{k}A_{t})-g^{kt}g^{ji}F_{kj}\partial_{i}A_{t}\nonumber\\
&=&-(g^{tt}g^{ik}-g^{tk}g^{it})F_{tk}\partial_{i}A_{t}-g^{kt}g^{ji}F_{kj}\partial_{i}A_{t}\\
&=&F^{it}\partial_{i}A_{t} \cong \frac{1}{\sqrt{-g}}\partial_{i}(\sqrt{-g}F^{it}A_{t}),
\end{eqnarray}
which is consistent with the general fact that these two currents are equal on shell up to a total divergence\cite{TWZ}. Here in the last step (and hereafter) $\cong$ means equality on shell. Holographically, this total divergence gives a boundary contribution $\int\mu\rho\sqrt{-g}d^{d-1}\vec{x}$ to the flux across $\Sigma$, which in the equilibrium case is just the difference between the free energy and the grand potential. Specifically, the flux of the energy current is the generalized free energy $F$ and that of the Noether current is the generalized grand potential $\Omega$:
\begin{eqnarray}
F&=&-\int T_{t}^{t}\sqrt{-g}d^{d}x,\qquad\Omega=\int H\sqrt{-g}d^{d}x,\nonumber\\
F&\cong&\Omega+\int\mu\rho\sqrt{-g}d^{d-1}\vec{x}.\label{relation}
\end{eqnarray}

In the above case, the potential density is
\begin{equation}\label{V}
V=-L(\partial_{t}=0)=\frac{1}{2}(g^{tt}g^{ik}-g^{tk}g^{it})\partial_{k}A_{t}\partial_{i}A_{t}-g^{it}g^{jk}\partial_{k}A_{t}F_{ij}+\frac{1}{4}g^{ik}g^{jl}F_{kl}F_{ij},
\end{equation}
i.e. the part of $L$ independent of any time derivative. Then the kinetic energy density is
\begin{equation}\label{kinetic}
H-V=-\frac{1}{2}(g^{tt}g^{ik}-g^{tk}g^{it})\partial_{t}A_{i}\partial_{t}A_{k},
\end{equation}
which is positive definite. The static grand potential $\Omega_{s}$ is still defined by (\ref{Omega_s}), as well as the corresponding density
\begin{equation}
\omega_s=\int V\sqrt{-g}dz,
\end{equation}
which again satisfy
\begin{equation}
\omega\ge\omega_s
\end{equation}
and (\ref{inequality}), respectively, due to the positivity of the kinetic energy density (\ref{kinetic}). A caveat here is that both the potential density (\ref{V}) and the kinetic energy density (\ref{kinetic}) are gauge dependent, so certain gauge fixing will be helpful. In particular, under the Eddington-Finkelstein coordinates, since $g^{tt}=0$ and $g^{ti}=0$ unless $i=z$, if we take the radial gauge $A_z=0$ as usual, it is obvious that the kinetic energy density (\ref{kinetic}) completely vanishes even for non-static configurations, similar to the scalar field case.

For the coupled Maxwell-scalar case
\begin{equation}
L=-g^{\mu\nu}(\partial_{\mu}-iA_{\mu})\phi(\partial_{\nu}+iA_{\nu})\phi^{*}-m^{2}|\phi|^{2}-\frac{1}{4}F^{\mu\nu}F_{\mu\nu},
\end{equation}
similarly we have
\begin{eqnarray}
-T_{t}^{t}&=&-(g^{t\nu}[D_{\nu}\phi D_{t}^{*}\phi^{*}+D_{\nu}^{*}\phi^{*}D_{t}\phi]-[D_{\rho}\phi D^{\rho*}\phi^{*}+m^{2}|\phi|^{2}]+\frac{1}{2}F^{ti}F_{ti}-\frac{1}{4}F^{ij}F_{ij})\nonumber\\
&=&-g^{tt}|D_{t}\phi|^{2}-\frac{1}{2}(g^{tt}g^{ik}-g^{tk}g^{it})F_{tk}F_{ti}+g^{ij}D_{i}\phi D_{j}^{*}\phi^{*}+m^{2}|\phi|^{2}+\frac{1}{4}g^{ik}g^{jl}F_{kl}F_{ij}.
\end{eqnarray}
The potential density is
\begin{eqnarray}
V=-L(\partial_{t}=0)&=&2g^{ti}\operatorname{Re}[-iA_{t}\phi D_{i}^{*}\phi^{*}]+g^{ij}D_{i}\phi D_{j}^{*}\phi^{*}+(m^{2}+g^{tt}A_{t}^{2})|\phi|^{2}\nonumber\\
&&+\frac{1}{2}(g^{tt}g^{ik}-g^{tk}g^{it})\partial_{k}A_{t}\partial_{i}A_{t}-g^{it}g^{jk}\partial_{k}A_{t}F_{ij}+\frac{1}{4}g^{ik}g^{jl}F_{kl}F_{ij},
\end{eqnarray}
and the Hamiltonian density
\begin{eqnarray}
H&=&-g^{tt}|\partial_{t}\phi|^{2}+2g^{ti}\operatorname{Re}[-iA_{t}\phi D_{i}^{*}\phi^{*}]+g^{ij}D_{i}\phi D_{j}^{*}\phi^{*}+(m^{2}+g^{tt}A_{t}^{2})|\phi|^{2}\nonumber\\
&&-\frac{1}{2}(g^{tt}g^{ik}-g^{tk}g^{it})(\partial_{t}A_{i}\partial_{t}A_{k}-\partial_{i}A_{t}\partial_{k}A_{t})-g^{kt}g^{ji}F_{kj}\partial_{i}A_{t}+\frac{1}{4}g^{ik}g^{jl}F_{kl}F_{ij}.
\end{eqnarray}
Then the kinetic energy density
\begin{equation}\label{kinetic_density}
H-V=-g^{tt}|\partial_{t}\phi|^{2}-\frac{1}{2}(g^{tt}g^{ik}-g^{tk}g^{it})\partial_{t}A_{i}\partial_{t}A_{k}
\end{equation}
is again positive definite. Under the Eddington-Finkelstein coordinates and the radial gauge, this kinetic energy density again vanishes even for non-static configurations. For our double scalar field model, the generalization of the above discussion is straightforward.

However, in holography, the above quantities generally diverge due to the asymptotic behavior approaching the AdS boundary, which need to be renormalized. In the standard procedure of holographic renormalization (see, e.g. \cite{Skenderis}), there will be a boundary counter term
\begin{equation}
B=\int L_{B}\sqrt{-\bar{g}}d^{d-1}\bar{x}
\end{equation}
added to the original action in order for the on-shell action and other quantities to be finite when the boundary tends to the AdS conformal boundary. This counter term leads to a ``boundary energy momentum tensor''
\begin{equation}\label{boundary_EM}
\bar{T}^{\mu\nu}=\frac{2}{\sqrt{-\bar{g}}}\frac{\delta B}{\delta\bar{g}_{\mu\nu}}=2\frac{\partial L_{B}}{\partial\bar{g}_{\mu\nu}}+\bar{g}^{\mu\nu}L_{B},
\end{equation}
which gives a boundary contribution
\begin{equation}
\omega_{B}=-\bar{T}_{t}^{t}\sqrt{-\bar{g}}
\end{equation}
to the grand potential density $\omega$. The static part of (\ref{boundary_EM}) gives a boundary contribution to the static grand potential density $\omega_s$, too.

Anyway, the landscape picture for holographic systems is just like that in the scalar field case. However, in the case involving a gauge field, it turns out that only the holographic boundary condition of fixed total particle number is possible for the dynamic evolution\cite{TWZ}, i.e. the holographic systems can be sourceless for the free energy $F$ instead of the grand potential $\Omega$. Accordingly, only the free energy has the property of monotonic decreasing
\begin{equation}\label{decrease}
\frac{dF}{dt}\le 0
\end{equation}
during a dynamical process without driving. The full equations of motion have a form similar to (\ref{full}), where all the time derivative terms are on the left hand side and the right hand side is a functional gradient of $\Omega_s$. But the potential landscape should be considered as the landscape of the static free energy\footnote{\label{about_F_s} Note that $F_s$ is not $F(\partial_t=0)$, though they are equal for static, on shell configurations. Generally, from the relation (\ref{relation}) and the definition (\ref{F_s}) of $F_s$ we have
\begin{equation}
	F\cong\mathbf{T}+F_s
\end{equation}
with $\mathbf{T}$ the total kinetic energy, which is the integral of (\ref{kinetic_density}) for the coupled Maxwell-scalar case.}
\begin{equation}\label{F_s}
F_s:=\Omega_s+\int\mu\rho\sqrt{-g}d^{d-1}\vec{x},
\end{equation}
with the static equations of motion
\begin{equation}\label{static_F}
\delta F_s=0.
\end{equation}
Actually, the form of the above equation is the same as $\delta\Omega_s=0$, while the only difference is the boundary condition (fixing the total particle number (\ref{number}) versus fixing $\mu$). So the dynamics is just a point particle rolling on the landscape with friction (dissipation). Eventually, it will tend to stop at some local minimum of the landscape, i.e. a static state satisfying (\ref{static_F}) and $\delta^2 F_s\ge 0$. In particular, under the Eddington-Finkelstein coordinates, since the kinetic energy always vanishes, the dynamics is extremely clear as the point particle keeping rolling down the landscape with all the decrease of its potential energy directly dissipated.

In the above discussion, we have considered the most general inhomogeneous configurations. To make things simpler, we can also consider the landscape of free energy (grand potential) for the homogeneous configurations, where all the spatial derivatives (other than $z$) are dropped from the quantities in the above discussion, and then we do not need to distinguish between $F_s$ (or $\Omega_s$) and their densities $f_s=\omega_s+\mu\rho$ (or $\omega_s$). In particular, sometimes we would like to explore the spacial variation of the local state in an inhomogeneous static configuration while ignoring the spacial gradient, e.g. as in the intuitive discussion of the domain wall configuration in Sec.~\ref{wall}. In this case, since the chemical potential $\mu$ is the same everywhere in this static system (see the discussion around (\ref{eq:F_vary}) below), it is especially convenient to consider this simpler landscape of grand potential density $\omega_s$. So our discussion here offers a concrete description of the physical picture in the previous two subsections.

In either case, we can explore the landscape by the following two ways:
\begin{itemize}
	\item The global way: finding extremal points on the landscape by solving the static equations of motion
	\item The local way: investigating linear perturbations, i.e. quasi-normal modes (QNM), around the extremal points
\end{itemize}
In principle, we can find all the extremal points on the landscape, i.e. all the static configurations, by solving the equations of motion, though it is rather difficult in practice. Then we should find out if those extremal points are local minimums or just saddle points.\footnote{To determine the global minimum from the local ones, one just needs to compare the values of their grand potentials.} Without friction (dissipation), whether the extremal points are local minimums or saddle points is directly related to the linear stability of the corresponding static configurations, as we know from fundamental courses of physics. But in the following we will argue that even in the presence of dissipation it is also the case, i.e. we can learn from the QNM whether the extremal points are local minimums or not.

For simplicity, we use the scalar field case to illustrate our argument, keeping in mind that it also works for more general cases. We start from an extremal point on the free energy landscape, which means $\delta F_s=0$ there, and consider the linear perturbation around that point. Suppose the dominant mode (the quasi-normal mode with largest imaginary part) is $\omega$, so the perturbation $\delta\phi$ of the real scalar field takes the form
\begin{equation}
\delta\phi(t)=f e^{-i\omega t}+f^{*}e^{i\omega^{*}t}.
\end{equation}
From the previous discussion we know
\begin{equation}\label{F_stable}
\delta F=-\int g^{tt}(\partial_{t}\phi)(\partial_{t}\delta\phi)\sqrt{-g}d^{d}x+\delta F_{s}=0,
\end{equation}
and
\begin{equation}\label{F_second}
	\delta^2 F\ge\delta^2 F_s=\int (\frac{1}{2}g^{ij}\partial_{i}\delta\phi\partial_{j}\delta\phi+\frac{1}{2}m^{2}\delta\phi^{2})\sqrt{-g}d^d x=:(\delta\phi,\delta\phi)_s,
\end{equation}
where we have defined the second order static free energy as an inner product on the functional space of $\delta\phi$.\footnote{Generally, the inner product is given by the Hessian of the static free energy at the extremal point that we start with. In the free scalar field case above, the Hessian is constant, so the inner product does not depend on the background fields that are perturbed.} (In the case involving gauge fields, the relations similar to (\ref{F_stable}) and (\ref{F_second}) only hold on shell, as mentioned in Footnote \ref{about_F_s}, but that is not a problem because the linear perturbations here do be on shell.) So we have
\begin{eqnarray}
\delta^2 F&\ge&(\delta\phi,\delta\phi)_s=(f e^{-i\omega t}+f^{*}e^{i\omega^{*}t},f e^{-i\omega t}+f^{*}e^{i\omega^{*}t})_s\nonumber\\
&=&(f,f)_s e^{-2i\omega t}+2(f,f^{*})_s e^{2\operatorname{Im}(\omega) t}+(f^{*},f^{*})_s e^{2i\omega^{*}t}\nonumber\\
&=&2(f,f^{*})_s e^{2\operatorname{Im}(\omega) t}+2\operatorname{Re}[(f,f)_s e^{-2i\omega t}],
\end{eqnarray}
where we have used the fact that the inner product is symmetric. Thus, $\delta^2 F$ cannot be monotonically decreasing in the course of time for the case of a positive definite $(,)_s$ if $\operatorname{Im}(\omega)>0$, contradicting the fact (\ref{decrease}) when turning off the external sources (that is just the condition of considering QNM).

Conversely, if $(,)_s$ is not positive definite (but is of full rank), we can choose an initial $\delta\phi$ such that
\begin{equation}
(\delta\phi,\delta\phi)_s<0.
\end{equation}
In this case, at least one of the quasi-normal frequencies $\omega$ should have a positive imaginary part, since otherwise the expansion
\begin{equation}
\delta\phi(t)=\sum_{i}(f_{(i)}e^{-i\omega_{(i)}t}+f_{(i)}^{*}e^{i\omega_{(i)}^{*}t})
\end{equation}
clearly shows that $\delta\phi$ will tend to zero eventually, again contradicting the fact (\ref{decrease}).

In sum, there is or is not an unstable mode if $(,)_s$ is indefinite (but of full rank) or positive definite, respectively. In between, $(,)_s$ will be degenerate (actually positive semi-definite), where it has a zero mode and $\omega=0$ becomes a quasi-normal frequency. Therefore, in this mechanism, the system loses its stability always by a quasi-normal mode crossing the real axis at exactly zero.

For the s+s holographic superfluid model considered in this paper, we have mentioned that there are three most relevant equilibrium states involved in the first order phase transition, the stable state, the meta-stable state and the unstable state, among which the first two are (local) minimums in the landscape while the last one is a saddle point. By numerical calculation of the quasi-normal modes, we show in Appendix \ref{sec:QNM} that the first two are indeed dynamically stable and the last one is dynamically unstable.

\section{Mixture configuration and domain wall in 1D}\label{sec:DomainWall}

\subsection{The domain wall from a generalized thermodynamic point of view}\label{thermodynamics}

Since the holographic superfluid model provides a description, valid
at all scales, of certain strongly coupled superfluid systems, it serves as a very powerful tool to study dynamic (far from equilibrium) processes
of inhomogeneous configurations, even out of local equilibrium\citep{Liu,DNTZ,LTZ}.
In general cases without local equilibrium, it is expected that most
thermodynamic quantities cannot be well defined for the system, or in other words, if the (local) thermodynamic quantities can be defined in some way at all, most (local) thermodynamic relations cannot hold\citep{Reichl}.
But in the context of holographic investigations, there have been
hints that some generalization of non-equilibrium thermodynamics (hydrodynamics)
can be applied to such systems\citep{TWZ}. Systems with first order
phase transitions have inhomogeneous (mixture) configurations of different
phases, which should be the ideal arena to carefully investigate to
what extent a generalized thermodynamic description works for such
systems. Therefore, we shall try to make it clear in this section.
First, we propose a general formalism as follows.

In principle, a complete formalism should involve backreaction of
the matter fields onto the bulk geometry, but that will render the
problem extremely complicated\citep{TWZ}. Here, for simplicity, we
stay in the probe limit, tailoring the discussion to the holographic
superfluid model in this paper. In the probe limit, the temperature of a holographic model is fixed due to the fixed bulk spacetime.

For a static configuration at finite temperature, we have
\begin{equation}
-I_{\mathrm{bulk}}(\beta,\mu)=\beta\Omega_s
\end{equation}
with $I_{\mathrm{bulk}}$ the (renormalized) bulk on-shell Euclidean
action, $\beta$ the inverse temperature and $\Omega_s$ the static grand
potential defined in (\ref{Omega_s}), which is the grand potential of the boundary system by the holographic duality. Besides the inverse temperature $\beta$ that is fixed
to be homogeneous in the probe limit, the chemical potential
$\mu$ and the grand potential are originally defined in global equilibrium,
which can be generalized to the case of local equilibrium if the spatial
variation of the configuration is gentle enough. In order to cope with
local structures like domain walls, which obviously break local equilibrium
due to the small scales characterizing them, now $\Omega_s$ is taken as the (generalized) grand potential of the boundary system
even for arbitrary static configurations, i.e. beyond local equilibrium,
where the (generalized) chemical potential $\mu(\vec{x})=A_{t}(\vec{x})|_{z=0}$
can vary acutely in space. For the variation of the bulk on-shell
action, we have
\begin{equation}
\delta I_{\mathrm{bulk}}=\beta\int\rho(\vec{x})\delta\mu(\vec{x})d^{d-1}\vec{x}
\end{equation}
with $\rho(\vec{x})$ the local particle number density, so the
grand potential $\Omega_s$ satisfies
\begin{equation}
\frac{\delta\Omega_s}{\delta\mu(\vec{x})}=-\rho(\vec{x})
\end{equation}
as a generalization of the usual thermodynamic relation. Then the (generalized)
free energy $F_s$ is given by the following functional Legendre transform (as the time-independent version of (\ref{F_s})):
\begin{equation}\label{Legendre}
F_s=\Omega_s+\int\mu(\vec{x})\rho(\vec{x})d^{d-1}\vec{x}
\end{equation}
with the transformed on-shell action
\begin{equation}
-\tilde{I}_{\mathrm{bulk}}(\beta,\rho)=\beta F_s,
\end{equation}
which yields the variation
\begin{equation}
\delta F_s=\int\mu(\vec{x})\delta\rho(\vec{x})d^{d-1}\vec{x}.\label{eq:F_vary}
\end{equation}
Under the boundary condition that there is no particle exchange between
the system and the environment or there is no boundary, e.g. periodic
boundary conditions, which means that the total particle number
\begin{equation}\label{number}
N=\int\rho(\vec{x})d^{d-1}\vec{x}
\end{equation}
is a constant, a static configuration should satisfy (\ref{static_F}). Combined with (\ref{eq:F_vary})
and
\[
\delta\int\rho(\vec{x})d^{d-1}\vec{x}=0,
\]
the condition (\ref{static_F}) gives the result that $\mu$ is a
constant (independent of $\vec{x}$). We call it the chemical balance
condition for static configurations without local equilibrium, which
is a generalization of the corresponding condition in the ordinary
thermodynamics and will be confirmed in our numerical evolution of
configurations containing domain walls.

In the absence of local equilibrium, e.g. at the domain wall, the
standard local thermodynamic relation
\begin{equation}
\omega=-p\label{eq:Gibbs}
\end{equation}
(the Gibbs-Duhem relation) with $\omega$ the local grand potential
density (i.e. the grand potential per unit volume) and $p$ the
pressure is not expected to hold, if the pressure can be reasonably defined at all\cite{XDEMTX}. However, away from the wall, the relation
(\ref{eq:Gibbs}) should hold due to the restoration of local equilibrium, similar to the case discussed in \cite{XDEMTX} (with solitons instead of domain walls). Actually, in the probe limit here, we cannot calculate the pressure from the holographic stress-energy tensor, so cannot really discuss the relation (\ref{eq:Gibbs}), but can instead define the pressure $p$ using this relation away from the wall and check the validity of its consequences, as will be shown in the following subsection and Sec.~\ref{sec:Bubble}. We hope that future investigations of our model beyond the probe limit would provide further verification on this point.

Now consider the surface tension of the domain wall. The surface tension
coefficient $\sigma$ is defined as the external work $W$ done (under
certain conditions) to enlarge the domain wall by a unit area. Besides
the isothermal condition, the certain conditions for our static inhomogeneous
configurations here also include the chemical balance condition described
above. Under those conditions, the external work is just the increase
of the grand potential due to the appearance of the
domain wall:
\begin{equation}
W=\Omega-\Omega_{0}\label{eq:work}
\end{equation}
with $\Omega_{0}$ the grand potential of the corresponding
homogeneous system (without the domain wall). In the 1D domain wall
case, the pressures on different sides of the wall should be equal to the same value $p$, which can be called the mechanical balance condition, otherwise a variation of the position of the wall would result in a change of the total grand potential and so the system cannot be in a static state. Suppose $\omega_{1}$ and
$\omega_{2}$ are the grand potential densities of the two phases
separated by (and far away from) the domain wall, respectively. Then
we have
\begin{equation}
\omega_{1}=-p=\omega_{2}.
\end{equation}
Actually, we see $\Omega_{0}=-pV$. At the position of the wall,
$\omega$ will be significantly different from $-p$, which is the
real contribution to (\ref{eq:work}). So $\sigma$ is reduced to
the following integral:
\begin{equation}
\sigma=\int_{x_{1}}^{x_{2}}(\omega+p)dx \label{eq:sigma}
\end{equation}
with $x_{1}$ and $x_{2}$ the left and right edges of the domain
wall, respectively. The above arguments will be verified in the following
numerical evolution.

\subsection{Numerical evolution of the mixture configuration and domain wall in 1D} \label{sec:DomainWall_Num}

In order to evolve the system nonlinearly, we switch to the Eddington-Finkelstein coordinates, under which the metric of the bulk black hole becomes
\begin{equation}\label{Eddington}
  ds^{2}=-\frac{1}{z^{2}}\left(-f\left(z\right)dt^{2}-2dtdz+d\vec{x}^{2}\right).
\end{equation}
We have adopted the radial gauge $A_{z}=0$ when working in the Schwarzschild coordinates, which is not satisfied after the coordinate transformation. To preserve the radial gauge, a $U\left(1\right)$ gauge transformation is required:
\begin{align}
  A_{\mu}  & \to A_{\mu}+\partial_{\mu}\alpha,\label{eq:gauge-trans-A} \\
  \Psi_{1} & \to e^{ie_1\alpha}\Psi_{1},                               \\
  \Psi_{2} & \to e^{i\alpha}\Psi_{2},                                  \\
  \alpha   & = -\int \frac{A_t}{f}dz.\label{eq:gauge-trans-alpha}
\end{align}
As a result, $\Psi_1$ and $\Psi_2$ should not be taken real anymore. Moreover, we should recover the spatial direction $x$, so the equations of motion are
\begin{align}
  2\left(\partial_{t}-ieA_{t}\right)\partial_{z}\psi_{1}-ie\partial_{z}A_{t}\psi_{1}-\partial_{z}\left(f\partial_{z}\psi_{1}\right)-\left(\partial_{x}-ieA_{x}\right)^{2}\psi_{1}+\left(z+\lambda_{12}\left|\psi_{2}\right|^{2}\right)\psi_{1}                                 & =0 \label{eq:EOM_psi1}  \\
  2\left(\partial_{t}-iA_{t}\right)\partial_{z}\psi_{2}-i\partial_{z}A_{t}\psi_{2}-\partial_{z}\left(f\partial_{z}\psi_{2}\right)-\left(\partial_{x}-iA_{x}\right)^{2}\psi_{2}+\left(z+\lambda_{12}\left|\psi_{1}\right|^{2}\right)\psi_{2}                                    & =0 \label{eq:EOM_psi2}  \\
  \partial_{z}^{2}A_{t}-\partial_{z}\partial_{x}A_{x}+2\mathrm{Im}\left(e\psi_{1}^{*}\partial_{z}\psi_{1`}+\psi_{2}^{*}\partial_{z}\psi_{2}\right)                                                                                                                             & =0 \label{eq:EOM_At}    \\
  \partial_{t}\left(\partial_{z}A_{t}+\partial_{x}A_{x}\right)-f\partial_{z}\partial_{x}A_{x}-\partial_{x}^{2}A_{t}\nonumber                                                                                                                                                                             \\
  +2\mathrm{Im}\left(ef\psi_{1}^{*}\partial_{z}\psi_{1}-e\psi_{1}^{*}\partial_{t}\psi_{1}+f\psi_{2}^{*}\partial_{z}\psi_{2}-\psi_{2}^{*}\partial_{t}\psi_{2}\right)+2A_{t}\left(\left|e\psi_{1}\right|^{2}+\left|\psi_{2}\right|^{2}\right)                                    & =0\label{eq:constraint} \\
  2\partial_{t}\partial_{z}A_{x}-\partial_{z}\left(f\partial_{z}A_{x}\right)-\partial_{z}\partial_{x}A_{t}-2\mathrm{Im}\left(e\psi_{1}^{*}\partial_{x}\psi_{1}+\psi_{2}^{*}\partial_{x}\psi_{2}\right)+2A_{x}\left(\left|e\psi_{1}\right|^{2}+\left|\psi_{2}\right|^{2}\right) & =0 \label{eq:EOM_Ax}
\end{align}
Here, we have made replacements $\Psi_{i}=z\psi_{i}$ ($i=1,2$). We will adopt the same nonlinear time evolution scheme as in (for example) \cite{Du:2015zcb}, where \eqref{eq:constraint} is taken to be a constraint equation (see Appendix~\ref{sec:time_evolution} for more detailed discussion about time evolution). We impose Dirichlet boundary conditions for $\psi_1$ and $A_{\mu}$ at $z=0$, while a Neumann-like boundary condition is adopted for $\psi_{2}$\cite{LTZ}:
\begin{equation}
  \left.\partial_{t}\psi_{2}\right|_{z=0}=\left.\left(\partial_{z}\psi_{2}+iA_{t}\psi_{2}\right)\right|_{z=0}\label{eq:boundary-cond-psi2}.
\end{equation}
This boundary condition can be easily derived from the coordinate transformation of $\left.D_{z}\psi_{2}\right|_{z=0}=0$ in the Schwarzschild coordinates to the Eddington-Finkelstein coordinates. The two different boundary conditions for the two scalar fields correspond to the two different quantizations, respectively, taken in our model in Sec.~\ref{model}. For the $x$ direction, we impose periodic boundary conditions for all fields.

\begin{figure}
  \subfloat[$\left< O_1 \right>$ as a function of $x$]{\includegraphics[width=0.4\textwidth] {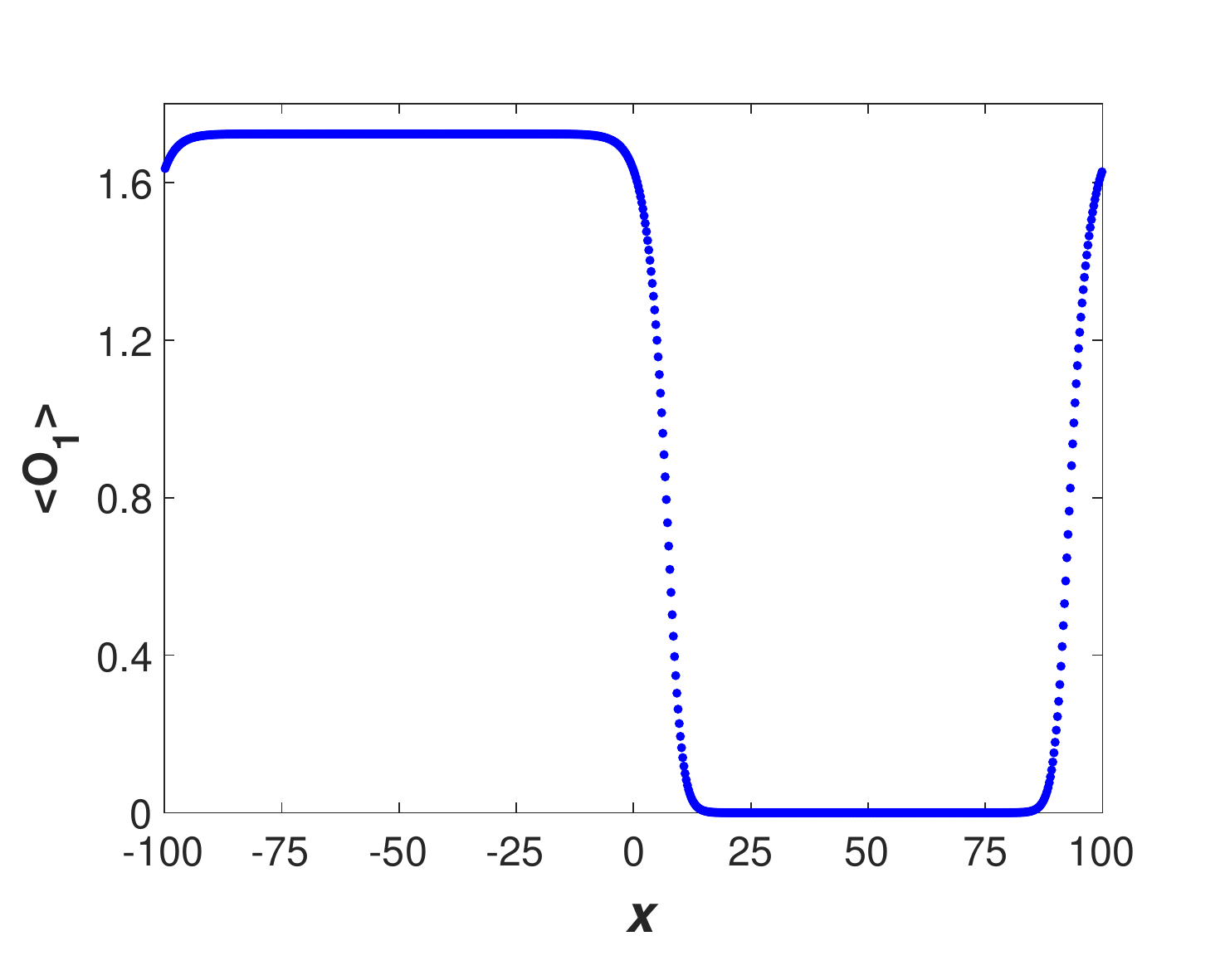}}
  \qquad
  \subfloat[$\left< O_2 \right>$ as a function of $x$]{\includegraphics[width=0.4\textwidth] {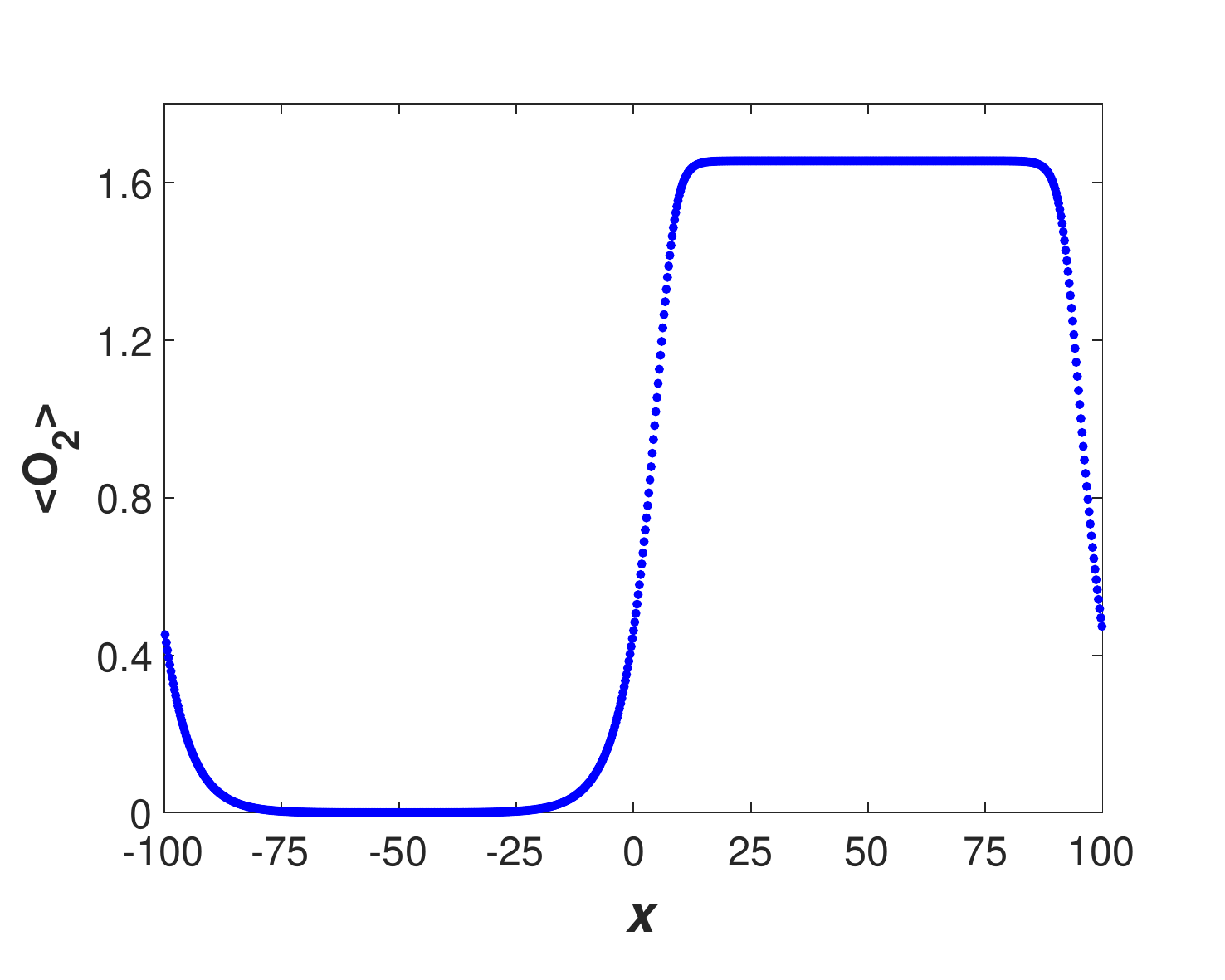}}

  \subfloat[$\rho$ as a function of $x$]{\includegraphics[width=0.4\textwidth] {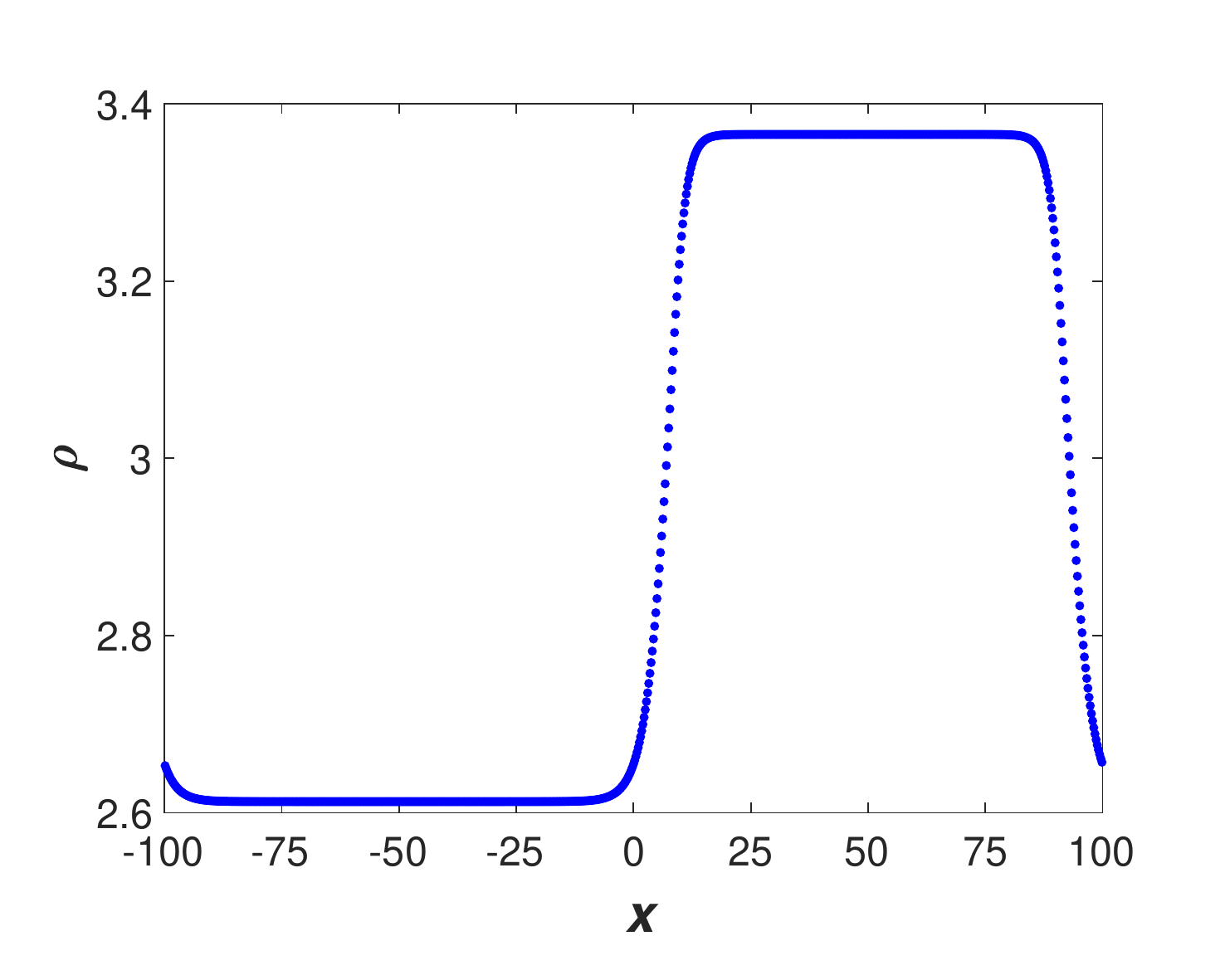}}
  \qquad
  \subfloat[Grand potential density as a function of $x$]{\includegraphics[width=0.4\textwidth] {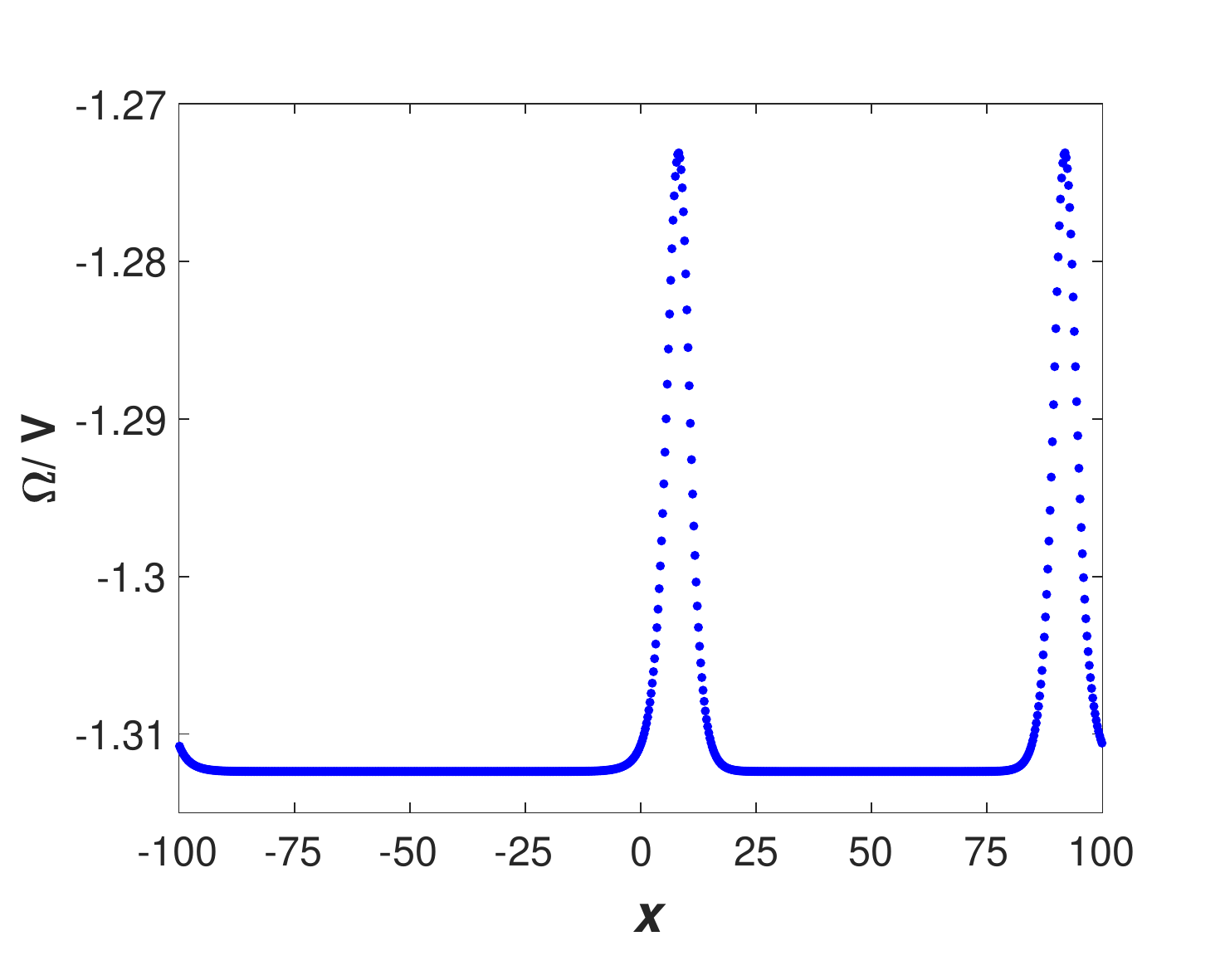}}
  \caption{\label{fig:1D_Pic_lmd_0.4} Physical quantities at $\lambda_{12}=0.4$ and $\mu=1.4351$, where $x \in \left[-100,100\right]$.}
\end{figure}

Formation of domain wall structures can be observed from nonlinear time evolution after perturbing the coexisting state by varying $\rho$. The form of the initial perturbation is\footnote{See appendix \ref{sec:QNM} for more detailed discussion about the perturbation form.}
\begin{equation}
  \delta\rho=A\sin\left(\frac {2\pi}{L}x\right).\label{eq:pert_rho}
\end{equation}
Here, $L$ is the length of the $x$ direction and $A$ is the amplitude for the perturbation. In our nonlinear time evolution, the coexisting state breaks down even when $A$ is very small ($\approx 10^{-12}$), which confirms us with the fact that this coexisting state is unstable. After a long time evolution, the 1D domain wall structure forms finally. Fig.~\ref{fig:1D_Pic_lmd_0.4} shows a typical result when $\lambda_{12}=0.4$ and $\mu=1.4351$. In this plot,  distributions of order parameters $\left< O_1 \right>$, $\left< O_2 \right>$, particle number density $\rho$ and grand potential density $\omega$ are given as functions of $x$, where the explicit form of grand potential density can be found in Appendix \ref{sec:GrandP}. In all these figures, it is obvious to find sharp structures at the center of domain wall. By employing the numerical integral, $\sigma$ defined in equation (\ref{eq:sigma}) can be calculated:
\begin{equation}
  \sigma=0.2634.\label{eq:tension}
\end{equation}

\begin{figure}
  \centering
  \subfloat[\label{fig:sigmas_diff_lmds}$\sigma$ as a function of $\lambda_{12}$.] {\includegraphics[width=0.32\textwidth]{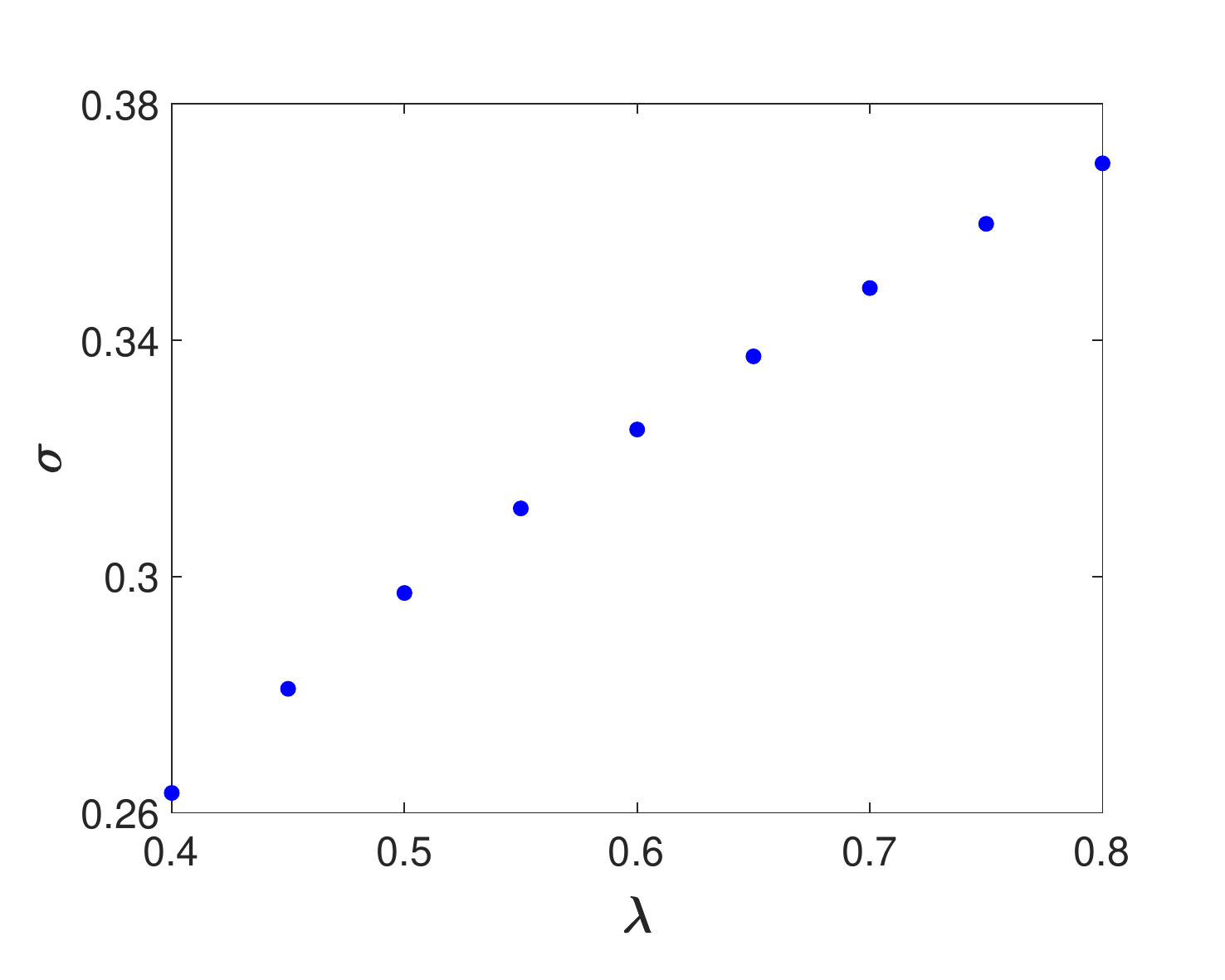}}
  \subfloat[\label{fig:Heights_diff_lmds}Relative heights of grand potential.]{\includegraphics[width=0.32\textwidth]{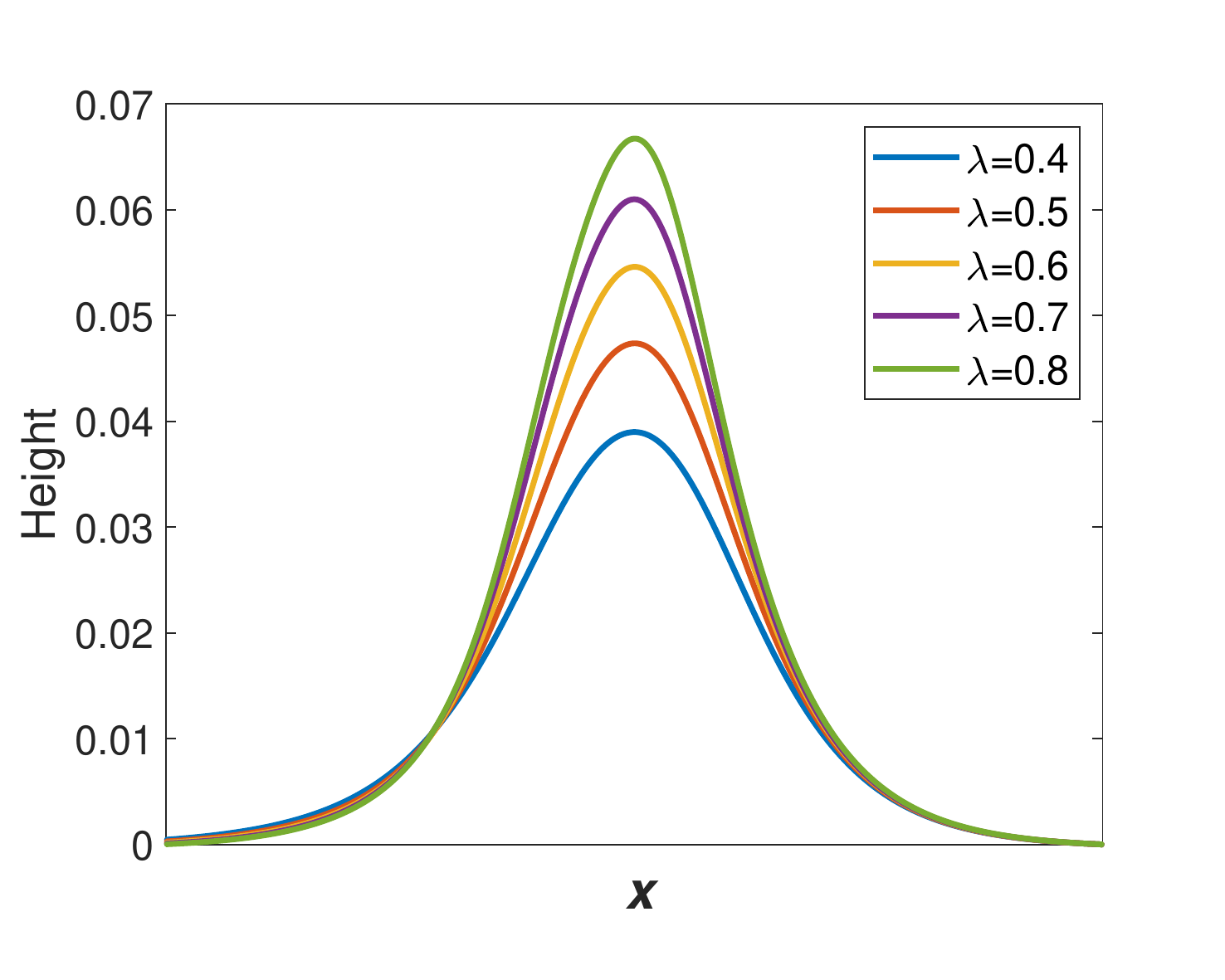}}
  \subfloat[\label{fig:Thickness_diff_lmds}The thickness of domain wall as a function of $\lambda_{12}$.]{\includegraphics[width=0.32\textwidth]{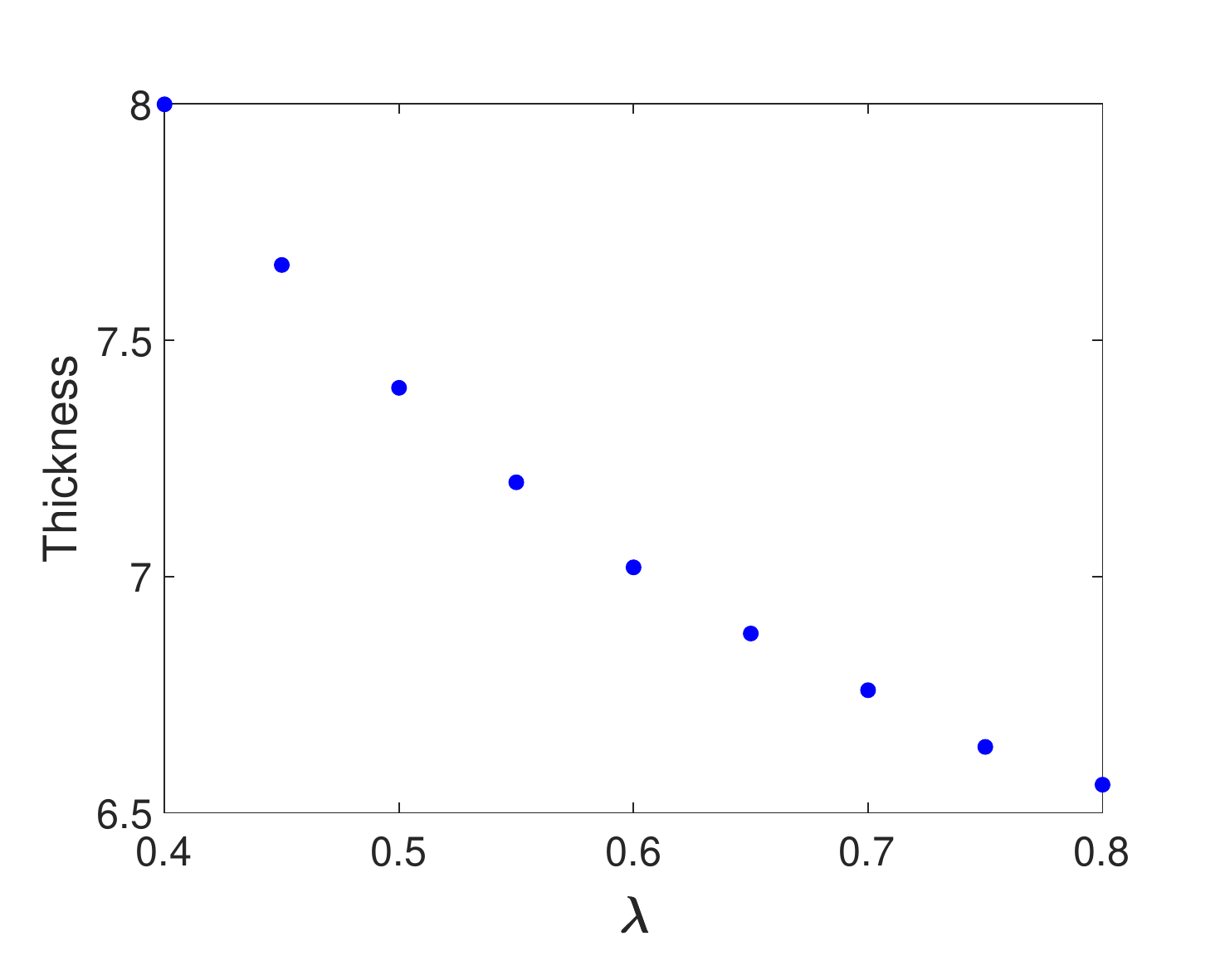}}
  \caption{\label{fig:1D_Tension} Physical quantities of the domain wall as functions of $\lambda_{12}$ when $\mu=1.4351$.}
\end{figure}

Similarly, we can evolve the system nonlinearly with different values of $\lambda_{12}$. Our results are shown in Fig.~\ref{fig:1D_Tension}, where $\mu=1.4351$, the same as the last paragraph. In Fig.~\ref{fig:sigmas_diff_lmds}, the tension $\sigma$ is calculated, in which an approximately linear relation can be witnessed between $\sigma$ and $\lambda_{12}$. In order to estimate the thickness of a domain wall, we calculate the relative height of the grand potential density $\omega$, where the ($\lambda_{12}$-independent) background value of $\omega$ far from the domain wall has been subtracted, as shown in Fig.~\ref{fig:Heights_diff_lmds}. To calculate the thickness of the domain wall, we define the region where relative heights are greater than $30\%$ of their maximum  to be wall areas. The results are shown in Fig.~\ref{fig:Thickness_diff_lmds}, from which we find that the thickness of the wall decreases with $\lambda_{12}$.

\section{Bubble nucleation and stabilization in 2D}\label{sec:Bubble}

In the previous section, we have investigated inhomogeneous holographic configurations from the generalized thermodynamic point of view, and then concretely calculated the surface tension of the domain wall from the final mixture state of a simple 1D evolution of our holographic superfluid model.

In this section, we will test in the 2D evolution the validity of our generalized thermodynamic picture and, in particular, the concrete value of the surface tension calculated in the 1D setup. In the 2D case, the domain wall is an effectively 1D object separating different phases, which can have certain shapes (and, in higher dimensions, also topologies) as bubbles. Then the surface tension just manifests itself as making the domain wall of least length (least area in the more familiar 3D case), i.e. making the bubbles round. Knowing this fact, we can easily write down the equilibrium conditions for such static round bubble configurations from the generalized thermodynamic point of view and then calculate the bubble size, which can be compared with the final bubble configuration of a 2D dynamic evolution of our model.

\subsection{Static 2D bubble configuration from thermodynamic and mechanical balance}\label{sec:calculate-r}

Now we deduce the equilibrium conditions for a static (round) bubble configurations. Without loss of generality, we only consider the case that Phase 1 is in the bubble. The volume of Phase 1 is
\begin{equation}
V_{1}=\pi r^{2}\label{eq:volume}
\end{equation}
with $r$ the radius of the bubble. The total volume $V$ and particle number $N$
are both fixed, so we have
\begin{eqnarray}
V_{1}+V_{2}&=&V,\label{eq:constrain_V}\\
\rho_{1}V_{1}+\rho_{2}V_{2}&=&N.\label{eq:constrain_N}
\end{eqnarray}
The equations of state read
\begin{eqnarray}
\rho_{1}&=&\rho_{1}(\mu),\label{eq:EOS_1}\\
\rho_{2}&=&\rho_{2}(\mu).\label{eq:EOS_2}
\end{eqnarray}
Here $\rho_{1}(\mu)$ and $\rho_{2}(\mu)$ are given functions characterizing
the properties of the two phases, respectively, which can be obtained numerically for our model, as shown in Appendix~\ref{sec:EOS}. The chemical potential $\mu$ is
the same for the phases in and outside the bubble, due to the chemical
balance condition proved in the last section. As usual in textbooks, we also have the mechanical balance
condition for the bubble:
\begin{equation}
p_{1}(\mu)-p_{2}(\mu)=\frac{\sigma}{r}.\label{eq:balance}
\end{equation}
Here $p_{1}(\mu)$ and $p_{2}(\mu)$ are also equations of state in our model, which are actually plotted in Fig.~\ref{FreeE} by virtue of $\omega=-p$ in equilibrium. So, in total, we have 6 equations (\ref{eq:volume}-\ref{eq:balance})
with 6 unknown variables $(r,V_{1},V_{2},\mu,\rho_{1},\rho_{2})$, which can be solved straightforwardly.

\subsection{Numerical evolution of bubble nucleation and stabilization} \label{sec:Bubbles_Num}

We will quench a meta-stable state locally. From numerical evolution, we find that proper quenches can trigger a (local) first order phase transition, eventually resulting in a stable bubble configuration under certain conditions as described in the previous subsection.

We will consider 2-dimensional configurations, so both the $x$ and $y$ directions must be included into the equations of motion. Boundary conditions for the $z$ direction are the same as the domain wall formation in the 1D case. ($A_y$ has same boundary condition as $A_x$.) For both the $x$ and $y$ directions, we impose the periodic boundary conditions. Without loss of generality, the meta-stable state that we are going to quench is the state with a single $\psi_{2}$ condensation. The amplitude of this local quench should be large enough to overcome the potential barrier, as described by the physical picture in Sec.~\ref{sec:model}.

\begin{figure}
  \subfloat[$t=200$]{\includegraphics[width=0.33\textwidth]{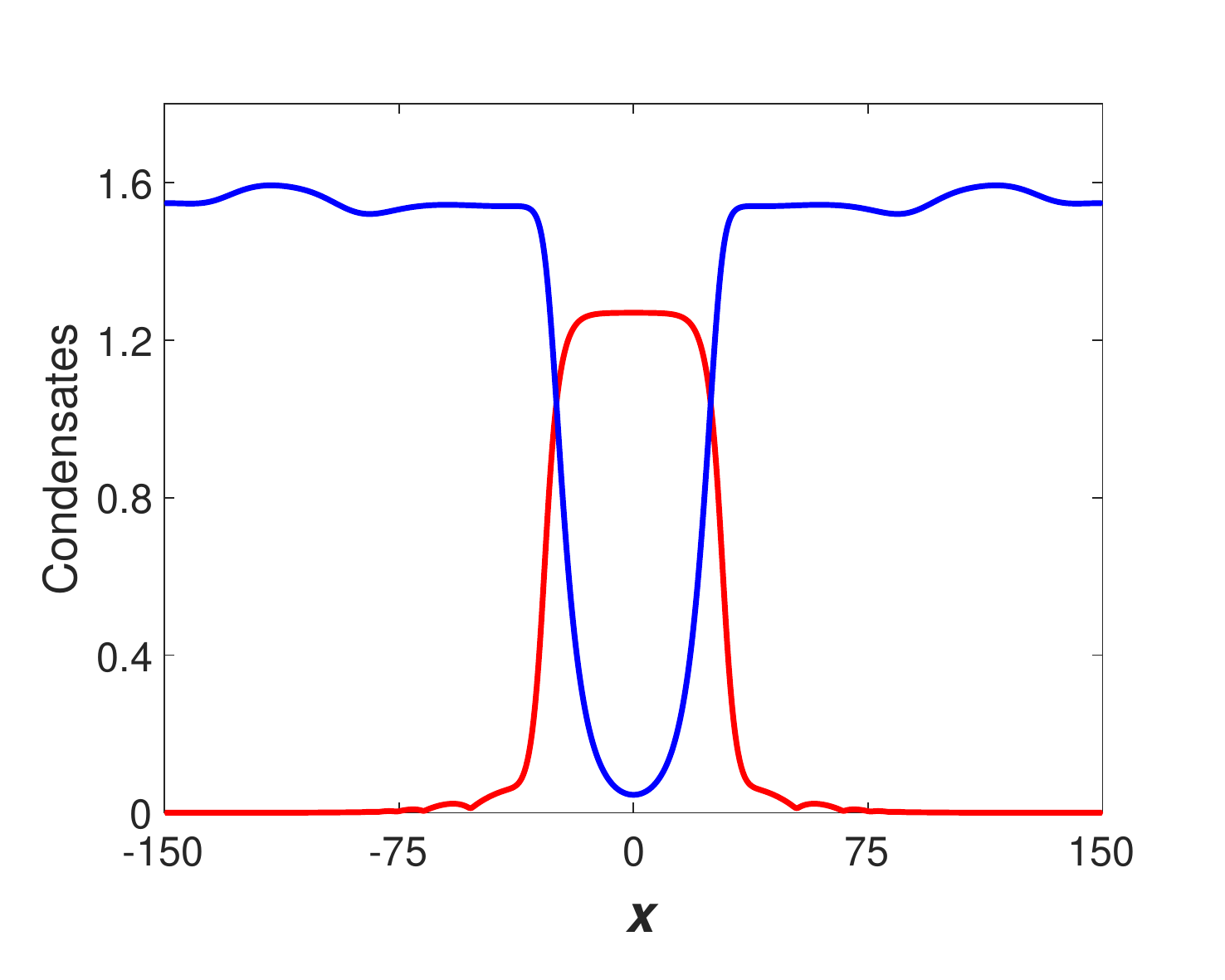}}
  \subfloat[$t=600$]{\includegraphics[width=0.33\textwidth]{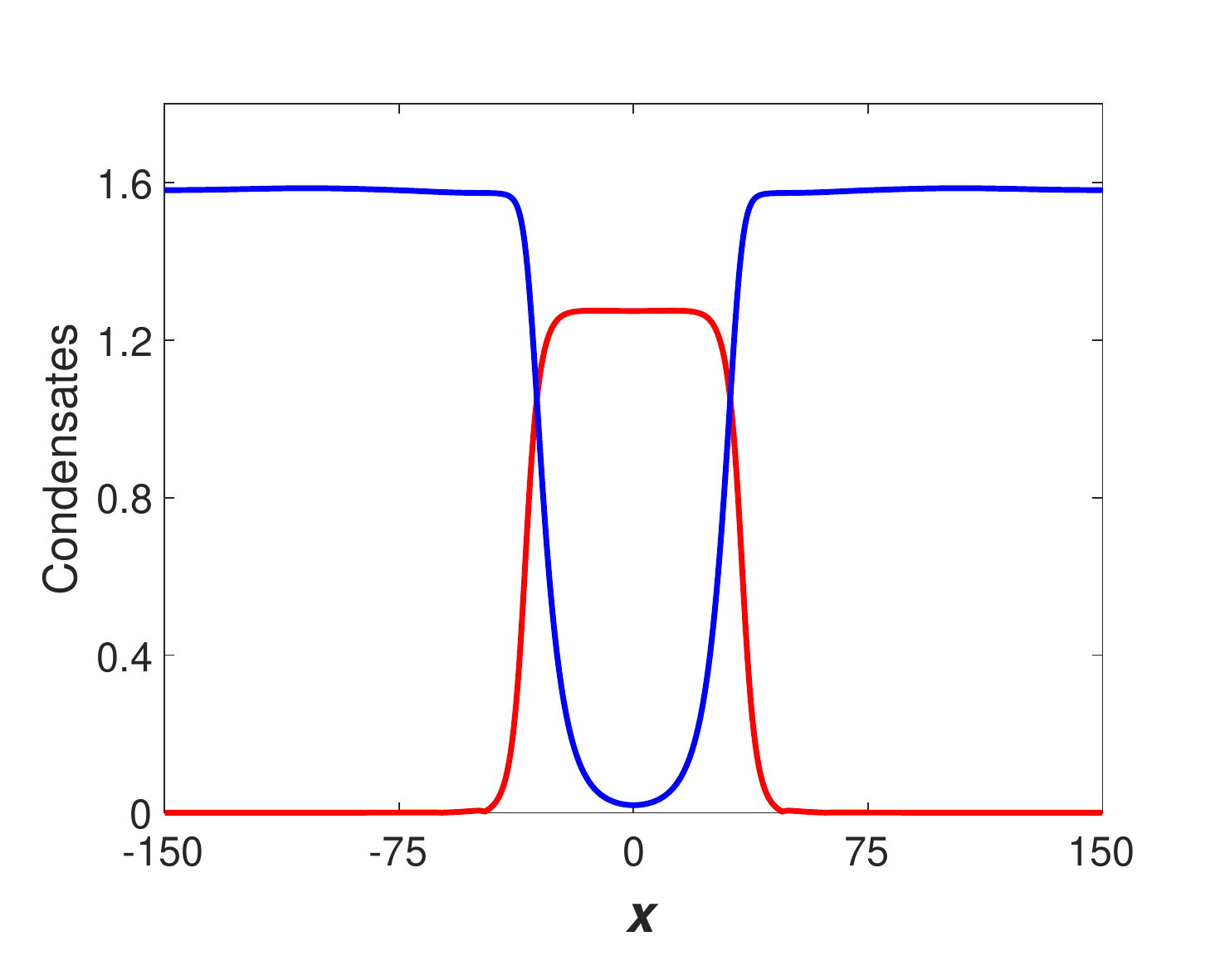}}
  \subfloat[$t=1000$]{\includegraphics[width=0.33\textwidth]{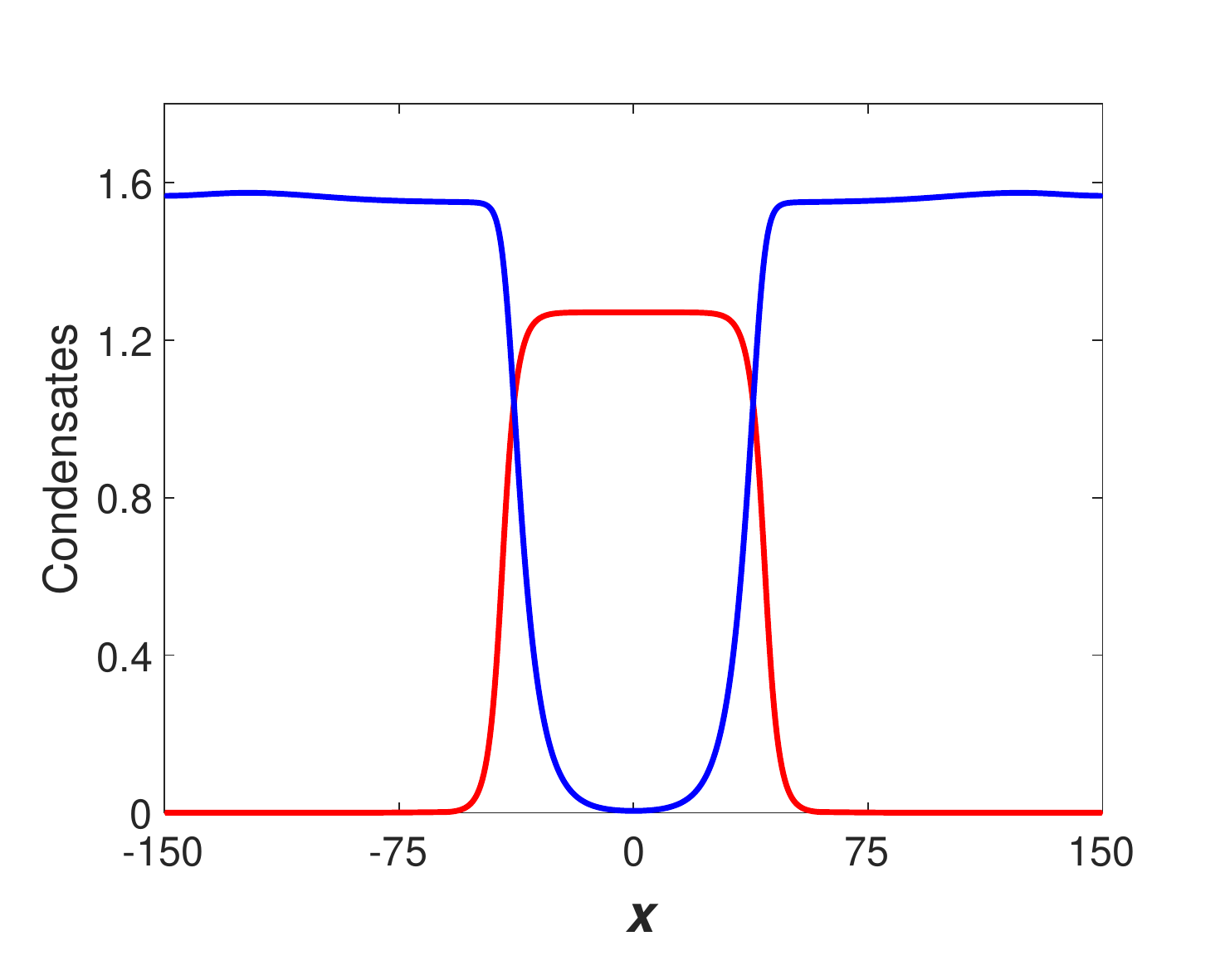}}

  \subfloat[$t=3000$]{\includegraphics[width=0.33\textwidth]{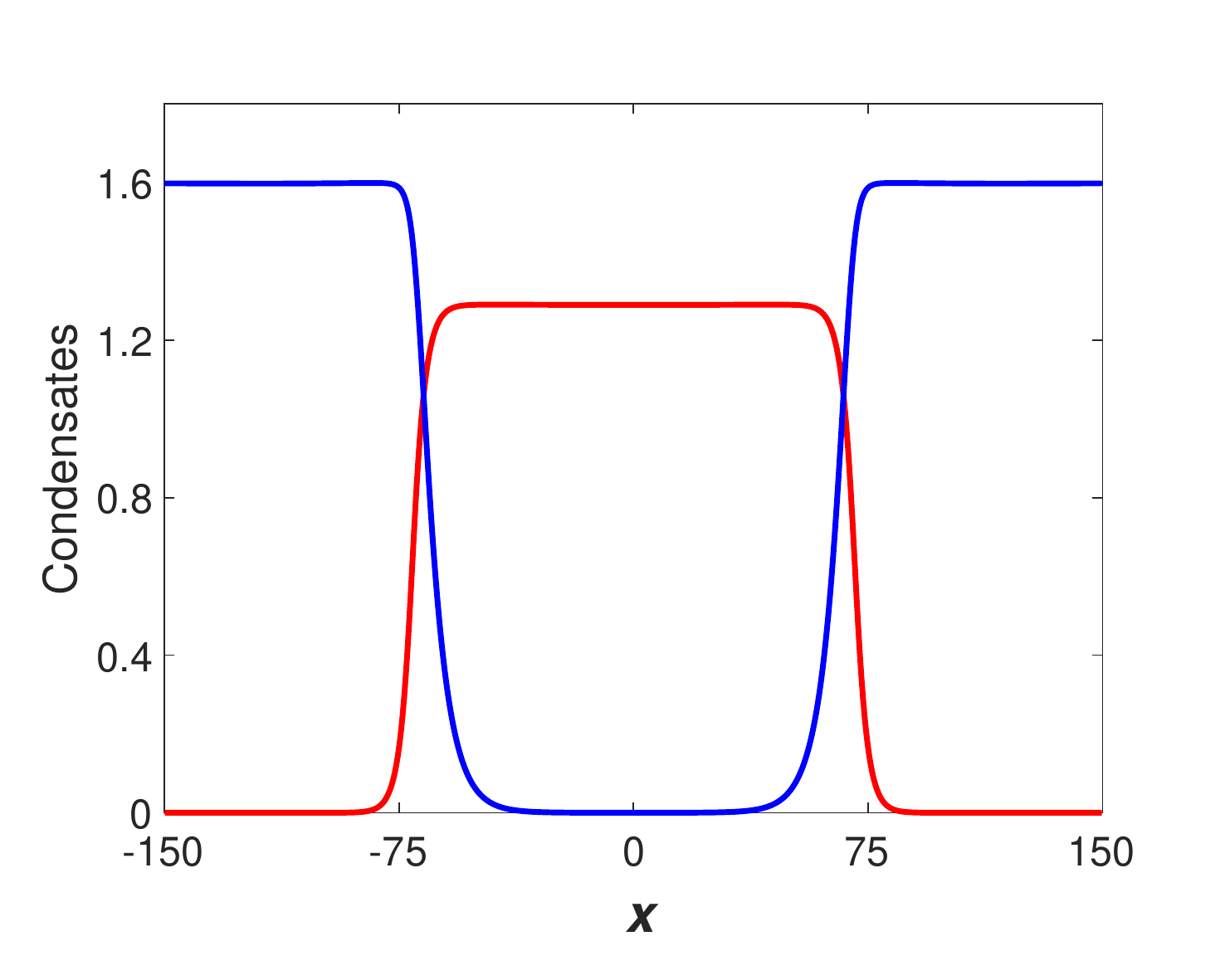}}
  \subfloat[$t=6000$]{\includegraphics[width=0.33\textwidth]{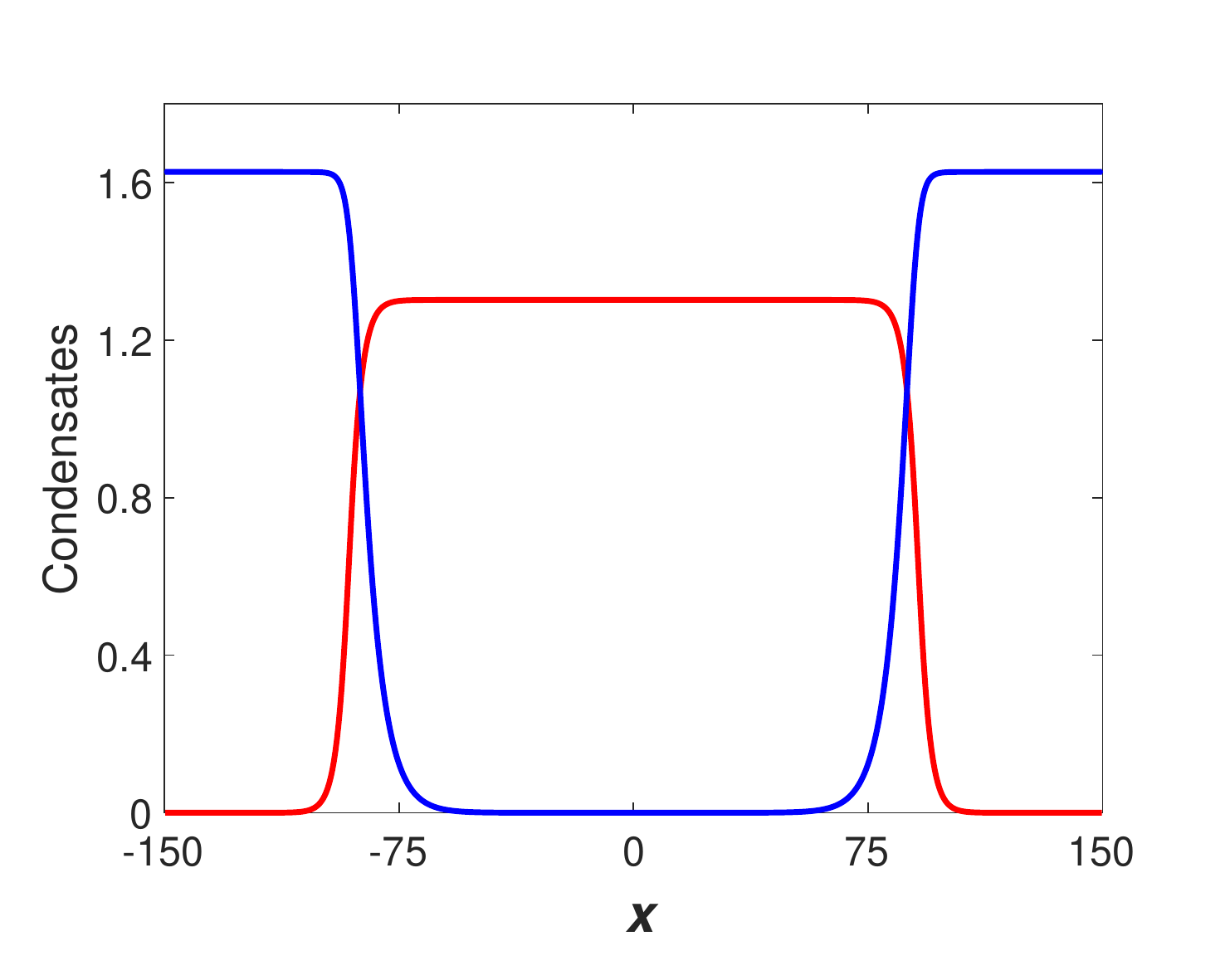}}
  \subfloat[$t=12000$]{\includegraphics[width=0.33\textwidth]{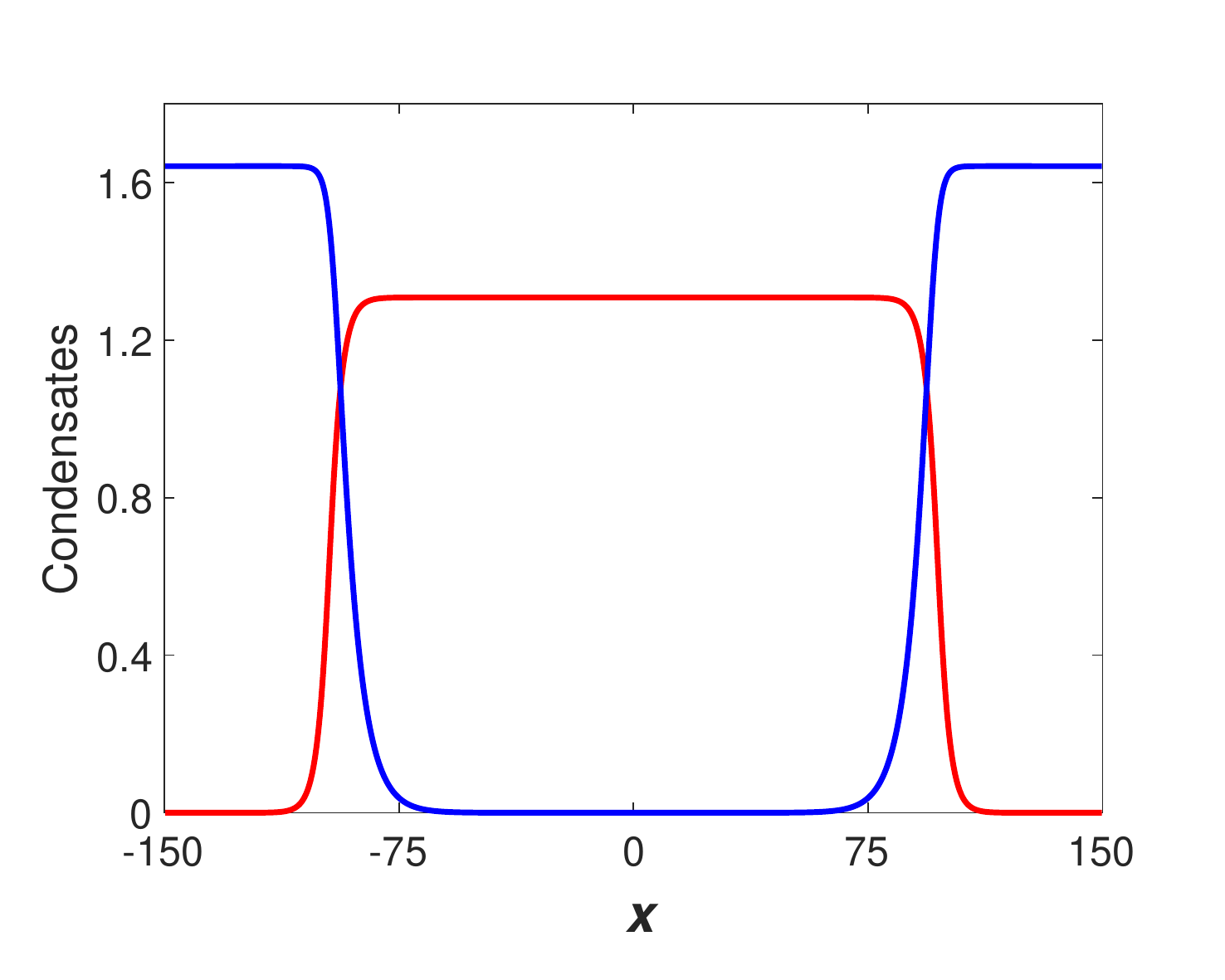}}
  \caption{\label{fig:Condensates12}Cross sections of condensates at different time $t$. In each subplot, red line denotes the condensate of $O_1$, and blue line denotes the condensate of $O_2$.}
\end{figure}

\begin{figure}
  \subfloat[\label{fig:r_DissP}Radius $r$ calculated from dissipation.] {\includegraphics[width=0.4\textwidth]{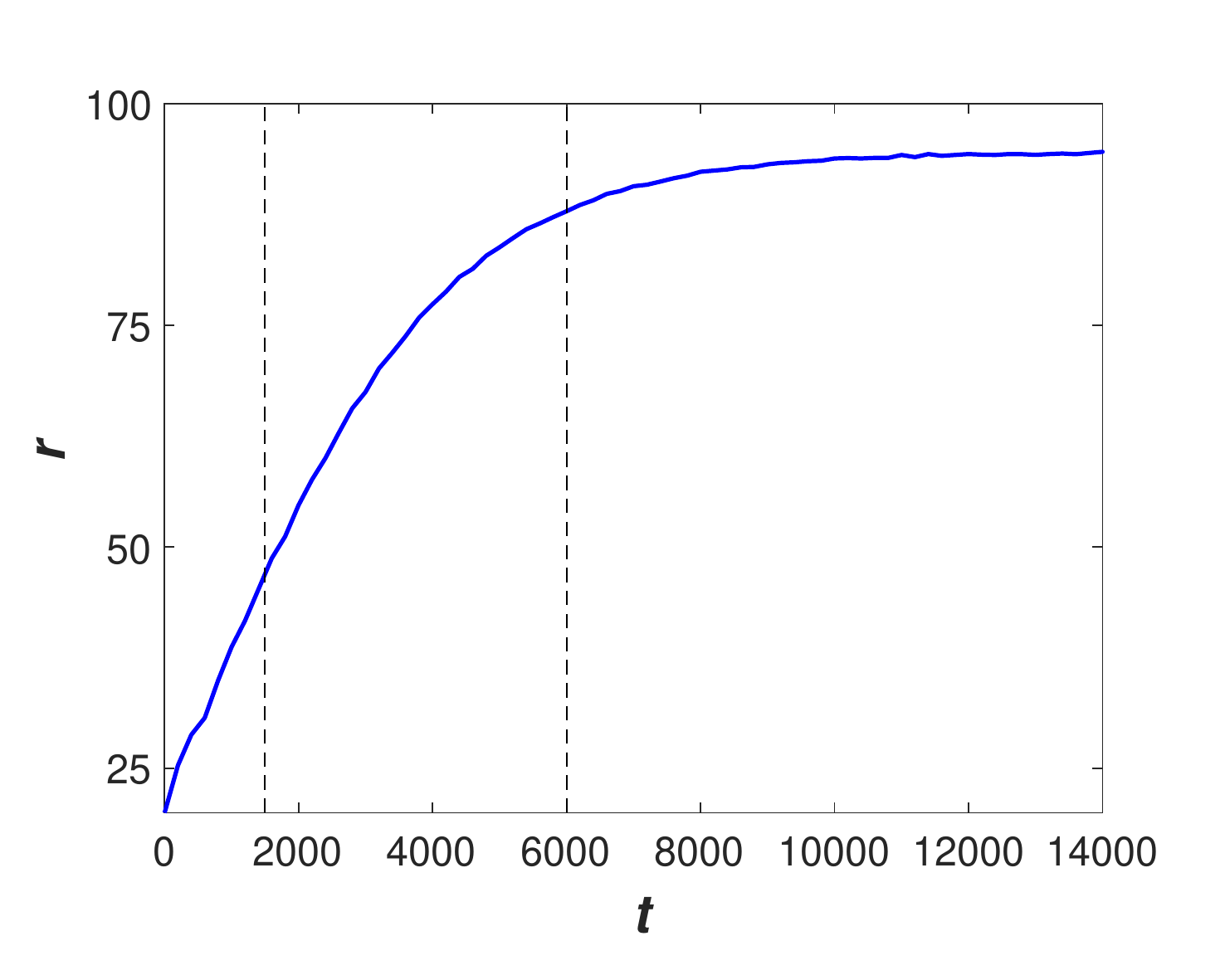}}
  \qquad
  \subfloat[\label{fig:r_GrandP}Radius $r$ calculated from grand potential density.] {\includegraphics[width=0.4\textwidth]{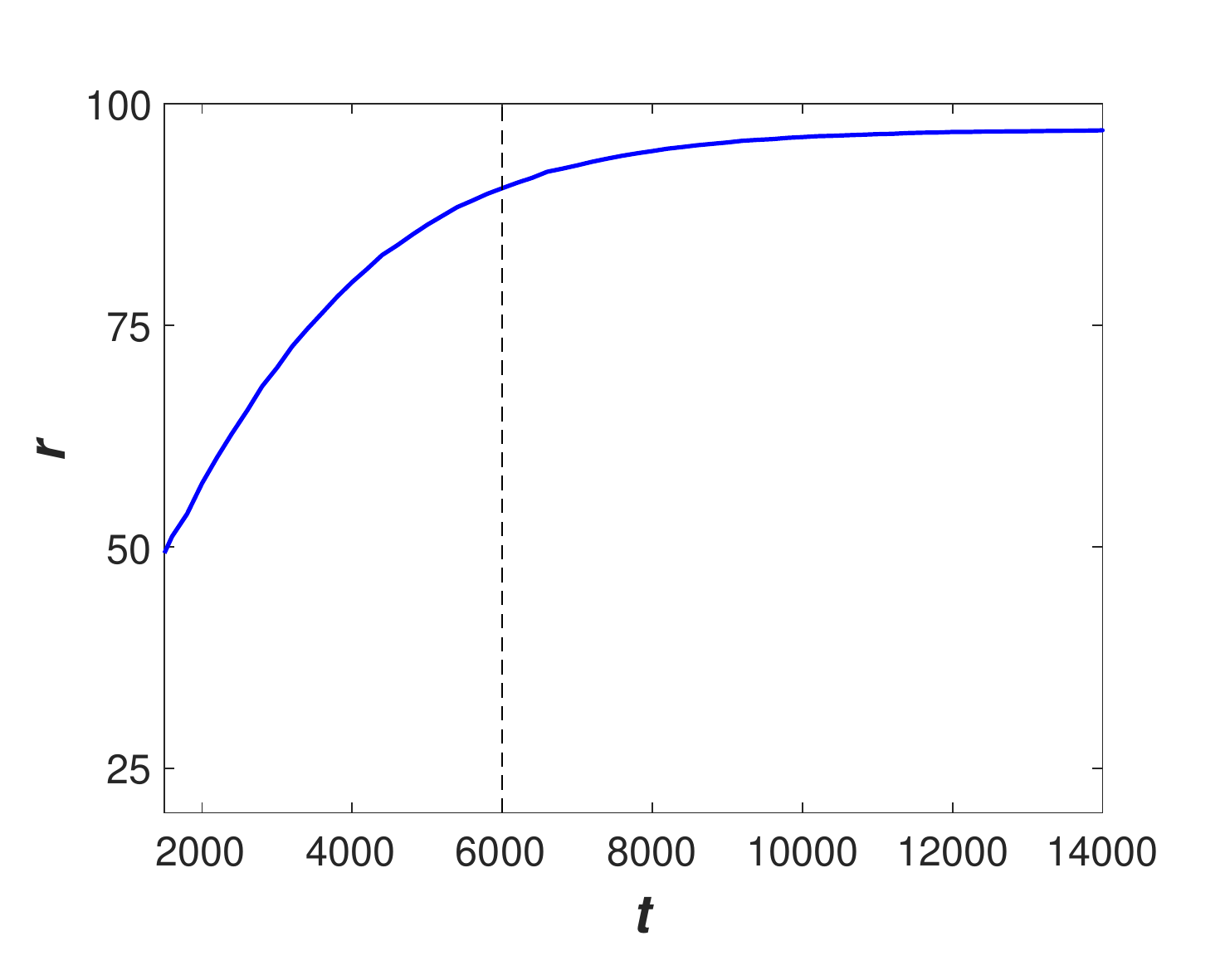}}
  \caption{Radius of the bubble as a function of time.}
\end{figure}

We leave the concrete form of the quench and details about time evolution in Appendix~\ref{sec:time_evolution}, and show the results of evolution directly. In Fig.~\ref{fig:Condensates12}, cross sections of the two condensates are plotted at different time $t$, where the growth of the bubble in the time evolution can be clearly witnessed.\footnote{Movies for time evolution are available at {[}http://people.ucas.edu.cn/\textasciitilde ytian?language=en\# 171556{]}, where evolutions of most concerned physical quantities, including $\left< O_1 \right>$, $\left< O_2 \right>$, density $\rho$ and grand potential density, can be found.} The time evolution can be divided into two stages: $t<1500$ and $t>1500$, which correspond to the bubble formation and expansion processes respectively. To characterize this formation process more precisely, we introduce a locally defined energy dissipation\cite{Liu}:
\begin{equation}
  Q \equiv \int \sqrt{-g} d^2x \left. T^z_t \right|_{z=1},
\end{equation}
where $T^M_N$ is energy momentum tensor. In (holographic) superfluids at finite temperature, dissipations occur mainly around moving local structures\cite{Liu}, so we expect that this dissipation reaches its maximum at the domain wall position. Thus, the distance between the position of maximum energy dissipation and origin can be adopted as the radius of the bubble. Moreover, the moment when distributions of all fields, including the dissipation distribution, becomes round can be defined as the bubble formation time, and this is the reason why we defined the evolution when $t<1500$ as the formation process of the bubble. Fig.~\ref{fig:r_DissP} shows the evolution of radius $r$ calculated from the dissipation. In this plot, the bubble expands all the time when $t<6000$ with its speed slowing down gradually. Then when $t>6000$, the radius $r$ tends to a constant. As the system approaches a stable state, the dissipation becomes rather small ($<3\times 10^{-9}$ at the final stage of our evolution), which in practice makes it not a proper method to determine the bubble size by the maximum of dissipation. However, as shown in Fig.~\ref{fig:r_DissP}, it is possible to obtain a rough size of the bubble from the dissipation even when $t<1500$ (in contrast to the grand potential density discussed below), so this method enables us to know more about the bubble radius at the initial stage.

\begin{figure}
  \subfloat[Modulus of $\left\langle \mathcal{O}_{1}\right\rangle$] {\includegraphics[width=0.4\textwidth]{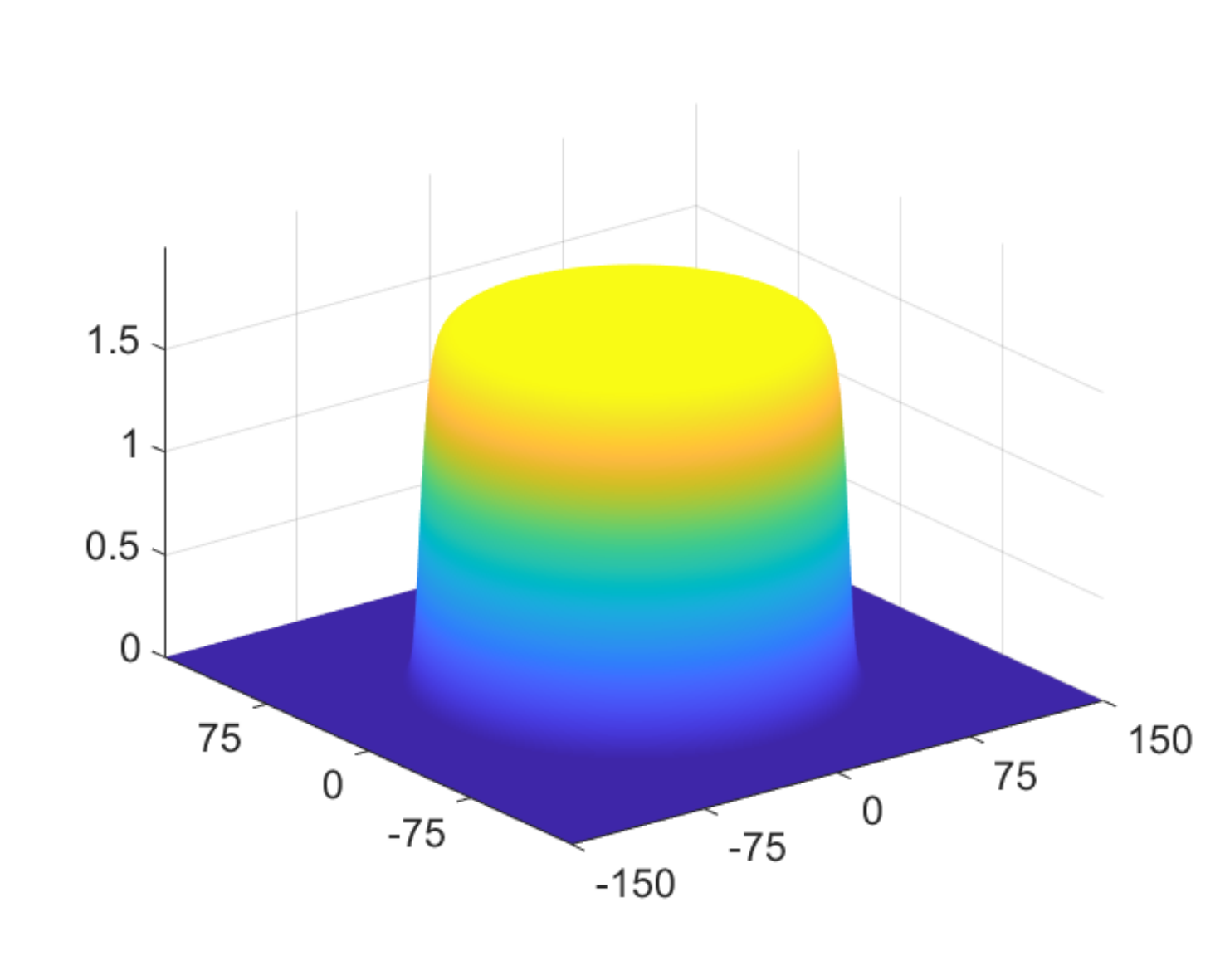}}
  \qquad
  \subfloat[Modulus of $\left\langle \mathcal{O}_{2}\right\rangle $] {\includegraphics[width=0.4\textwidth]{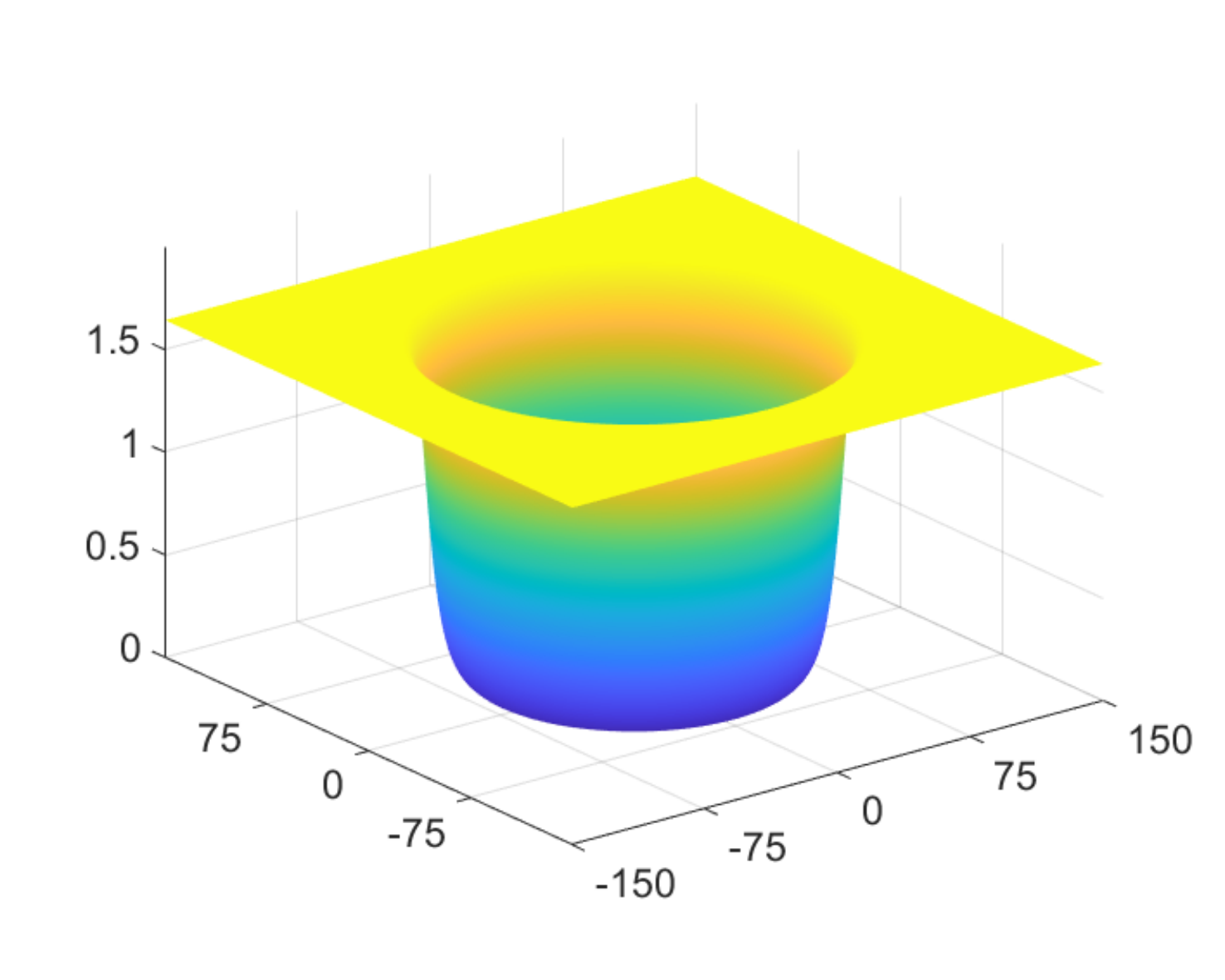}}

  \subfloat[Density $\rho$] {\includegraphics[width=0.4\textwidth]{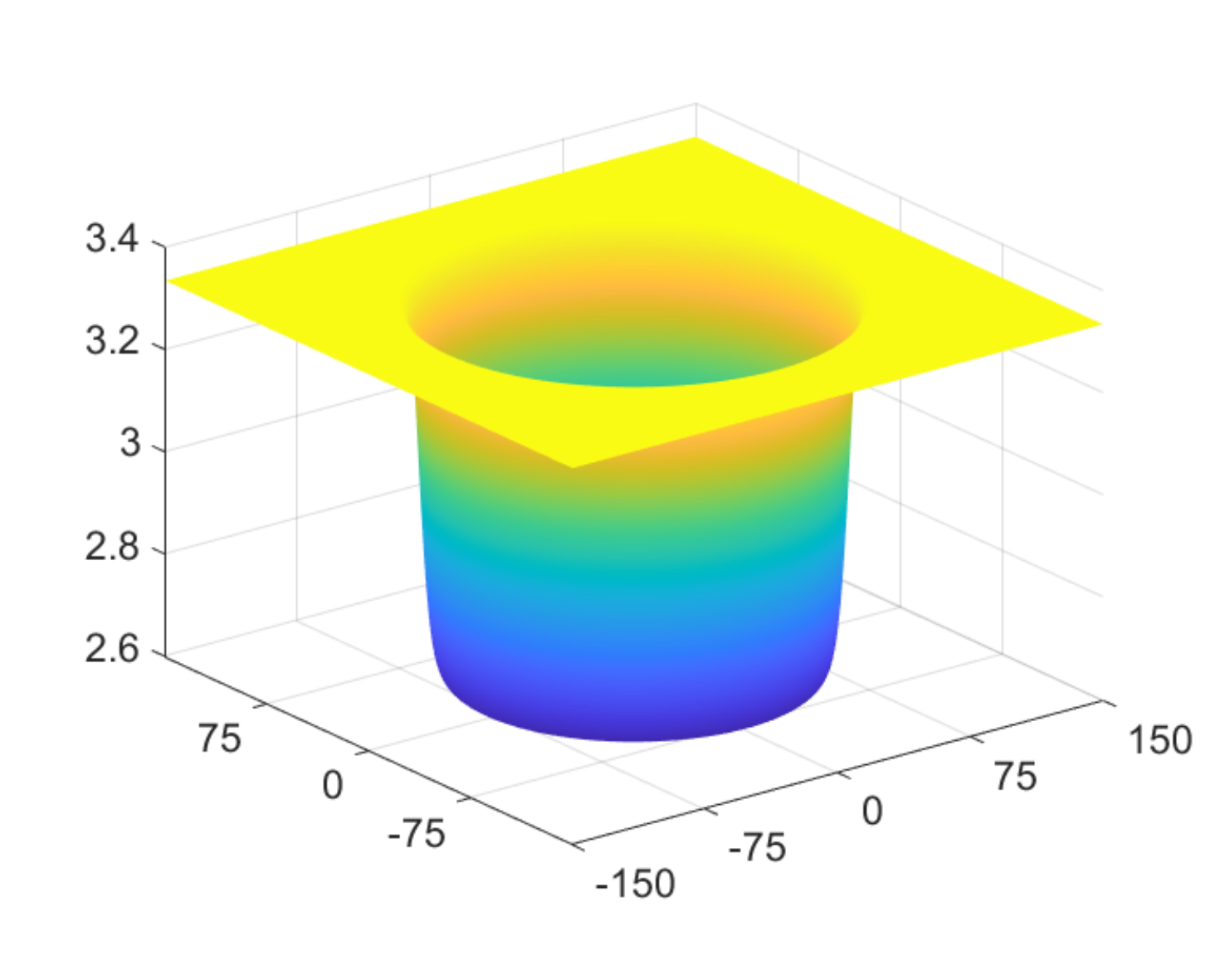}}
  \qquad
  \subfloat[\label{fig:Bubble-Grand-thermal-potential}Grand potential density $\omega$] {\includegraphics[width=0.4\textwidth]{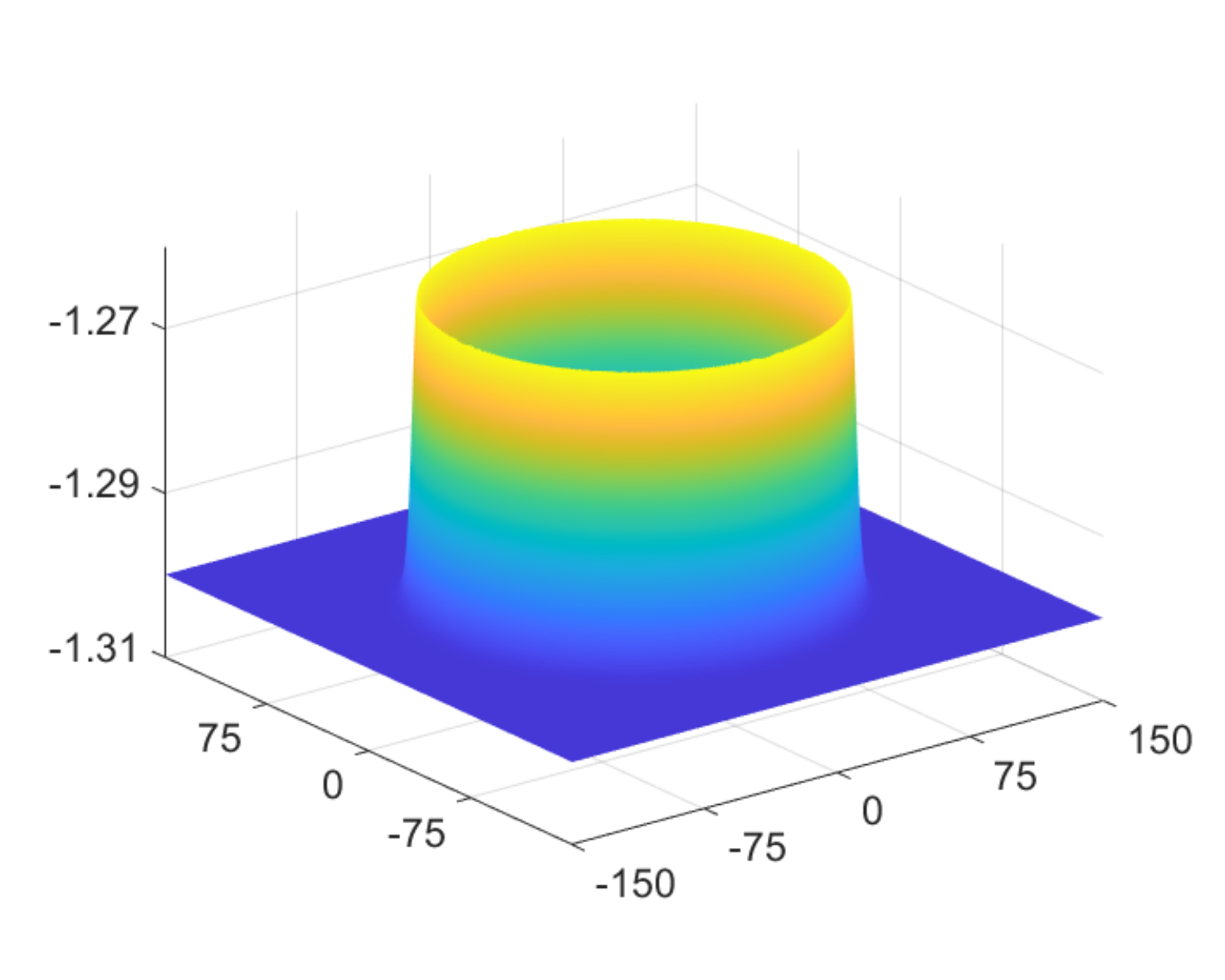}}
  \caption{\label{fig:Bubbles}Physical quantities as functions of spatial directions when the bubble stabilizes. In this plot, $\lambda_{12}=0.4$ and $\bar{\rho}=3.1$.}
\end{figure}

\begin{figure}
  \includegraphics[width=0.4\textwidth]{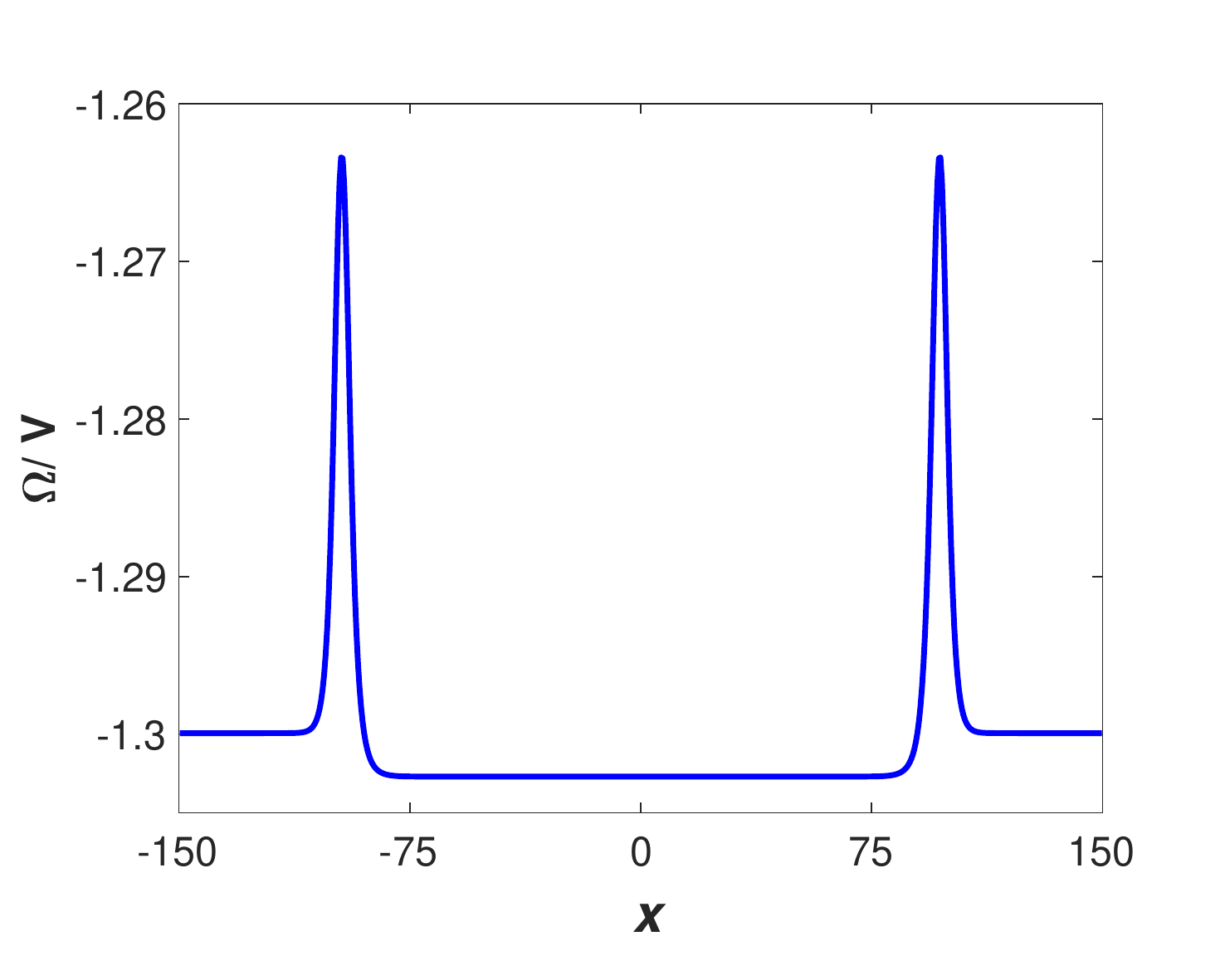}
  \caption{\label{fig:Bubble-GP-SD} The cross section of grand potential density $\omega$ at $y=0$.}
\end{figure}

Aside from the dissipation, the evolution of the radius $r$ can be determined by studying the dynamical grand potential density (see Appendix~\ref{sec:GrandP} for the concrete form of the grand potential). In this way, the radius $r$ is calculated by the distance between positions where the grand potential density reaches its maximum and the center of the system. Fig.~\ref{fig:r_GrandP} shows the result, where we can find a similar shape to that calculated by dissipation in Fig.~\ref{fig:r_DissP} when $t>1500$. The curve in Fig.~\ref{fig:r_GrandP} extends smoothly to the final state when the system stabilizes, so it is a better choice to describe the evolution of radius when $t$ is large. And we are able to read the radius of the bubble from Fig.~\ref{fig:r_GrandP} at the final stage:
\begin{equation}
  r=97.11.\label{eq:r}
\end{equation}
Moreover, configurations of the final stable bubble are plotted in Fig.~\ref{fig:Bubbles}, where, similar to Fig.~\ref{fig:1D_Pic_lmd_0.4}, sharp structures at the domain wall position are witnessed. In Fig.~\ref{fig:Bubble-GP-SD}, the cross section of the grand potential density at $y=0$ is plotted, in which the grand potential density inside the bubble is obviously smaller than that outside and so agrees with equation (\ref{eq:balance}).

\begin{figure}
  \subfloat[$t=200$]{\includegraphics[width=0.33\textwidth]{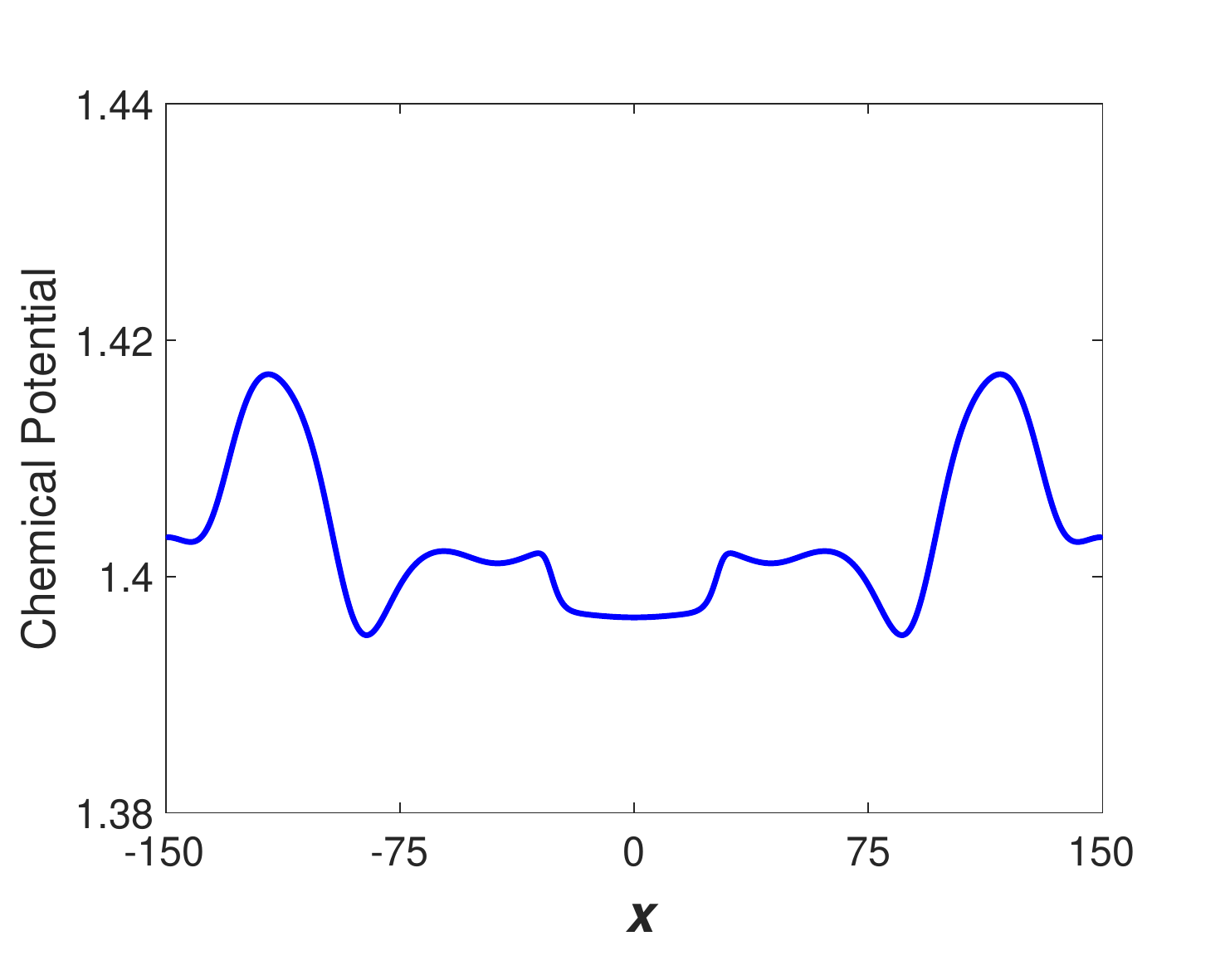}}
  \subfloat[$t=600$]{\includegraphics[width=0.33\textwidth]{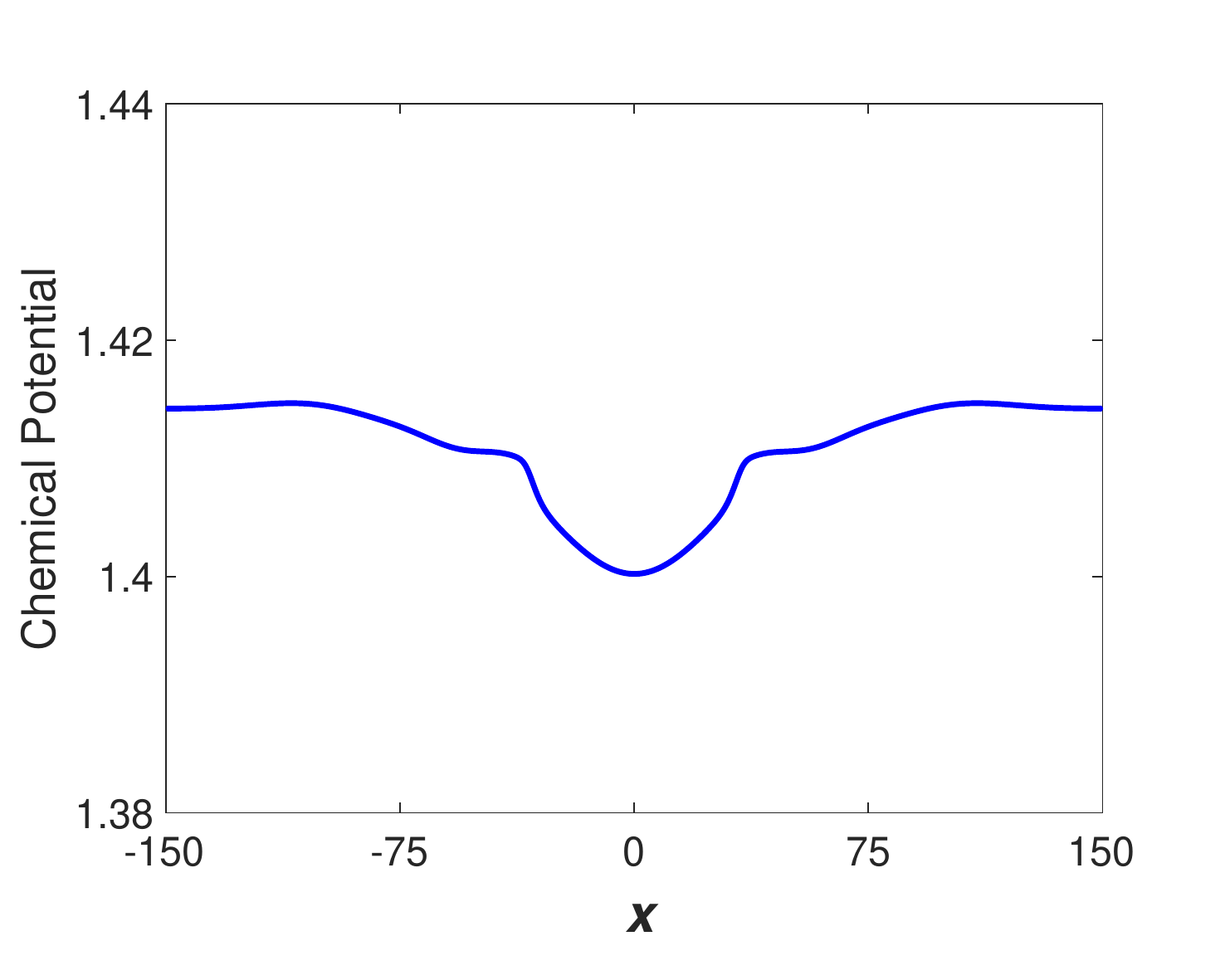}}
  \subfloat[$t=1000$]{\includegraphics[width=0.33\textwidth]{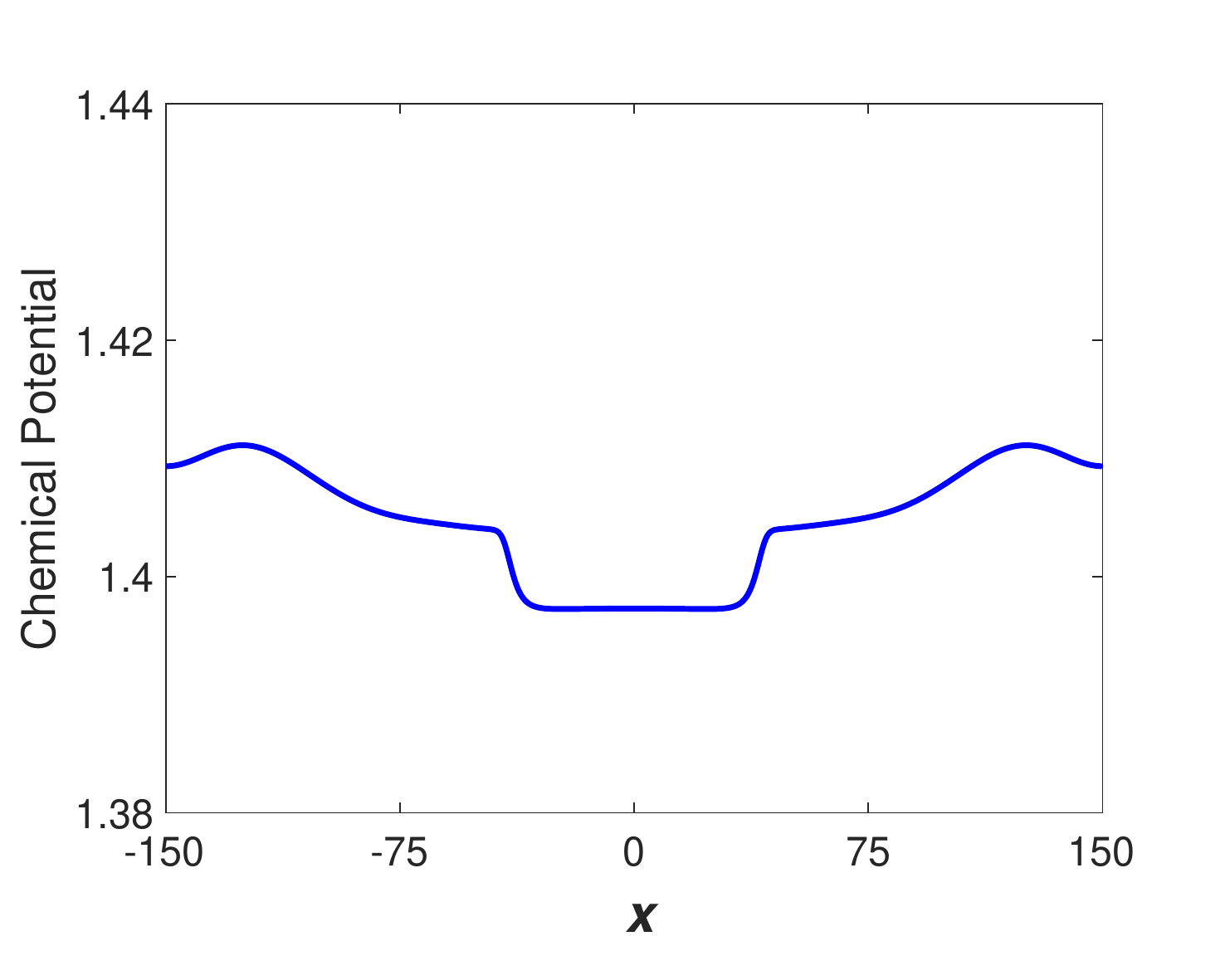}}

  \subfloat[$t=3000$]{\includegraphics[width=0.33\textwidth]{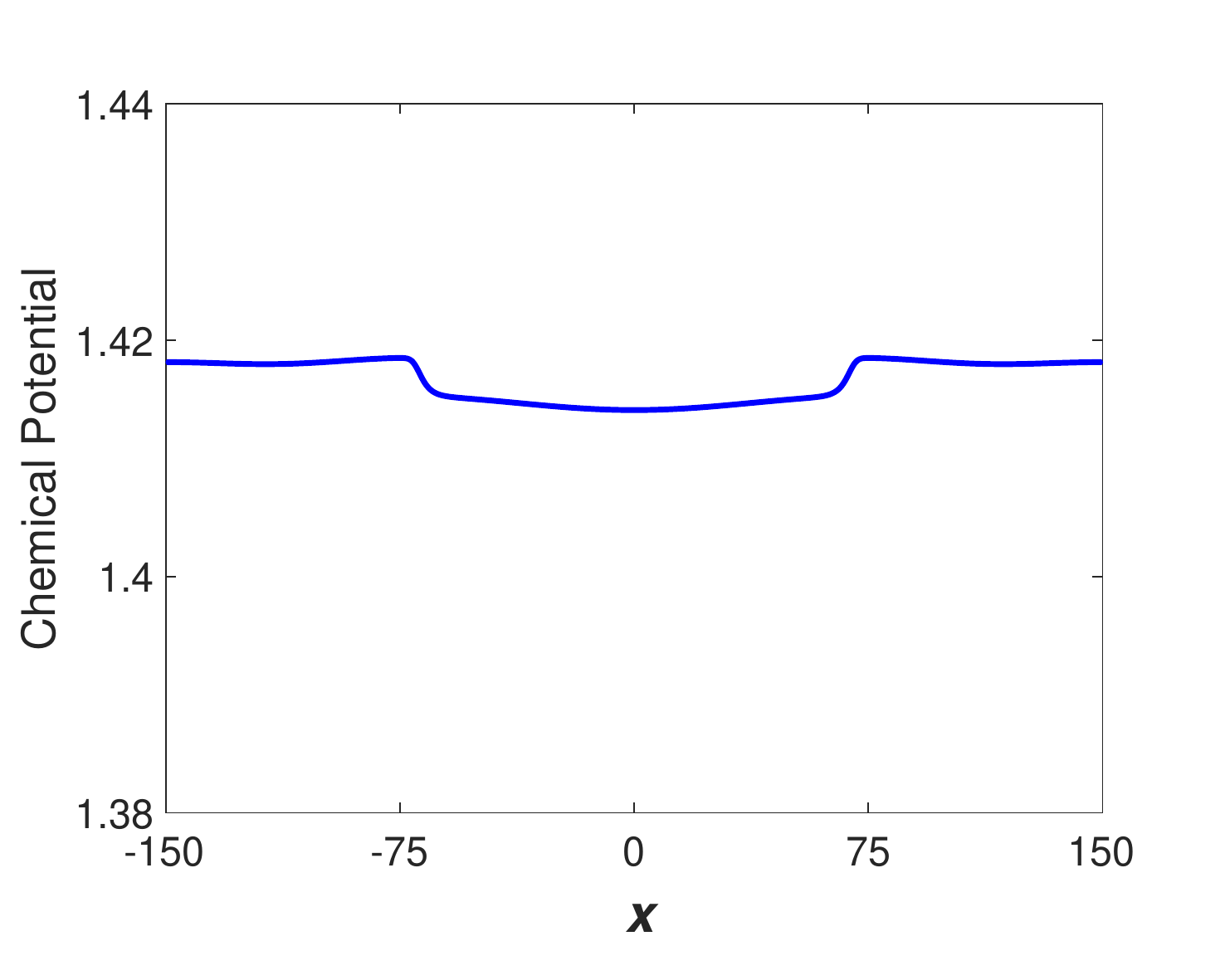}}
  \subfloat[$t=6000$]{\includegraphics[width=0.33\textwidth]{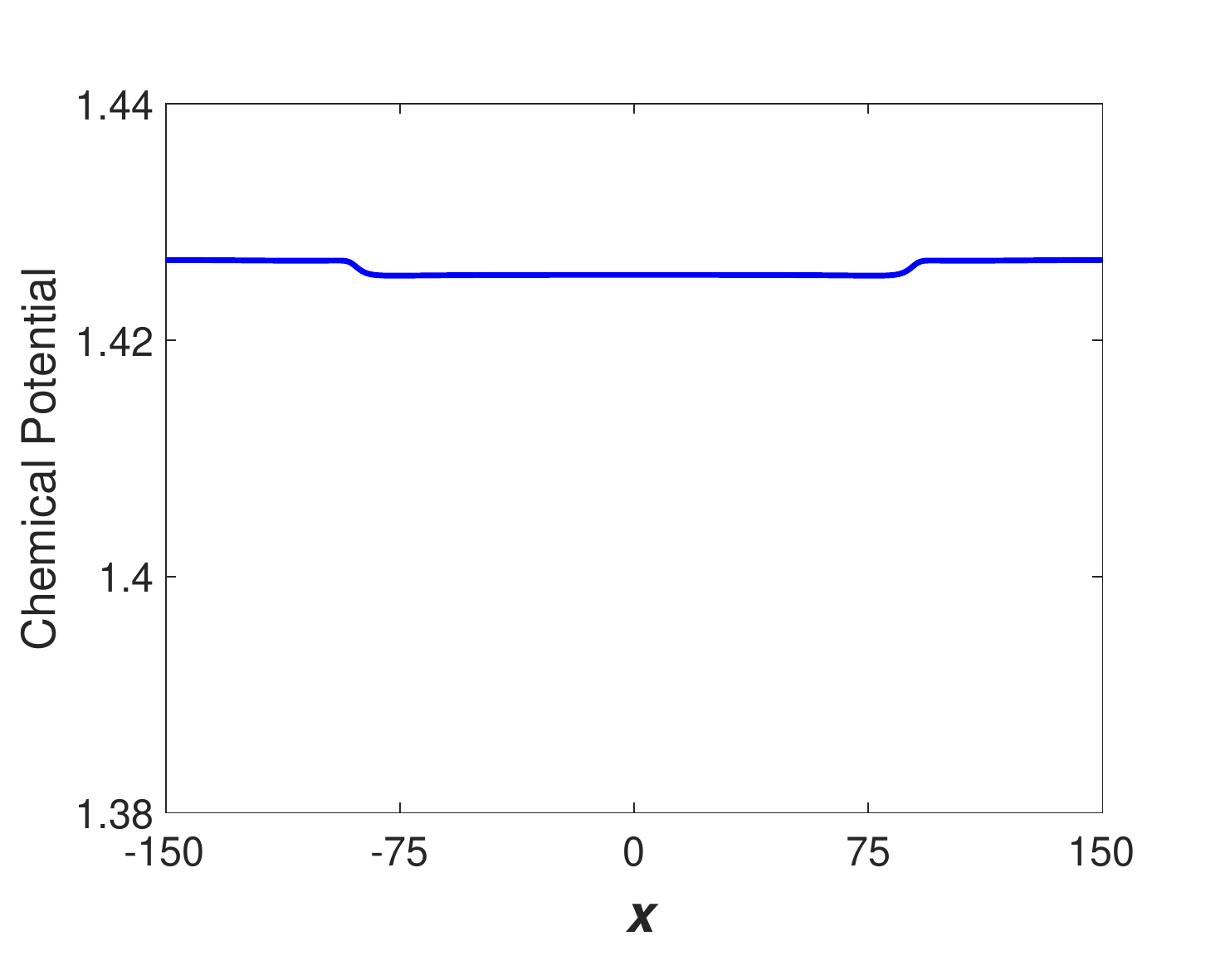}}
  \subfloat[\label{fig:ChmP-12000}$t=12000$]{\includegraphics[width=0.33\textwidth]{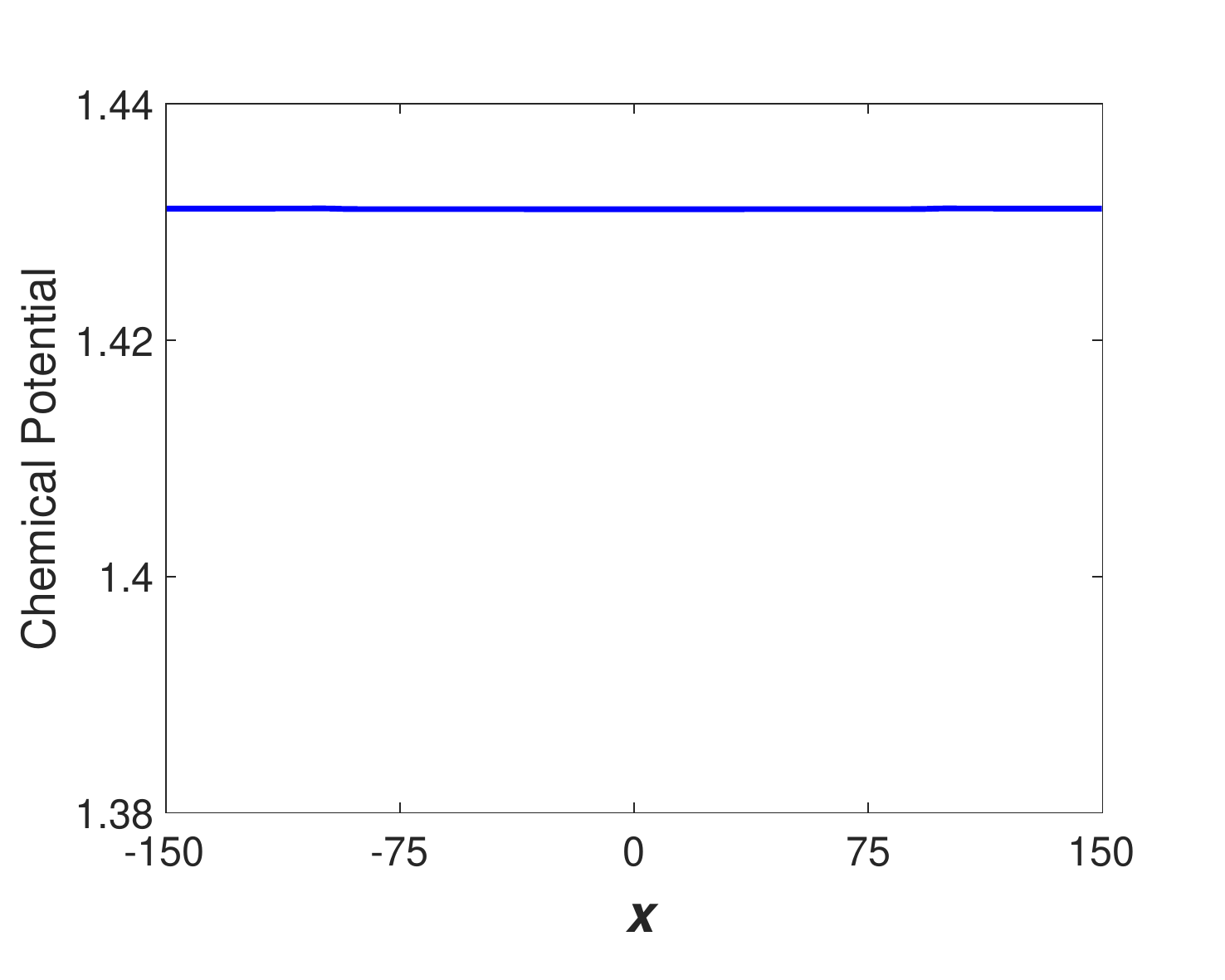}}
  \caption{\label{fig:ChmP}Cross sections of chemical potential $\mu$ at different time $t$. In each subplot, $\mu$ is shown as a function of $x$ and becomes flat when $t$ is large enough.}
\end{figure}

\begin{figure}
  \includegraphics[width=0.4\textwidth]{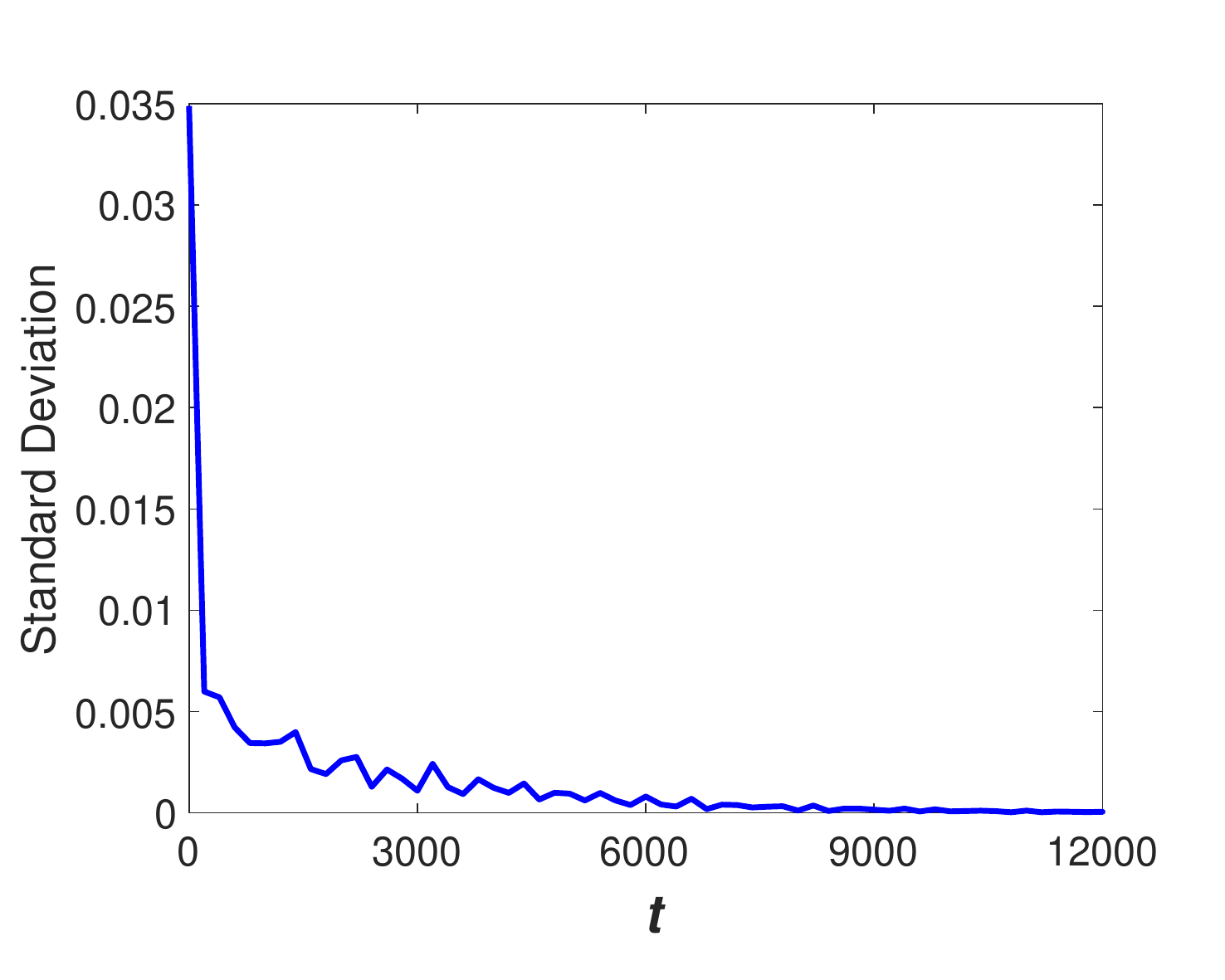}
  \caption{\label{fig:ChmP-DV}Standard deviation of chemical potential as a function of time $t$.}
\end{figure}

The requirement $A_t|_{z=1}=0$ in equation (\ref{horizon}) can be viewed as a gauge choice, in particular when we are considering dynamical evolution. Actually a more convenient (and more commonly used) gauge choice for dynamical evolution is fixing $A_t|_{z=0}$ to be a constant independent of $t$ and $\vec{x}$, since it is much easier for us to impose the source free condition on the conformal boundary under this gauge choice. For a dynamical process under the radial gauge $A_z=0$, where only the $z$-independent gauge degrees of freedom remain, the chemical potential can be defined locally as the gauge invariant expression
\begin{equation}
  \mu(t,\vec{x}) \equiv \left.A_t\right|_{z=0}(t,\vec{x})-\left.A_t\right|_{z=1}(t,\vec{x}).
\end{equation}
In Fig.~\ref{fig:ChmP}, cross sections of chemical potential at $y=0$ are plotted, which verifies theoretical prediction in Sec.~\ref{thermodynamics}, that the chemical potential must satisfy the chemical balance condition, if the system has stabilized. Numerically speaking, the static state is achieved if $\mu$ approaches a constant independent of $t$ and $\vec{x}$.
In our time evolution, this condition is specified as that the standard deviation of $\mu$  in the spatial directions is small enough. In Fig.~\ref{fig:ChmP-DV}, the standard deviation is plotted as a function of $t$, which is smaller than $1\times10^{-6}$ when the bubble stabilizes (see Appendix \ref{sec:time_evolution} for more discussion about Fig.~\ref{fig:ChmP-DV}.).

As we mentioned in Sec.~\ref{sec:calculate-r}, the radius of the bubble can be calculated directly.  To be specific, we substitute initial conditions ($\bar{\rho}=3.1$ and $V=300\times300$), pressures ($P_1(\mu)$ and $P_2(\mu)$ calculated in Fig.~\ref{FreeE}), equations of state ($\rho_1(\mu)$ and $\rho_2(\mu)$ calculated in Appendix \ref{sec:EOS})  and the surface tension $\sigma$ (calculated in equation (\ref{eq:tension})) into equations (\ref{eq:volume}-\ref{eq:balance}), from which another radius is obtained:
\begin{equation}
  r_b=95.49. \label{eq:r_b}
\end{equation}
This result from theoretical calculation matches very well with equation (\ref{eq:r}). There are mainly three factors for the deviation between $r_b$ and $r$. The first one is that we have taken the surface tension $\sigma$ at the critical chemical potential, which is not the true tension of the bubble here. The second one is the curvature effect of the domain wall, which can be estimated to be of the order
\begin{equation}
  \frac{\theta}{r_b}\approx\frac{\theta}{r}\approx 0.083,
\end{equation}
which is consistent with the order of our deviation here. And the third one comes from the fact that the thickness of the bubble is not considered in equations (\ref{eq:volume}-\ref{eq:constrain_N}), which results in deviations of volumes and particle numbers.

\section{Conclusion and discussion}\label{sec:C&D}

In this paper, we have studied the first order phase transition between two superfluid phases from the holographic point of view. First of all, taking a very simple holographic superfluid model as an example, we have established an intuitive picture for first order phase transitions, which can be made more precise by the detailed potential landscape discussion for our model. The generalized thermodynamic description, for systems out of (local) equilibrium, from holography enables us to study the domain wall structures of the system under first order phase transitions, either static or in real time dynamics.

Under the above theoretical framework, we have numerically constructed the 1D static domain wall configuration, with its surface tension computed from the integration of local grand potential density, and simulated the 2D dynamic bubble nucleation process, where the final radius of the expanding bubble can be related to the surface tension and the equations of state by the usual mechanical balance condition and the chemical balance condition, proved in this paper, for static configurations without local equilibrium. Our numerical results shows the consistency of our theoretical framework.

Our discussion in Sec.~\ref{landscape} seems to have something to do with the topic on the relation between thermodynamic and dynamical stabilities from a holographic perspective, and it does, but not in a straightforward way. The original conjecture by Gubser and Mitra in \cite{GM} about the equivalence between these two stabilities deals with thermodynamic stability in the sense of classical thermodynamics, i.e. from the positivity (or negativity) of the Hessian matrix of energy (or entropy). In this paper, our discussion is more related to thermodynamic (meta-)stability in a generalized sense, i.e. the positivity of the (functional) Hessian matrix of the generalized thermodynamic potentials with respect to field configurations, unlike the usual Hessian matrix with respect to conserved quantities (see, e.g. Eq. (8) in \cite{GM}). Specifically, we have shown that the thermodynamic (meta-)stability in such a generalized sense is equivalent to the dynamical stability (characterized by quasi-normal modes).

It is interesting to investigate further the properties of bubbles (domain walls) in such systems with first order phase transitions. An important aspect is the difference between domain walls and other local structures, especially other co-dimension one objects like solitons. Actually, solitons also have positive surface tension, but in contrast their effective mass (energy) is negative, which leads to an instability\cite{XDEMTX}. Instead, the domain walls separating different phases here are stable and tend to have the least length (area) under evolution, which implies that their effective mass should be positive. In order to confirm this inference, as well as to further investigate the relation (\ref{eq:Gibbs}), one should take into account the backreaction of the matter fields onto the bulk spacetime (just as has been done in \cite{XDEMTX}) in our model, which also leads to complicated inhomogeneous black hole configurations as higher dimensional generalizations of that in \cite{Janik:2017ykj}.

\begin{acknowledgments}
  YT would like to thank Hongbao Zhang, Hong Liu and Shan-Quan Lan for useful discussions. He would also like to thank the Center for Theoretical Physics, Massachusetts Institute of Technology for the hospitality. ZYN would like to thank Shao-Jing Qin, Zhi-Yuan Xie, Mao-Xin Liu and Shao-Jiang Wang for useful discussions. ZYN would also like to thank the organizers of ``International Conference on Multi-Condensate Superconductivity and Superfluidity in Solids and Ultra-cold Gases'' for their hospitality. XL would like to thank Zhongshan Xu and Chuan-Yin Xia for useful discussions on numerical computations. This work is partially supported by NSFC with Grant No.11675015, 11975235, 11565017, 11881240248 and 11965013. YT is partially supported by the grants (No.14DZ2260700) from Shanghai Key Laboratory of High Temperature Superconductors. He is also supported by the ``Strategic Priority Research Program of the Chinese Academy of Sciences'' with Grant No.XDB23030000. ZYN is partially supported by Yunnan Ten Thousand Talents Plan Young \& Elite Talents Project.
\end{acknowledgments}

\appendix

\section{Quasi-normal modes}\label{sec:QNM}
In order to perform the analysis of quasi-normal modes numerically, we start from the equations of motion in Eddington-Finkelstein coordinates, under which equations (\ref{eq:EOM_psi1}-\ref{eq:EOM_Ax}) are obtained as we referred before. In contrast to Sec.~\ref{sec:DomainWall_Num}, when studying the stabilities of the homogeneous solutions, we do not need to consider the $x$ direction, so equations (\ref{eq:EOM_psi1}-\ref{eq:EOM_Ax}) get simplified:
\begin{align}
  2\left(\partial_{t}-i e A_{t}\right) \partial_{z} \psi_{1}-i e \partial_{z} A_{t} \psi_{1}-\partial_{z}\left(f \partial_{z} \psi_{1}\right)+\left(z+\lambda_{12}\left|\psi_{2}\right|^{2}\right) \psi_{1}=0, \label{eq:homo-t-psi1} \\
  2\left(\partial_{t}-i A_{t}\right) \partial_{z} \psi_{2}-i \partial_{z} A_{t} \psi_{2}-\partial_{z}\left(f \partial_{z} \psi_{2}\right)+\left(z+\lambda_{12}\left|\psi_{1}\right|^{2}\right) \psi_{2}=0,                            \\
  \partial_{z}^{2} A_{t}+2 \operatorname{Im}\left(e \psi_{1}^{*} \partial_{z} \psi_{1}+\psi_{2}^{*} \partial_{z} \psi_{2}\right)=0,                                                                                                   \\
  \partial_{t} \partial_{z} A_{t}+2 \operatorname{Im}\left(e f \psi_{1}^{*} \partial_{z} \psi_{1}-e \psi_{1}^{*} \partial_{t} \psi_{1}+f \psi_{2}^{*} \partial_{z} \psi_{2}-\psi_{2}^{*} \partial_{t} \psi_{2}\right)+2 A_{t}\left(\left|e \psi_{1}\right|^{2}+\left|\psi_{2}\right|^{2}\right)=0.\label{eq:homo-t-const}
\end{align}
Similar to the case when studying the time evolution, all fields above, including $\psi_1$, $\psi_2$ and $A_t$, are related to the fields in equations (\ref{eq:homogeneous-psi1}-\ref{eq:homogeneous-At}) through coordinates transformation and the $U(1)$ gauge transformation (\ref{eq:gauge-trans-A}-\ref{eq:gauge-trans-alpha}). Linearize equations (\ref{eq:homo-t-psi1}-\ref{eq:homo-t-const}), we will be left with equations of linear perturbations $\delta \psi_1$, $\delta\psi_2$ and $\delta A_t$. Since the homogeneous solutions in Sec.~\ref{sec:Homogeneous-Solution} are time translation invariant, the perturbations can be expanded as\cite{Du:2015zcb}
\begin{align}
  \delta \psi_{1} & =p_{1} e^{-i \omega t}+q_{1}^{*} e^{i \omega^{*} t}, \label{eq:qnm_psi1} \\
  \delta \psi_{2} & =p_{2} e^{-i \omega t}+q_{2}^{*} e^{i \omega^{*} t},                     \\
  \delta A_{t}    & =a_{t} e^{-i \omega t}+a_{t}^{*} e^{i \omega^{*} t},\label{eq:qnm_At}
\end{align}
where $p_i=p_i(z)$, $q_i=q_i(z)$,  ($i=1,2$) and $a_t=a_t(z)$ are functions of $z$. Substitute the perturbations (\ref{eq:qnm_psi1}-\ref{eq:qnm_At}) into the linearized forms of equations (\ref{eq:homo-t-psi1}-\ref{eq:homo-t-const}), we obtain
\begin{align}
  \left(f \partial_{z}^{2}+f^{\prime} \partial_{z}+2 i \omega \partial_{z}+2 i e A_{t} \partial_{z}+i e \partial_{z}A_{t}-z-\lambda_{12}\left|\psi_{2}\right|^{2}\right) p_{1}                                    & \nonumber                    \\
  -\lambda_{12} \psi_{1} \psi_{2}^{*} p_{2}-\lambda_{12} \psi_{1} \psi_{2} q_{2}+\left(i e \psi_{1} \partial_{z}+2 i e \partial_{z} \psi_{1}\right) a_{t}                                                         & =0, \label{eq:qnm-p1}        \\
  \left(f \partial_{z}^{2}+f^{\prime} \partial_{z}+2 i \omega \partial_{z}-2 i e A_{t} \partial_{z}-i e \partial_{z} A_{t}-z-\lambda_{12}\left|\psi_{2}\right|^{2}\right) q_{1}                                   & \nonumber                    \\
  -\lambda_{12} \psi_{1}^{*} \psi_{2}^{*} p_{2}-\lambda_{12} \psi_{1}^{*} \psi_{2} q_{2}-\left(i e \psi_{1}^{*} \partial_{z}+2 i e \partial_{z} \psi_{1}^{*}\right) a_{t}                                         & =0,                          \\
  \left(f \partial_{z}^{2}+f^{\prime} \partial_{z}+2 i \omega \partial_{z}+2 i A_{t} \partial_{z}+i \partial_{z} A_{t}-z-\lambda_{12}\left|\psi_{1}\right|^{2}\right) p_{2} \nonumber                                                            \\
  -\lambda_{12} \psi_{1}^{*} \psi_{2} p_{1}-\lambda_{12} \psi_{1} \psi_{2} q_{1}+\left(i \psi_{2} \partial_{z}+2 i \partial_{z} \psi_{2}\right) a_{t}                                                             & =0,                          \\
  \left(f \partial_{z}^{2}+f^{\prime} \partial_{z}+2 i \omega \partial_{z}-2 i A_{t} \partial_{z}-i \partial_{z} A_{t}-z-\lambda_{12}\left|\psi_{1}\right|^{2}\right) q_{2}                                       & \nonumber                    \\
  -\lambda_{12} \psi_{1}^{*} \psi_{2}^{*} p_{1}-\lambda_{12} \psi_{1} \psi_{2}^{*} q_{1}-\left(i \psi_{2}^{*} \partial_{z}+2 i \partial_{z} \psi_{2}^{*}\right) a_{t}                                             & =0, \label{eq:qnm-q2}        \\
  -\left(i e \psi_{1}^{*} \partial_{z}-i e \partial_{z} \psi_{1}^{*}\right) p_{1}+\left(i e \psi_{1} \partial_{z}-i e \partial_{z} \psi_{1}\right) q_{1}                                                          & \nonumber                    \\
  -\left(i \psi_{2}^{*} \partial_{z}-i \partial_{z} \psi_{2}^{*}\right) p_{2}+\left(i \psi_{2} \partial_{z}-i \partial_{z} \psi_{2}\right) q_{2}+\partial_{z}^{2} a_{t}                                           & =0, \label{eq:qnm-at}        \\
  \left(-i e f \psi_{1}^{*} \partial_{z}+i e f \partial_{z} \psi_{1}^{*}+2 e^{2} A_{t} \psi_{1}^{*}+e \omega \psi_{1}^{*}\right) p_{1}+\left(i e f \psi_{1} \partial_{z}-i e f \partial_{z} \psi_{1}\right) q_{1} & \nonumber                    \\
  +\left(2 e^{2} A_{t} \psi_{1}-e \omega \psi_{1}\right) q_{1}
  +\left(-i f \psi_{2}^{*} \partial_{z}+i f \partial_{z} \psi_{2}^{*}+2 A_{t} \psi_{2}^{*}+\omega \psi_{2}^{*}\right) p_{2}                                                                                       & \nonumber                    \\
  +\left(i f \psi_{2} \partial_{z}-i f \partial_{z} \psi_{2}+2 A_{t} \psi_{2}-\omega \psi_{2}\right) q_{2}
  +\left(-i \omega \partial_{z}+2 e^{2}\left|\psi_{1}\right|^{2}+2\left|\psi_{2}\right|^{2}\right) a_{t}                                                                                                          & =0.\label{eq:qnm-constraint}
\end{align}

We impose Dirichlet boundary conditions for $p_1$, $q_1$ and $a_t$:
\begin{equation}
  \left.p_1\right|_{z=0}=0,\quad	\left.q_1\right|_{z=0}=0,\quad
  \left.a_t\right|_{z=0}=0,
\end{equation}
where the first two boundary conditions for $p_1$ and $q_1$ are consistent with that for $\psi_1$, while the third for $a_t$ is a gauge fixing condition.
As for $p_2$ and $q_2$, we demand
\begin{equation}
  \left.\left(i \omega+\partial_{z}+i \mu\right) p_{2}\right|_{z=0}=0,\quad
  \left.\left(i \omega+\partial_{z}-i \mu\right) q_{2}\right|_{z=0}=0,
\end{equation}
which are linearized forms of the boundary condition (\ref{eq:boundary-cond-psi2}), and $\mu=\left.A_t\right|_{z=0}$ is the chemical potential. Moreover, the constraint equation (\ref{eq:qnm-constraint}) at $z=0$ brings another boundary condition for $a_t$:
\begin{equation}
  -i \omega \partial_{z} a_{t}+\left(\mu \psi_{2}^{*}+i \partial_{z} \psi_{2}^{*}\right) p_{2}+\left.\left(\mu \psi_{2}-i \partial_{z} \psi_{2}\right) q_{2}\right|_{z=0}=0.
\end{equation}
Here, as the reductions of equations (\ref{eq:qnm-p1}-\ref{eq:qnm-q2}) at $z=1$ contribute another four boundary conditions (natural boundary conditions), we have gotten enough boundary conditions to solve equations (\ref{eq:qnm-p1}-\ref{eq:qnm-at}).

\begin{figure}
  \subfloat[\label{fig:qnm-S1}Solution S1] {\includegraphics[width=0.4\textwidth]{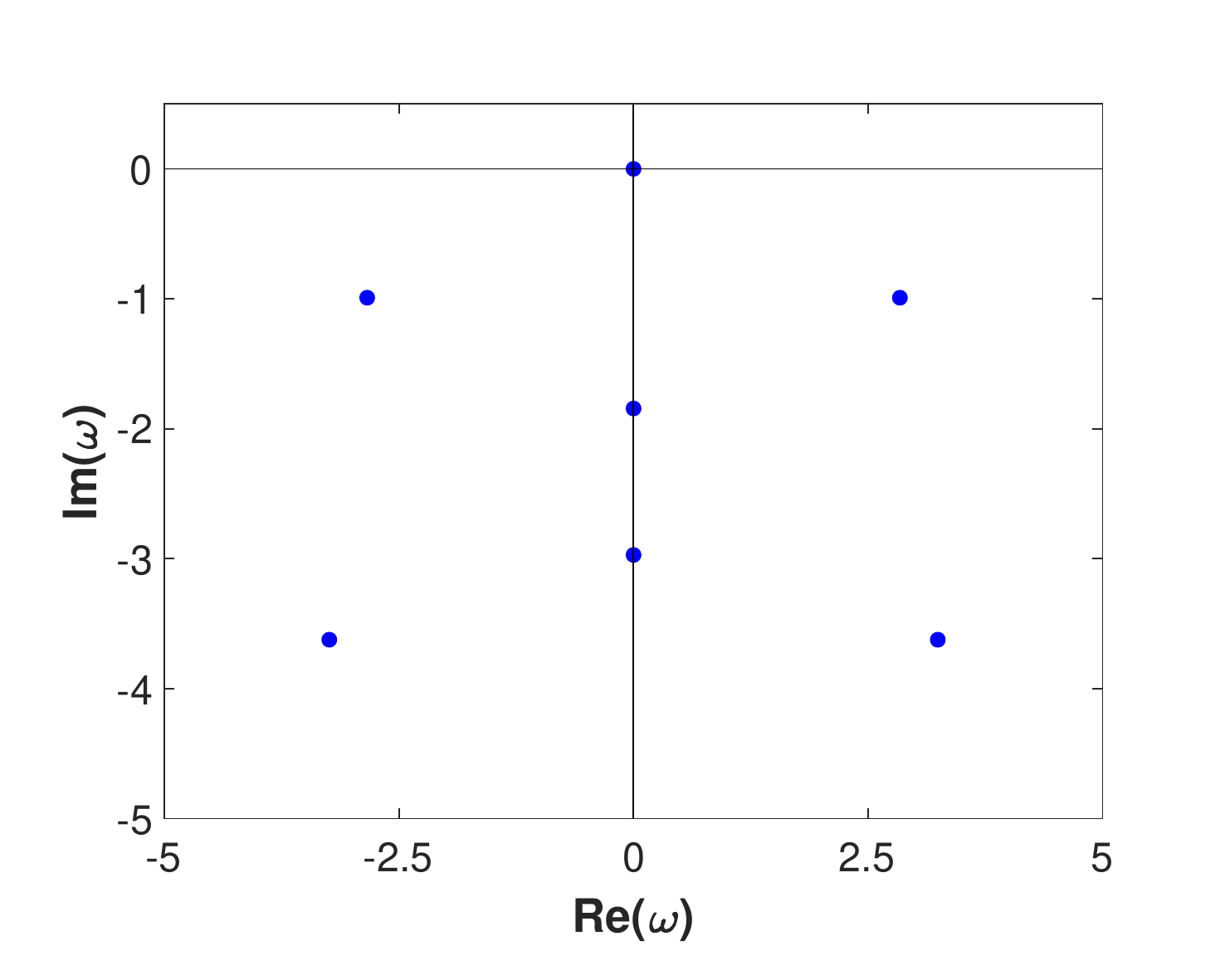}}
  \subfloat[\label{fig:qnm-S2}Solution S2] {\includegraphics[width=0.4\textwidth]{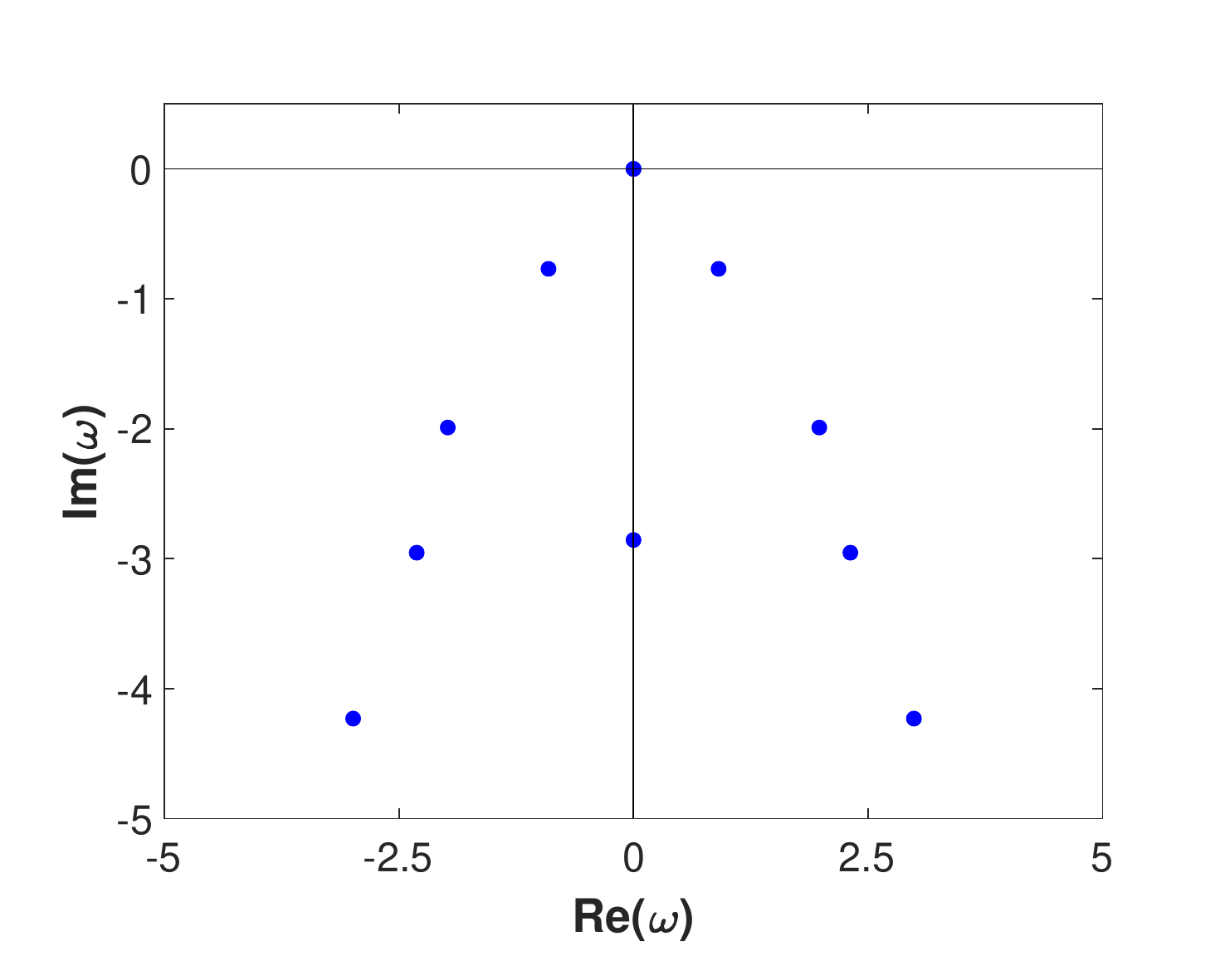}}\\
  \subfloat[\label{fig:qnm-S1+S2}Solution S1+S2] {\includegraphics[width=0.4\textwidth]{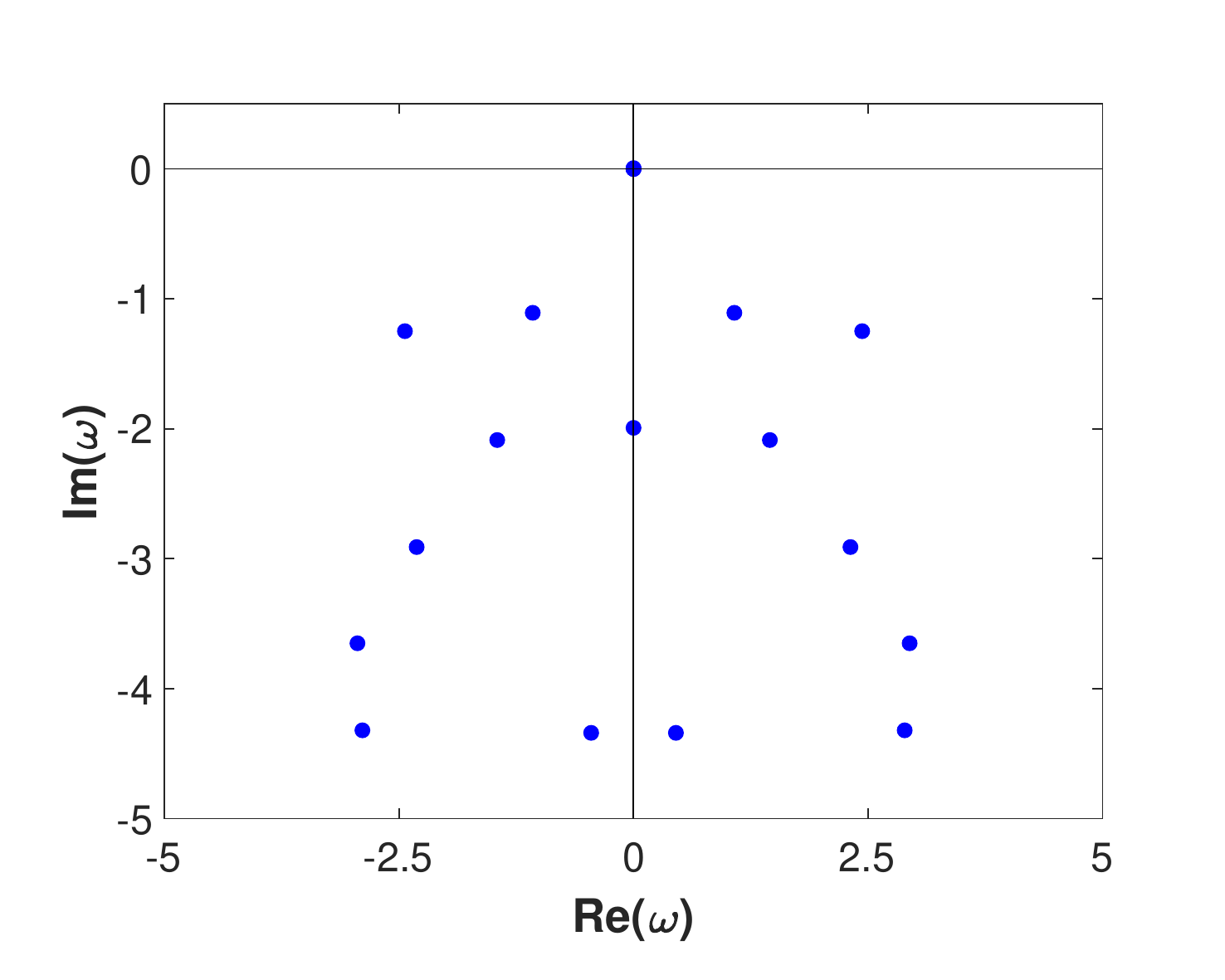}}
  \subfloat[\label{fig:qnm-S1+S2-2}Enlarged drawing of the central area of Solution S1+S2] {\includegraphics[width=0.4\textwidth]{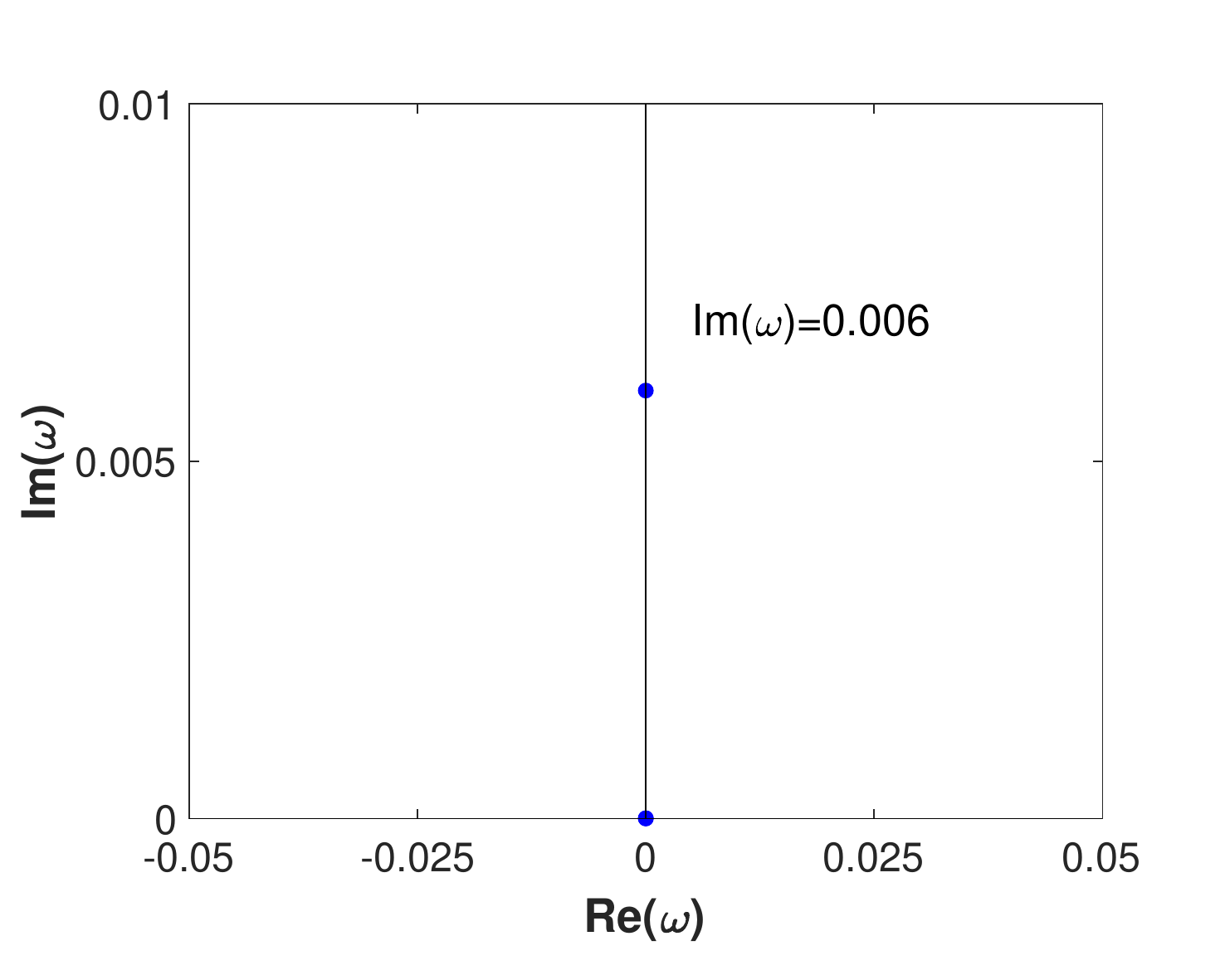}}
  \caption{\label{fig:qnm}Results of quasi-normal modes in the form of $\operatorname{Re}(\omega)$ vs. $\operatorname{Im}(\omega)$, when $\rho=2.94$ and $\lambda_{12}=0.4$.}
\end{figure}

Transform the three homogeneous solutions, Solution S1, Solution S2 and Solution S1+S2, into their proper forms as we mentioned in the last paragraph, and then submit them into equations (\ref{eq:qnm-p1}-\ref{eq:qnm-at}), $\omega$ can be solved numerically. And, from perturbations (\ref{eq:qnm_psi1}-\ref{eq:qnm_At}), it is obvious that if there exists an $\omega$ whose imaginary part $\operatorname{Im}(\omega)>0$, the corresponding solution is unstable, otherwise it is stable. Our results are shown in Fig.~\ref{fig:qnm}, where we choose $\rho=2.94$ and $\lambda_{12}=0.4$. In both Fig.~\ref{fig:qnm-S1} and Fig.~Fig.~\ref{fig:qnm-S2}, imaginary parts of all solutions of $\omega$ are equal or less than $0$, so it confirms us that these two solutions are stable or meta-stable. In Fig.~\ref{fig:qnm-S1+S2}, the Solution S1+S2 is studied, which seems to be stable as the former two solutions. However, when looking at the enlarged drawing of its central area (see Fig.~\ref{fig:qnm-S1+S2-2}), it is obvious that an $\omega$ with imaginary part greater than $0$ ($\approx0.0026$) exists, so the Solution S1+S2 is unstable.

\begin{figure}
  \centering
  \includegraphics[width=0.4\textwidth]{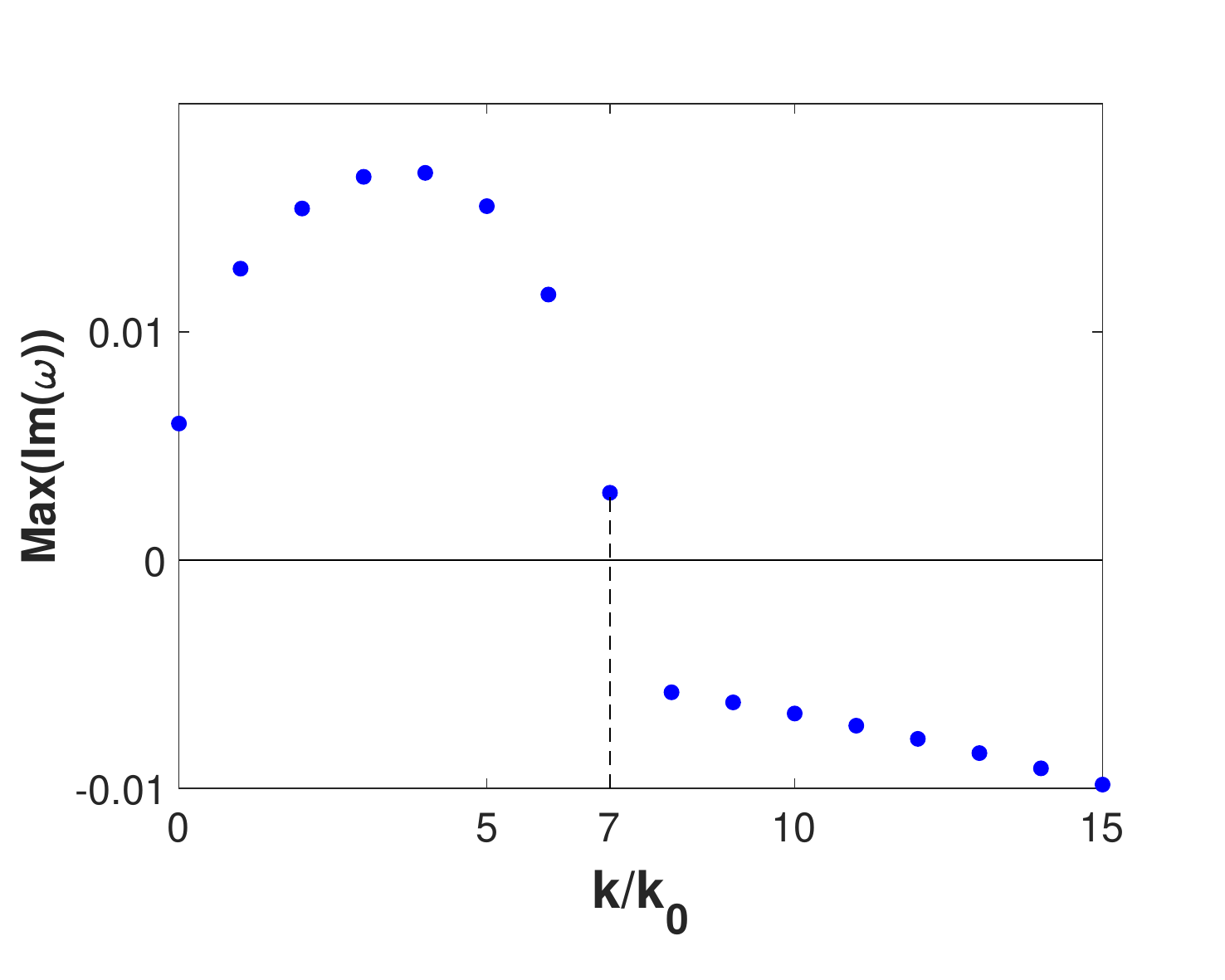}
  \caption{\label{fig:qnm-omega-vs-k}The maximum of $\operatorname{Im}(\omega)$ as a function of $k/k_0$ at $\rho=2.94$, where $k_0=2\pi/L$.}
\end{figure}

The instability from non-zero momentum can be considered similarly, and the perturbations are
  \begin{align}
    \delta \psi_{1} & =p_{1} e^{-i \omega t+ikx}+q_{1}^{*} e^{i \omega^{*} t-ikx}, \\
    \delta \psi_{2} & =p_{2} e^{-i \omega t+ikx}+q_{2}^{*} e^{i \omega^{*} t-ikx}, \\
    \delta A_{t}    & =a_{t} e^{-i \omega t+ikx}+a_{t}^{*} e^{i \omega^{*}t-ikx},  \\
    \delta A_{x}    & =a_{x} e^{-i \omega t+ikx}+a_{x}^{*} e^{i \omega^{*}t-ikx},
  \end{align}
  where $k=nk_0$, $k_0=2\pi/L$ and $n$ is integer.
  Substitute perturbations above into the perturbative form of equations (\ref{eq:EOM_psi1}-\ref{eq:EOM_Ax}), and we can similarly analyze instabilities for the three solutions. In Fig.~\ref{fig:qnm-omega-vs-k}, the maximum of $\operatorname{Im}(\omega)$ for Solution S1+S2 is plotted as a function of $k$. In this plot, the Solution S1+S2 is unstable under perturbations when $|k|\leq 7k_0$, where the absolute value sign enters because the system is symmetric under the transformation $k\to-k$.

\section{Equations of state}\label{sec:EOS}
\begin{figure}
  \centering
  \includegraphics[width=0.4\textwidth]{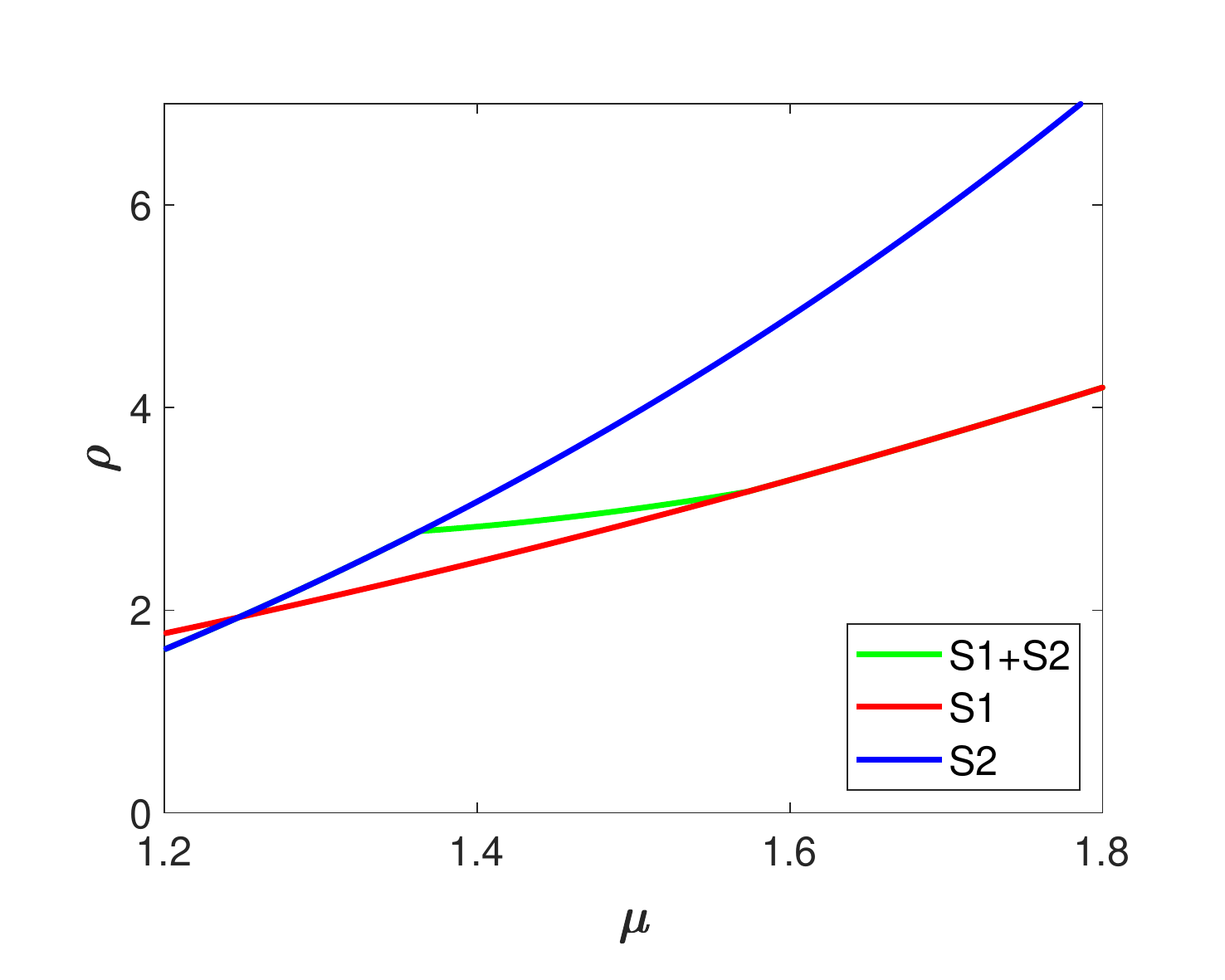}
  \caption{Equations of state for S1, S2 and S1+S2.}\label{fig:EoS}
\end{figure}

Once the chemical potential $\mu$ is given, we are able to solve static equations (\ref{eq:homogeneous-psi1}-\ref{eq:homogeneous-At}) numerically, from which density $\rho$ ($=-\left.\partial_z A_t\right|_{z=0}$) is obtained. In this way, we express $\rho$ in terms of $\mu$, and thus obtain numerical solutions to equations of state (\ref{eq:EOS_1}) and (\ref{eq:EOS_2}). Our results are shown in Fig.~\ref{fig:EoS}, where the blue, red and green curves correspond to equations of state for S1, S2 and S1+S2.

\section{\label{sec:time_evolution} Note on time evolutions}

In order to best illustrate our scheme for  time evolutions, let us take the case of the 1D domain wall as an example and compare ours\cite{Du:2015zcb,DNTZ} with a former one. For the former scheme, $A_t$ is calculated from equation (\ref{eq:constraint}), while $\psi_1$, $\psi_2$ and $A_x$ are obtained from  (\ref{eq:EOM_psi1}), (\ref{eq:EOM_psi2}) and (\ref{eq:EOM_Ax}) respectively. In this scheme, equation (\ref{eq:EOM_At}) is input as an initial condition and satisfied merely at $t=0$. This scheme works well, if no time accumulation error exists, which, however, cannot be avoided in time evolutions, so a larger deviation from equation (\ref{eq:EOM_At}) will be witnessed if a longer evolution is performed. In our scheme, all fields are obtained from the same equations except for $A_t$. We calculate $A_t$ from  equation (\ref{eq:EOM_At}) and take the restriction of equation (\ref{eq:constraint}) on the conformal boundary as a boundary condition for $A_t$. In this way, the largest deviation comes from the constraint of equation (\ref{eq:constraint}) at horizon, which is a space accumulation error and can be easily controlled. This new scheme is employed for evolutions of both 1D domain wall and 2D bubble.

There are many ways to obtain the 1D domain wall structure, such as perturbing the unstable states and quenching the meta-stable state, the method we adopt to obtain the 2D bubble in Sec.~\ref{sec:Bubbles_Num}. As we are interested in the final configurations of domain walls instead of their formation processes, obtaining them through perturbing the unstable states (Solution S1+S2) with the perturbation (\ref{eq:pert_rho}), corresponding to the unstable mode $k=\pm k_0$ ($k_0=2\pi/L$), is more efficient. As we mentioned before, $t$, $z$ and $x$ directions are needed to obtain the 1D domain walls. We adopt the Chebyshev spectrum for the non-periodic direction $z$ and Fourier spectrum for the periodic direction $x$. The lengths for directions $z$ and $x$ are $1$ and $200$, and numbers of grid points are $40$ and $800$ respectively. For the $t$ direction, the Runge-Kutta method is chosen, and the step $\Delta t=0.05$.

The formation process of 2D bubble is important, so we add a quench to the meta-stable state, the solution S2  in our case, by adding sharp structures to $\psi_1$ and $\psi_2$ locally:
\begin{align}
  \delta \psi_1 & =1-\cosh^{-1}\left( \left[\cos\left(\frac{\pi}{2 L_0^2}\left( x_0^2+y_0^2\right)+1\right] \right)^{n}\right), \\
  \delta \psi_2 & =\cosh^{-1}\left( \left[\cos\left(\frac{\pi}{2 L_0^2}\left( x_0^2+y_0^2\right)+1\right] \right)^{n}\right),
\end{align}
where we choose $n=10$, $x_0, y_0 \in (-L_0/2,L_0/2)$, $L=300$, and $L_0/L\approx0.13$. Similar to the 1D case, Chebyshev spectrum is chosen for the non-periodic direction $z$, and Fourier spectrums are chosen for the periodic directions $x$ and $y$. The numbers of grid points in $z$, $x$ and $y$ directions are $30$, $601$ $601$ respectively\footnote{In fact, it is not necessary to adopt so large a grid, because the time evolution works when the grid is larger than $25\times 300\times 300$, and so does the 1D case.}. And the step $\Delta t$ for the $t$ direction is chosen to be $0.08$. This rather large grid demands a powerful computing system, so we utilize the GPU computing, which enables us to complete a full evolution in several hours.

In Fig.~\ref{fig:ChmP-DV}, the standard deviation of chemical potential seems to oscillate numerically, which, however, appears due to the periodic boundary conditions in spatial directions. Because of the periodic boundary conditions, the whole system is in fact infinite numbers of repeated regions with length $L$. Thus, we add an inhomogeneous structure to each region at $t=0$ of time evolutions, which results in propagation of some wave-like structures when an evolution starts. Therefore, the standard deviation of chemical potential increases (decreases) when superpositions of waves from different regions strengthen (weaken) the amplitude. This causes the observed oscillation of the standard deviation of chemical potential.

\section{Concrete form of grand potential}\label{sec:GrandP}

The Noether current reads\cite{TWZ}
\begin{equation}
  J^{a} =\sum_{i=1}^{2}\left[\frac{\partial L_m}{\partial\left(\partial_{a} \Psi_{i}\right)} \mathcal{L}_{\xi} \Psi_{i}+\frac{\partial L_m}{\partial\left(\partial_{a} \Psi_{i}^{*}\right)} \mathcal{L}_{\xi} \Psi_{i}^{*}\right]+\frac{\partial L_m}{\partial\left(\partial_{a} A_{b}\right)} \mathcal{L}_{\xi} A_{b}-\xi^{a}L_m,
\end{equation}
where $\mathcal{L}_{\xi}$ is the Lie derivative with respect to $\xi=(\partial/\partial t)$ and $L_m$ the Lagrangian in (\ref{eq:Lagrangian}). The grand potential is defined as
\begin{align}
  \Omega= & \int \sqrt{(-g)} d^3x J^t
  +\int \mathrm{d}^{2} x \sqrt{-\gamma} n^{M}\left[\left(D_{2 M} \Psi_{2}\right)^{*} \Psi_{2}+\Psi_{2}^{*}\left(D_{2 M} \Psi_2\right)\right] \nonumber \\
          & +\int \mathrm{d}^{2} x \sqrt{-\gamma}\left(\left|\Psi_{2}\right|^{2}-\left|\Psi_{1}\right|^{2}\right),\label{eq:GrandP}
\end{align}
where the gauge $\left.A_t\right|_{z=1}=0$ should be adopted, the last two terms are counter terms, and $\gamma_{\mu\nu}$ is the induced metric. Expanding all the indices in (\ref{eq:GrandP}) and apply boundary equations for all the fields, we can reach the final form of the grand potential:
\begin{align}
  \Omega & =\int d^{3} x\left[\partial_{t} A_{x} F_{z x}+\partial_{t} A_{y} F_{z y}-\frac{1}{2} \partial_{t}\left(A_{x} F_{z x}+A_{y} F_{z y}\right)-\lambda_{12}\left|\psi_{1}\right|^{2}\left|\psi_{2}\right|^{2}\right.\nonumber                               \\
         & +2 \operatorname{Re}\left(\partial_{t} \psi_{1} \partial_{z} \psi_{1}^{*}\right)+e \operatorname{Im}\left(-A_{t} \psi_{1}^{*} \partial_{z} \psi_{1}+A_{x} \psi_{1}^{*} D_{1 x} \psi_{1}+A_{y} \psi_{1}^{*} D_{1 y} \psi_{1}\right)\nonumber            \\
         & \left.+2 \operatorname{Re}\left(\partial_{t} \psi_{2} \partial_{z} \psi_{2}^{*}\right)+\operatorname{Im}\left(-A_{t} \psi_{2}^{*} \partial_{z} \psi_{2}+A_{x} \psi_{2}^{*} D_{2 x} \psi_{2}+A_{y} \psi_{2}^{*} D_{2 y} \psi_{2}\right)\right]\nonumber \\
         & -\left.\int d^{2} x\left[\frac{1}{2}\left(-A_{t} F_{t z}+A_{x} F_{t x}+A_{y} F_{t y}\right)+\partial_{t}\left(\left|\psi_{1}\right|^{2}+\left|\psi_{2}\right|^{2}\right)\right]\right|_{z=1}\nonumber                                                  \\
         & +\left.\int d^{2} x\left\{\frac{1}{2}\left[-A_{t} F_{t z}+A_{x}\left(F_{t x}-F_{z x}\right)+A_{y}\left(F_{t y}-F_{z y}\right)-\partial_{z}\left|\psi_{2}\right|^{2}\right]+\partial_{t}\left|\psi_{2}\right|^{2}\right\}\right|_{z=0},
\end{align}
where we have adopted $\psi_i=\Psi_i/z$.

\end{document}